\documentclass[a4paper,fleqn,usenatbib,useAMS]{mnras}

\usepackage{graphicx}	% Including figure files
\usepackage{amsmath}	% Advanced maths commands
\usepackage{amssymb}	% Extra maths symbols
\usepackage{multicol}   % Multi-column entries in tables
\usepackage{bm}		% Bold maths symbols, including upright Greek
\usepackage{pdflscape}	% Landscape pages
\usepackage{subfig}     %for subfigers
\usepackage{xcolor}		% Colour text

\newcommand{\e}[1]{\times 10^{#1}}
\newcommand{\degree}{^{\circ}}

\usepackage[T1]{fontenc}
\usepackage{ae,aecompl}

\usepackage{newtxtext,newtxmath}

\usepackage{subfiles} 
\title[Astrocladistics of the Trojans]{Astrocladistics of the Jovian Trojan Swarms}

\author[T. R. Holt et al.]{
Timothy R. Holt,$^{1,2}$\thanks{E-mail: timothy.holt@usq.edu.au (TRH)}
Jonathan Horner,$^{1}$
David Nesvorn{\'y},$^{2}$
Rachel King,$^{1}$
\newauthor
Marcel Popescu,$^{3}$
Brad D. Carter,$^{1}$
and Christopher C. E. Tylor,$^{1}$
\\
$^{1}$Centre for Astrophysics, University of Southern Queensland, Toowoomba, QLD, Australia\\
$^{2}$Department of Space Studies, Southwest Research Institute, Boulder, CO. USA.\\
$^{3}$Astronomical Institute of the Romanian Academy, Bucharest, Romania. 
}

\date{Accepted XXX. Received YYY; in original form ZZZ}

\pubyear{2020}

\begin{document}
\label{firstpage}
\pagerange{\pageref{firstpage}--\pageref{lastpage}}
\maketitle

% Abstract of the paper
%250 words
\begin{abstract}
%200 words
The Jovian Trojans are two swarms of small objects that share Jupiter's orbit, clustered around the leading and trailing Lagrange points, L$_4$ and L$_5$. In this work, we investigate the Jovian Trojan population using the technique of astrocladistics, an adaptation of the `tree of life' approach used in biology. We combine colour data from \textit{WISE}, SDSS, \textit{Gaia} DR2 and {\tt MOVIS} surveys with knowledge of the physical and orbital characteristics of the Trojans, to generate a classification tree composed of clans with distinctive characteristics. We identify 48 clans, indicating groups of objects that possibly share a common origin. Amongst these are several that contain members of the known collisional families, though our work identifies subtleties in that classification that bear future investigation. Our clans are often broken into subclans, and most can be grouped into 10 superclans, reflecting the hierarchical nature of the population.  Outcomes from this project include the identification of several high priority objects for additional observations and as well as providing context for the objects to be visited by the forthcoming \textit{Lucy} mission. Our results demonstrate the ability of astrocladistics to classify multiple large and heterogeneous composite survey datasets into groupings useful for studies of the origins and evolution of our Solar system.

\end{abstract}

%Clear take home message. 

% Select between one and six entries from the list of approved keywords.
% Don't make up new ones.

%%%%%%%%%%%%%%%%%%%%%%%%%%%%%%%%%%%%%%%%%%%%%%%%%%

%%%%%%%%%%%%%%%%% BODY OF PAPER %%%%%%%%%%%%%%%%%%

% \section*{Notes to Co-Authors}
% \begin{itemize}
%     \item \thcoms{Just a place to put notes and will be removed prior to submission} 

%     %Figures: Fig. \ref{fig:1990VU1}\\
%     %Tables: Table \ref{tab:L4}\\
%     %Equation: equation \ref{Equ:Dest}\\
%     %Section: section \ref{SubSec: L4}\\
    
%     %Asteroid names: #### NAME (19XX ##$_{xx}$) eg. 7152 Euneus (1973 SH$_1$). 
    
%     %\item \jhcom{Note - in the main text, anything in {\bf bold red text}, that starts with Jonti: or JH: is a comment, rather than text to add}
% \end{itemize}

\section{Introduction}

At the Jovian Lagrange points, $60\degree$ ahead (L$_4$) and behind (L$_5$) the giant planet in its orbit, there are two swarms of small Solar system objects, collectively termed the Jovian Trojans. Members of the leading swarm, which librate around Jupiter's L$_4$ Lagrange point, are named after the Greek heroes in the Iliad \citep{Nicholson1961TrojanAsteroids}, with members of the trailing swarm being named for the Trojan heroes. The first Jovian Trojans, 588 Achilles (1906 TG), 617 Patroclus (1906 VY), 624 Hektor (1907 XM) and 659 Nestor (1908 CS) were discovered in the early 20th century \citep{Wolf1907588Achillies, Heinrich1907617Patroclus, Stromgren1908624Hektor, Ebell1909659Nestor}. In the decades that followed, the number of known Trojans grew slowly, as a result of ongoing work at the Heidelberg observatory \citep[e.g.][]{Nicholson1961TrojanAsteroids,Slyusarev2014JupiterTrojens}. With the advent of CCD imaging, in the later part of the 20th Century, the rate at which Trojans were discovered increased markedly, such that, by the end of the century, a total of 257 had been confirmed \citep{Jewitt2000JovTrojan}. 

\begin{figure}
	\includegraphics[width=\columnwidth]{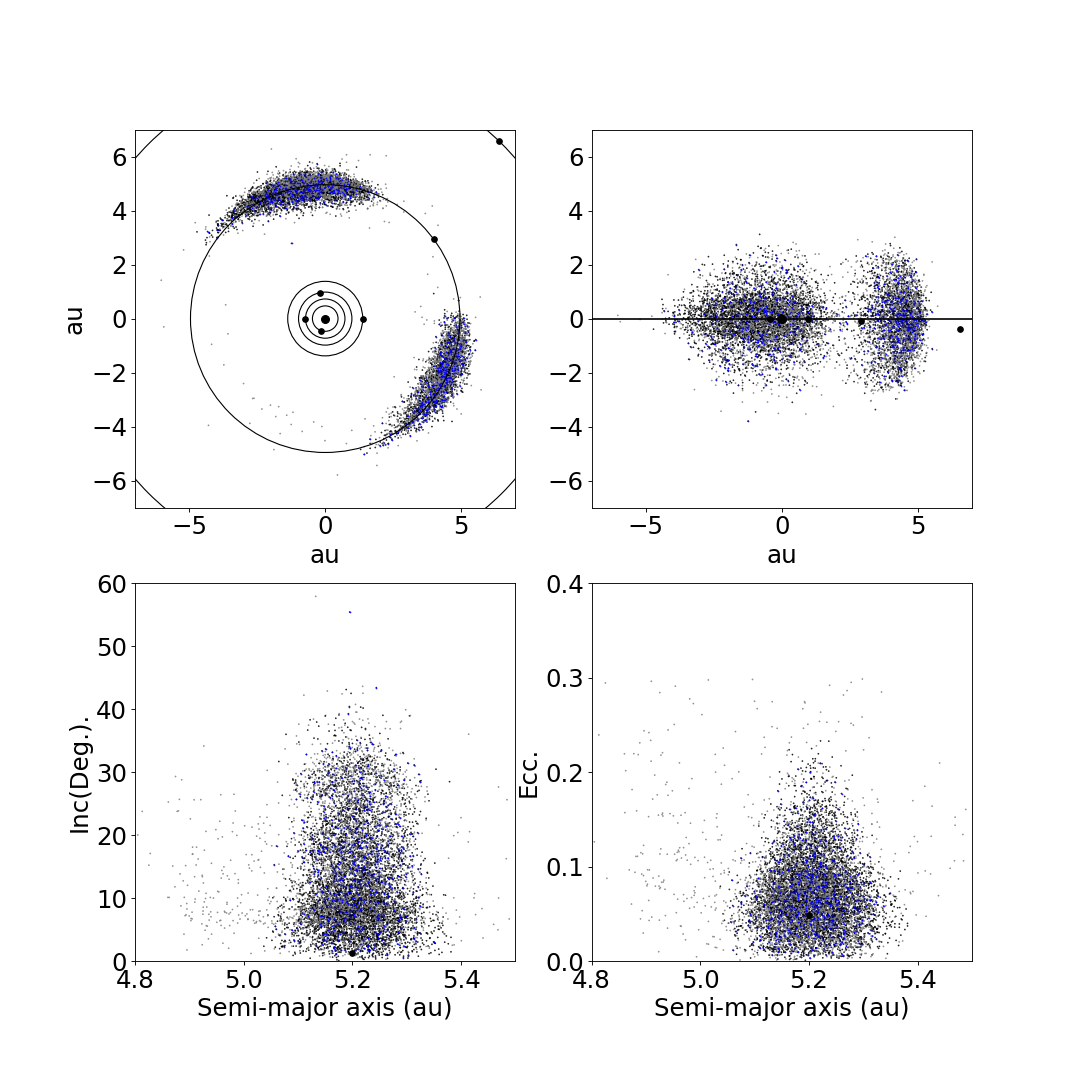}
    \caption{Distribution of the Jovian Trojans after \citet{Horner2020SSforExo}. The upper panels show the positions of the Trojans  relative to the planets on 2000 January 1, 00:00 in XY (left) and XZ (right) planes. The lower panels show the Trojans in semi-major axis vs inclination space (left) and semi-major axis vs eccentricity space (right). All data from NASA {\tt HORIZONS} \citep{Giorgini1996JPLSSdatabase}, access on 2020 October 13. Black points are initially stable objects, from the {\tt AstDyS} \citep{Knezevic2017AstDysTrojans} dataset. Grey points are potentially transient objects. Blue points identify those objects used in this work. }
    \label{Fig:TrojansSub}
\end{figure}

Over the last twenty years, the rate at which Jovian Trojans have been discovered has increased still further, as a result of new instrumentation and automated surveys coming online to scour the skies. As a result, more than 8700 Jovian Trojans have been discovered to date\footnote{Taken from the NASA-JPL {\tt HORIZONS} Solar system Dynamics Database \url{https://ssd.jpl.nasa.gov/} \citep{Giorgini1996JPLSSdatabase} taken 13th October, 2020 October 13}. We show the current distribution of objects around the Jovian Lagrange points in Fig. \ref{Fig:TrojansSub}. Whilst objects can still be temporarily captured to orbits within the Trojan clouds, without the destabilisation of the clouds caused by planetary migration, such captures are very short-lived \citep[e.g. 2019 LD$_2$][]{Hsieh20202019ld2, Steckloff20202019ld2, Bolin2020p2019ld2}. At any time, it is likely that there are a number of such `temporary Trojans', whose residence in the swarms can be measured in years, decades, or centuries at most \citep[e.g.][]{Horner2006CentaurCapTrojans}. 

It is now well established that the Jovian Trojans did not form in their current orbits \citep[see][for review]{EmeryAsteroidsIVJupTrojan}. Instead, they are thought to have been captured as a byproduct of the migration of Jupiter. Such capture would require some mechanism by which the Trojans could become trapped in such dynamically stable orbits. 

One leading theory to explain the capture of the Jovian Trojans is the proposed period of instability in the early Solar system \citet{Nesvorny2013TrojanCaptureJJ} that has come to be known as the `Nice' model \citep{Tsiganis2005NICEplanets,Morbidelli2010NiceReview, Levison2011Nice,Nesvorny2012JumpingJupiter, Nesvorny2018SSDynam}. The Nice model invokes a period of instability triggered by the slow migration of Jupiter and Saturn, in response to their interactions with the debris left behind from planet formation. Eventually, that migration drove the two planets into an unstable architecture, leading to a period of chaotic evolution for objects throughout the Solar system. During that period of instability, the Jovian Trojan clouds would also have been destabilised. As a result, some of the debris being flung around the system by the migrating giant planets would have experienced temporary capture to the Jovian Trojan clouds. As Jupiter and Saturn migrated away from the location of the instability, the Jovian Trojan clouds would have become stable once again, freezing in place those temporarily captured Trojans, making their capture permanent \citep{Roig2015JJSAstSims}. More recently, it has been suggested that the required instability in the outer Solar system may have been triggered by the ejection of a fifth giant planet \citep{Nesvorny2012JumpingJupiter,Deienno2017PlanetInstability} from the Solar system. This scenario has become known as the Jumping-Jupiter model, and has been invoked to explain a number of peculiarities in the distribution of Solar system small bodies, including the origin of the Jovian Trojans. 

A recent alternative to the scenarios painted above proposes instead that the Trojans were captured from the same region of the Solar system's protoplanetary disc as Jupiter, and were both captured and transported during the planet's proposed inward migration \citep{Pirani2019PlanetMigrationSSB}. A recent update to this \textit{in-situ} transport model \citep{Pirani2019TrojanInc} explains the observed excitation in the orbital inclinations of the Jovian Trojans, which is a natural byproduct of the chaotic evolution proposed in the Nice and Jumping-Jupiter models, by invoking mixing in the Jovian feeding region. Therefore, the observed inclinations are considered to be primordial in these simulations, and are preserved during transportation as Jupiter migrates. In contrast to the idea that the captured Trojans formed on inclined orbits, earlier studies of smooth, non-chaotic migration \citep[e.g.][]{Lykawka2010TrojanCapMigration} showed that Jupiter could capture a significant population of Trojans. The common feature of all of the proposed capture models, however, is that the capture of the  Jovian Trojans occurred during the Solar system's youth \citep{EmeryAsteroidsIVJupTrojan}. These two competing theories for the origins of the Trojans highlight the importance of the population in our understanding of the early Solar system.

\subsection{Taxonomy and wide Field Surveys}\label{intro:Tax}
The methods by which the Solar system's small bodies are classified, can be broken down into two broad categories. First, the objects are grouped based on their orbital parameters, in combination with any evidence of cometary activity, into broad dynamical clusters \citep[Near-Earth Asteroid; Main Belt Asteroid; Centaur etc. see][for review]{Horner2020SSforExo}. Those objects can then be further classified based on their visual and infrared spectra. This classification is useful as the resulting taxonomy can indicate that certain objects share a common origin. 

Building on an original taxonomy by \citet{Tholen1984Taxonomy, Tholen1989Taxonomy}, the modern iteration of this observationally-motivated categorisation is based on the works of \citet{Bus2002AsteroidTax} and \citet{DeMeo2009AsteroidTax,DeMeoAsteroidsIVTaxonomy}, and is collectively termed the Bus-DeMeo taxonomy \citep[see ][ for summary]{DeMeoAsteroidsIVTaxonomy}. In this taxonomy, spectra are used to place objects into categories known as 'types'. Each type reflects a major compositional category, for example, the C-types are the most numerous and correspond to Carbonaceous chondrite meteorites. Since the Bus-DeMeo taxonomy requires spectral information in order to classify asteroids, its use is naturally limited to those objects bright enough for such data to have been obtained -- either through wide-field surveys, or targeted observations. As a result, to date less than 1 per cent of the Trojan population have been officially classified under this scheme. In the initial \citet{Tholen1984Taxonomy, Tholen1989Taxonomy} data-set, 22 Trojans were classified, with a further 12 in the Small Solar system Object Spectral Survey (S$^3$OS$^3$) \citep{Lazzaro2004s3os2asteroids}. In these initial surveys, D-types (85.29 per cent) were found to dominate the population. This is consistent with the dynamical modeling, as the D-types are thought to have formed in the outer solar system \citep{Morbidelli2005TrojanCapture,Levison2009Contamination} and those found in the Main belt are interlopers \citep{DeMeo2014MBDtypes}.

Two large members of the Trojan population, 617 Patroclus (1906 VY) and 588 Achilles (1906 TG), were initially classified as P-type objects, though in recent years, that category (P-type) has been degenerated into the X-types \citet{Bus2002AsteroidTax, DeMeo2009AsteroidTax}. For this work, we substitute any members of the `P-type' from their original works into the X-types, including the hybrid `DP-type' (now DX-type) and `PD-type' (now XD-type). Amongst the small number of Trojans classified in those initial studies, the population was found to include another X-type, 3451 Mentor (1984 HA$_1$), and an Xc-type, 659 Nestor (1908 CS), as well as two C-types, namely 4060 Deipylos (1987 YT$_1$) and 1208 Troilus (1931 YA).

Following these initial spectral surveys, \citet{Bendjoya2004JTSpectra} investigated 34 Trojans spectrally between 0.5 and 0.9 $\mu m$, finding again that the majority were D-type (70.6 per cent), with several X-types (11.7 per cent) and C-types (5.8 per cent). There were two objects, 7641 (1986 TT$_6$) and 5283 Pyrrhus (1989 BW), that showed a negative slope and were not classified, although 7641 (1986 TT$_6$) was later classified as as D-type by \citet{Hasselmann2012SDSSTaxonomy}, based on new observations. In a larger set of visual spectral surveys, \citet{Fornasier2004L5TrojanSpec, Fornasier2007VisSpecTrojans} examined a further 80 Jovian Trojans, and added their classifications. Though these classifications comprise a total of just 2.14 per cent of the Trojan population, they can still provide indications of the compositional distribution of the population as a whole.

In recent years, a number of studies have begun to gather data on the colour and physical properties of the Trojans. Wide-band surveys can give indications of taxonomic classification, circumventing the need for full spectra to be obtained of object. Several studies have investigated the colours of the Jovian Trojans \citep[e.g.][]{Emery2003TrojanIR, Dotto2006TrojanSurface}. Once again, the initial observations were limited in number, yielding data for less than 100 objects in the Jovian swarm in the infrared \citep{Emery2003TrojanIR, Emery2006SpectraJovianTrojan, Emery2011JovTrojanIRComp}, visual \citep{Fornasier2004L5TrojanSpec,Dotto2006TrojanSurface,Fornasier2007VisSpecTrojans} and broadband \textit{UBVRI} \citep{Karlsson2009UVBRIJovTrojan}. As in the prior studies, these initial surveys found that the majority of objects studied were best classified as D-types.

With the current generation of large ground-based facilities and space telescopes, recent years have seen a significant increase in the numbers of Trojans being observed and given preliminary classifications. \citet{Grav2012JupTrojanWISE} observed 557 Trojans at infrared wavelengths, using two \textit{Wide-field Infrared Survey Explorer} (\textit{WISE}) filters. In doing so, they confirmed the prevalence of D-types in the Trojan population, with such objects dominating both the L$_4$ and L$_5$ swarms, independent of the size of the Trojans studied \citep{Grav2011JupTrojanWISEPrelim}. \citet{Grav2011JupTrojanWISEPrelim, Grav2012JupTrojanWISE} noted that the population in the \textit{WISE} dataset was quite heterogeneous, with a mean albedo of $0.07 \pm 0.03$. In the visual five-band Sloan Digital Sky Survey (SDSS) catalogue \citep{Carvano2010SDSSAstTax, Hasselmann2012SDSSTaxonomy}, a total of 461 Trojans have been classified. Unlike previous surveys, the catalogue includes a measure of the confidence in the assigned taxonomy. Of the 461 objects in the SDSS dataset, only 106 have significantly high confidence value, greater than 50, to be considered valid classifications. In using this dataset to make inferences about asteroid taxonomy as across the Solar system, \citet{DeMeo2013SDSSTaxonomy, Demeo2014SSEvolAstComp} noted that again, the Jovian Trojans are heterogeneous in comparison to other populations. 

%Source of canonical taxonomic classification (tax_c) Tholen1989:Tholen (1989); Bendjoya2004: Bendjoya et al. (2004); Fornasier2004 (Fornasier et al. 2004); Lazzaro2004: Lazzaro et al.(2004); Fornasier2007: Fornasier et al. (2007); Hasselmann2012: Hasselmann et al. (2012).

In summary, taking data from each of these data-sets \citep{Tholen1989Taxonomy,Bendjoya2004JTSpectra,Fornasier2004L5TrojanSpec,Fornasier2007VisSpecTrojans, Lazzaro2004s3os2asteroids}, including those Trojans classified in the SDSS catalogue with a confidence score of greater than 50 \citep{Hasselmann2012SDSSTaxonomy}, there is a canonical set of 214 Trojans that are classified under the Bus-DeMeo taxonomy\footnote{The taxonomy is included in the online datasets, available from the Github repository for this study \url{https://github.com/TimHoltastro/holt-etal-2021-Jovian-Trojan-astrocladistics.git}}. As other authors have noted \citep{Grav2012JupTrojanWISE,Hasselmann2012SDSSTaxonomy,EmeryAsteroidsIVJupTrojan,DeMeo2013SDSSTaxonomy}, 72.2 per cent are classified as D-type, which is a much higher fraction than is seen in the Main Belt \citep{DeMeoAsteroidsIVTaxonomy,DeMeo2013SDSSTaxonomy,DeMeo2014MBDtypes} and in the Hilda \citep{Wong2017HildaColor} populations. The remainder of the Trojans classified to date in the canonical set are split between the C-types (10.8 per cent) and X-types (16.5 per cent). 

The current generation of surveys are laying the groundwork for our future exploration of the Trojan population. A NASA discovery class mission, \textit{Lucy}, is set to visit six Jovian Trojans between 2025 and 2033. One of the justifications for this mission is the diversity of taxonomic classes found in the population \cite{Levison2017Lucy}, with the mission visiting two C-types, two D-types and two X-types. The mission will also visit 3548 Eurybates (1973 SO), a C-type and the parent body of a collisional family. In combination with the \textit{Lucy} mission, in the coming decades, several relevant observational surveys coming online including the Vera Rubin Observatory, with the Legacy Survey of Space and Time \citep[Rubin Obs. LSST][]{LSST2009ScienceBook}, the James Web Space telescope \citep[\textit{JWST}][]{Rivkin2016JWSTAsteroids}, \textit{Twinkle} \citep{Savini2018TWINKLE} and \textit{Nancy Grace Roman Space Telescope} \citep[(\textit{RST}, formally \textit{WFIRST})][]{Milam2016WFIRSTSSS}. We explore these in further depth, with a specific focus on how they relate to the Jovian Trojans and our work, in section \ref{Sec:FutureSurvey}.

\begin{figure}
	\includegraphics[width=\columnwidth]{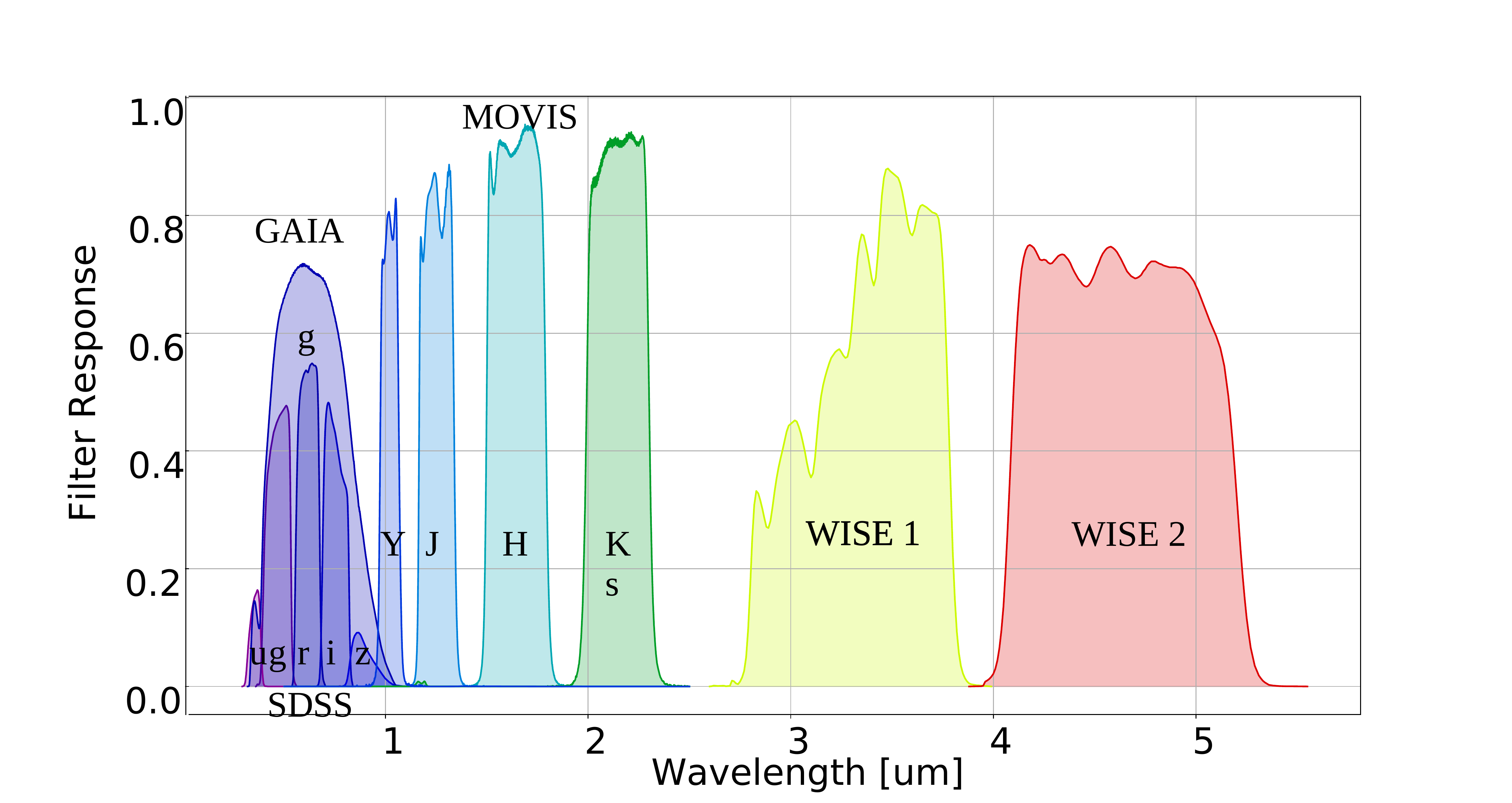}
    \caption{Wavelengths of the filters surveys used in this study. }
    \label{Fig:Filters}
\end{figure}

\subsection{Clustering Methods}
Contemporary studies of the Jovian Trojans have attempted to identify groups of objects within the population that share common dynamical properties. The most effective of these models to date has been the Hierarchical Clustering Method \citep[HCM;][]{Zappala1990HierarchicalClustering1}. Several collisional families have already been identified in the Trojan population using this method, despite the number of known Trojans being some two orders of magnitude smaller than the known population of the main belt \citep[e.g.][]{Milani1993JovianTrojFamilies,Beauge2001TrojanFamilies, Nesvorny2015AsteroidFamsAIV}. Another family identification method, uses the size dependent drift pattern due to the Yarkovsky effect \citep{Bottke2006YarkovskyAsteroids} to identify ancestral dynamic families in main-belt Asteroids \citep{Walsh2013PrimativeFam,Bolin2017YarkovskiAsteroidFam,Bolin2018VshapeVelo,Deienno2020Vshape}. The technique, while useful in the Main belt, has reduced usefulness in the Trojans. This is due to the dependence of the Yarkovsky effect on the Solar flux. At the 5.2au mean semi-major axis of the Jovian Trojans, the mean Yarkovsky effect is minimal, particularly for Trojans over 1km in diameter \citep{Wang2017YarkovskyTrojanModel,Hellmich2019TrojanYarkovski}.

The HCM is a technique that uses Gauss' equations to find groups in proper element (semi-major axis, eccentricity and inclination) parameter space \citep{Zappala1990HierarchicalClustering1}. The rationale behind these calculations is that the dispersal velocities of the clusters created by the collisional disruption of an object would be similar to the escape velocities of the parent body. The unique dynamical situation of the Jovian Trojans makes the identification of dynamical families using the traditional HCM difficult. Despite this, several collisional families are thought to be present amongst the Trojan population \citep[e.g.][]{Nesvorny2015AsteroidFamsAIV}. More modern dynamical analysis of the Jovian Trojans has identified a total of six canonical families \citep{Broz2011EurybatesFamily,EmeryAsteroidsIVJupTrojan,Vinogradova2015TrjoanFamilies, Nesvorny2015AsteroidFamsAIV, Rozehnal2016HektorTaxon, Vinogradova2015TrjoanFamilies}. The individual members and numbers in each work are inconsistent, and for this work we follow the canonical six families found in \citet{Nesvorny2015AsteroidFamsAIV}, with their associated members. There are two other modern sets that could be considered, \citet{Rozehnal2016HektorTaxon} or \citet{Vinogradova2015JupTrojanMass}. \citet{Vinogradova2015JupTrojanMass} found families in the L$_4$swarm using HCM with independently derived proper elements, though questioned the existence of any families in the L$_5$swarm. \citet{Rozehnal2016HektorTaxon} is incorporated into the canonical set \citet{Nesvorny2015AsteroidFamsAIV}, with several exceptions in the population. In our discussion, we note where these differ from the canonical set \citep{Nesvorny2015AsteroidFamsAIV}. Initial imaging surveys suggested that there is some spectral conformity within these dynamical families in the Jovian Trojans \citep{Fornasier2007VisSpecTrojans}. More recent observational data has brought this into question \citep{Roig2008JovTrojanTaxon}, with a heterogeneity being seen in the colours of the identified family members.

The disadvantage of the HCM system is that it only identifies recent family breakups, with the vast majority of objects considered `background'. Another issue with HCM is the issue of `chaining', where families are identified with interlopers included due to near proximity in phase space. In an attempt to overcome some of these issues \citet{Rozehnal2016HektorTaxon} offer an expansion to the HCM developed by \citet{Zappala1990HierarchicalClustering1}. This new `randombox' method uses Monte Carlo simulations to gain statistics on the probability that the identified clusters are random in parameter space. \citet{Carruba2007HCMFreq} also tried using elements in the proper frequency domain instead of orbital element space to overcome some of the issues of the HCM. The inclusion of `background objects' can be further mitigated by the inclusion of colours \citep{Parker2008AsteroidFamSDSS}, albedo \citep{Carruba2013AsteroidFamilies}, and taxonomy into the family identification pipeline \citep{Milani2014AsteroidFamilies, Radovic2017AsteroidFamProgram}, though these methodologies have focused on the Main-belt families. 

Though these methods do improve some of the faults identified in HCM, they still suffer from the issues inherent to the method. In order to use the HCM, a complete parameter space is required. This restricts the dataset in one of two ways, due to the limited information available for most small Solar system bodies. For the majority of family identification work, \citep[for review, see ][]{Nesvorny2015AsteroidFamsAIV}, only the dynamical elements are used. In order to expand the technique to include photometric information, albedo and colours, the number of objects needs to be restricted. For example, \citet{Carruba2013AsteroidFamilies} used a subset of only 11,609 main belt asteroids, out of the approximately 60,000 available in the Sloan Digital Sky survey (SDSS) \citep{Ivezic2002SDSSAsteroidFam}, 100,000 from \textit{WISE} \citep{Masiero2011MainBeltWISE}, and over 400,000 for which proper elements were available at the time. In the main-belt, \citet{Milani2014AsteroidFamilies} similarly attempted to combine together the AstDys database consisting of $\sim$340,000 asteroids, the \textit{WISE} \citep{Masiero2011MainBeltWISE} database consisting of $\sim$95,000 asteroids and the SDSS database \citep{Ivezic2002SDSSAsteroidFam} consisting of $\sim$60,000 asteroids into family classifications.

%\subsection{Cladistics}
%\label{LitCladistics}
In order to overcome some of the issues inherent in the HCM, as well as incorporating disparate colour surveys, in this work, we apply a technique called `cladistics' to the Jovian Trojan swarms. Cladistics is traditionally used to examine the relationships between biological organisms, and has played an important role in the study of our own history as a species. The namesake of the \textit{Lucy} mission, a near complete \textit{Australopithecus afarensis}, was used in some of the first hominid cladistical investigations \citep{Johanson1979EarlyHominid, Chamberlain1987EarlyHominid}, and continues to be an important resource for studies into human origins today \citep{ParinsFukuchi2019HomininFossil}. 

The premise of the cladistical method is that characteristics are inherited through descent. It is then inferred that organisms with similar characteristics are related to one another. As cladistics was originally developed to incorporate incomplete fossil records \citep{Hennig1965PhylogeneticSystem}, not all characteristics need to be known in order for a cladistical analysis to be carried out. This allows for the use of a larger number of characteristics and organisms, without needing to truncate the dataset due to missing values. Whilst cladistics can account for these unknown characteristics, the more that is known about an object/organism, the more confidence that can be placed in the analysis. Minimising missing data in the analysis would also decrease the number of equality parsimonious trees, trees that minimises the number of changes, produced during the analysis. The result of a biological cladistical analysis is a hierarchical dendritic tree, the `Tree of Life' \citep[e.g.]{Darwin1859Origin,Hennig1965PhylogeneticSystem,Hug2016TreeLife}, in which those organisms that are most closely related to one another other are joined by the shortest branch lengths. The advantage of cladistics over other analytical techniques is that it allows the use of multiple characteristics from disjointed datasets, including those that are unknown in some objects. 

The application of cladistics in an astronomical context is analogous to the biological framework, in that it facilitates the identification of groups of objects that likely share a common origin. For example, the members of collisional families are expected to cluster together, due to similarities in their orbital and physical elements. The previously identified collisional families can thus be used to comment on the cladistical methodology. The technique has already been used in a growing body of work called `astrocladistics' \citep{FraixBurnet2006DwarfGalaxies}. Astrocladistics has been used to study a wide range of astronomical objects, including galaxies \citep[e.g.,][]{FraixBurnet2006DwarfGalaxies}, gamma-ray bursts \citep[e.g.,][]{Cardone2013GRBClads} and stellar phylogeny \cite{Jofre2017StarsClads}. Within the planetary sciences, \citet{Holt2018JovSatSatsClad} used the technique to investigate the satellite systems of Jupiter and Saturn. 

\subsection{This work}
This is the first time that astrocladistics has been applied to large Solar system survey datasets. The extension of the technique presented in \citet{Holt2018JovSatSatsClad} to these large datasets could greatly improve our understanding of the relationships between Solar system objects. By increasing the number of Solar system objects that can be studied using astrocladistics, this project will help to establish the method as a valid analytical tool for the planetary science community. To do this, we combine proper orbital elements \citep{Knezevic2017AstDysTrojans}, \textit{WISE} albedos \citep{Grav2012JupTrojanWISE}, SDSS colours \citep{Hasselmann2012SDSSTaxonomy}, \textit{G}-band colour from the \textit{Gaia} DR2 \citep{Spoto2018GaiaDR2} datasets and the Moving Objects from VISTA Survey ({\tt MOVIS}) near-infrared colours \citep{Popescu2018MOVISTaxa}, into a single cladistical analysis. As a result, this paper will provide a methodological basis for future astrocladistical studies in the planetary sciences. 

In section \ref{Sec:Methods} we present an overview of the methodology of our work, and describe how astrocladistics is applied in the context of the Jovian Trojan population. Section \ref{Sec:Results} shows the results of the Jovian Trojan L$_4$ and L$_5$ swarm taxonomic analysis, including the dendritic trees and a discussion of the previously identified collisional families. As part of our analysis, we identify multiple objects of interest, presented in section \ref{Sec: targets}. The implications for the targets of the \textit{Lucy} mission are discussed in section \ref{SubSec:lucy}. 
%In the next decade, several new surveys and astronomical facilities will come online, including The Vera Rubin Observatory \citep[Rubin Obs. LSST][]{LSST2009ScienceBook}, the James Web Space telescope \citep[\textit{JWST}][]{Rivkin2016JWSTAsteroids}, the \textit{Twinkle} space telescope \citep{Savini2018TWINKLE} and the \textit{Nancy Grace Roman Space Telescope} \citep[\textit{RST}][]{Milam2016WFIRSTSSS}, which will yield new data that will enable the classification of a far greater proportion of the known Jovian Trojans. 
In section \ref{Sec:FutureSurvey}, we discuss the implications of our work in the context of the next generation of wide-field surveys that are coming online in the next decade. Finally, we draw our conclusions in section \ref{Sec: Conclusion}.

\section{Datasets and methods}
\label{Sec:Methods}
Here we present an overview of the cladistical methodology used in a planetary science context. For a more detailed overview of the techniques involved, we direct the interested reader to \citet{Holt2018JovSatSatsClad}. 

\subsection{Matrix and Characteristics}
\label{SubSec:Method:Matrix}
Each analysis begins with the creation of a 2-d matrix that contains all known information about the objects of interest -- in this case, the Jovian Trojans. Individual object are allocated a row in that matrix. The columns of the matrix contain information on a different characteristic of the objects studied -- including their physical properties and orbital elements. 

The great advantages of using the cladistical methodology is that it can take a wide and disparate set of characteristics for a group of objects, and can cope with incomplete datasets. To illustrate the breadth of characteristics that can be incorporated into a cladistical study, in this work we bring together the proper elements of the Jovian Trojans, retrieved from {\tt AstDyS} \citep{Knezevic2017AstDysTrojans}, geometric albedos from NASA {\tt HORIZONS} \citep{Giorgini1996JPLSSdatabase}, simulated libration properties, the \textit{WISE} albedos \citep{Grav2011JupTrojanWISEPrelim}, SDSS \citep{Carvano2010SDSSAstTax} colours, \textit{Gaia} DR2 \textit{G}-band color \citep{Spoto2018GaiaDR2} and {\tt MOVIS} colours \citep{MOrate2018AstColMOVIS, Popescu2016MPMOVIS, Popescu2018MOVISTaxa}. 

Due to the unique dynamics of the Jovian Trojans, the instantaneous osculating orbital elements can not be used for taxonomic proposes \citep[e.g.,][]{Beauge2001TrojanFamilies,Broz2011EurybatesFamily}. The {\tt AstDyS} database \citep{Knezevic2017AstDysTrojans} provides a set of robust proper elements, in semi-major axis, eccentricity and inclination, for the Jovian Trojans. Those proper elements are generated from the results of $1\e6$ year simulations, from which the osculating elements of the target objects are output at regular intervals. The resulting database of osculating elements are then processed using Fourier Transform analysis. This technique removes oscillations due to planetary perturbations, with the final result being three proper elements ($\Delta a_p$: proper delta in semi-major axis to the Jovian mean (5.2~au), $e_p$; proper eccentricity; $\textrm{sin} i_p$: sine of the proper inclination. By moving from an instantaneous value for the objects orbit to one that has been modified to take account of the periodic motion of the Trojans around the Lagrange points, these proper elements provide a much more accurate insight into a given object's provenance. Two objects with a common origin would be expected, in the absence of any major chaotic scattering events, to have similar proper elements, but might, at any given instant, be at a different part of their libration cycle, and hence have markedly different osculating elements. These proper elements can therefore, unlike the osculating elements, inform us about long term orbital relationships in the population. 

The Jovian Trojans are unique in that they are trapped in 1:1 mean-motion resonance with Jupiter, which means that their proper semi-major axes lie very close to that of Jupiter, approximately 5.2~au. The proper semi-major axis for the Trojans is therefore expressed, in this work, as a distance from the 5.2~au baseline ($\delta a_p$). An additional benefit to using these elements is that, due to the requirement that the object's exhibit $1\e6$ years of stable osculations around their host Lagrange point in simulations of their dynamical evolution, the {\tt AstDyS} database represents a dataset of Jovian Trojans that are at least relatively dynamically stable, and should exclude any objects that have otherwise been mis-classified, such as objects temporarily captured from the Jupiter family comet and Centaur populations \citep[e.g.][]{Horner2006CentaurCapTrojans}. It should be noted that stability for $1\e6$ years does not equate to, or even imply, stability on timescales comparable to the age of the Solar system \citep{Levison1997JupTrojanEvol,Tsiganis2005ChaosJupTrojans, Horner2012AnchisesThermDynam, DiSisto2014JupTrojanModels,DiSisto2019TrojanEscapes,Holt2020TrojanStability}. We use this stability level to exclude objects temporarily captured near the Jovian Lagrange points, such as P/2019 LD$_2$ \citep{Hsieh20202019ld2,Steckloff20202019ld2, Bolin2020p2019ld2}.

In addition to the proper elements obtained from {\tt AstDyS}, we also include information on the libration of the Jovian Trojans around their host Lagrange point. To obtain these libration values, we performed $1\e{4}$ year integrations of the orbital evolution of the Trojans under the influence of the Sun and four giant planets, using the {\tt REBOUND} WHFAST integrator \citep{Rein2012REBOUND}. For these integrations, we used a timestep of 0.3954 years, and wrote out the instantaneous orbital elements of all objects simulated every 10 years. From these, we were able to calculate the amplitude of libration, as well as the mean angle in the Jovian reference frame.

Similar physical properties, such as albedos and colours, would also be suggestive of analogous formation scenarios. We chose to not include mass, or any properties related to mass, as characteristics in the analysis. Their inclusion could hide any relationships between a massive object and any daughter objects, the result of collisions resulting in families. We do include visual geometric albedo \citep{Giorgini1996JPLSSdatabase}, as this represents analogous physical properties. 

In \citet{Holt2018JovSatSatsClad}, the presence or absence of various chemical species were used as characteristics in the cladistical analysis. This information requires detailed spectral analysis, which is only currently available for two Trojans, 624 Hektor (1907 XM) \citep{Marchis2014Hektor, Perna2018HektorAgamemnonSpec} and 911 Agamemnon (1919 FD) \citep{Perna2018HektorAgamemnonSpec}, although it is likely that this situation will change in the coming decade as a result of both the \textit{Lucy} mission and observations with the \textit{James Webb Space Telescope}. As a proxy for composition, broadband colours can be used in astrocladistics, as has been undertaken by \citet{FraixBurnet2010EarlyGalx} in their studies of galaxies. 

Several of the Jovian Trojans have been imaged by large all-sky surveys, with data available from the Sloan Digital Sky Survey (SDSS) \citep{Szab2007JovTrojanSloneDSS}, the \textit{Wide-field Infrared Survey Explorer} (\textit{WISE}) \citep{Grav2012JupTrojanWISE}, \textit{Gaia} DR2 \citep{Spoto2018GaiaDR2} and {\tt MOVIS} \citep{Popescu2016MPMOVIS}. The wide range of wavelengths represented by these datasets are shown in Fig. \ref{Fig:Filters}. We include these colours as characteristics in our analysis, in addition to the dynamical dataset described above. In total, combining the dynamical and observational data, this results in a maximum of 17 characteristics being included for each Trojan studied in this work. Each of these characteristics, along with their coefficient of determination ($R^2$) and ranges, are presented in appendix \ref{App:char}. 

Once all characteristics are collated for our objects of interest, they are binned to give each object a unique integer value for each characteristic. This was carried out using a Python program developed for \citet{Holt2018JovSatSatsClad} \footnote{Available from \url{ https://github.com/TimHoltastro/holt-etal-2021-Jovian-Trojan-astrocladistics.git}.}. The binning of the data has multiple benefits.  The primary reason for binning is the requirement the cladistical methodology to have whole numbers for analysis, representing character states. This has the added benefit of normalising each of the independent datasets. By normalising the datasets, the binning program also reduces the heterogeneity seen in the colours of the population \citep{Grav2011JupTrojanWISEPrelim,Grav2012JupTrojanWISE,DeMeo2013SDSSTaxonomy,DeMeo2014MBDtypes}, mitigating some of the effects of the `information content' \citep{Milani2014AsteroidFamilies} from each catalogue. The maximum number of bins for each characteristic, is set at 15, though if a co-efficient of determination ($R^2$) of greater than 0.99 is reached, a smaller number is used, shown in appendix \ref{App:char}. Those characteristics with a smaller number are then weighted, to standardise their contribution to the analysis. All binned characteristics have $R^2$ values larger than 0.95. The binned matrices are then imported into {\tt Mesquite} \citep{Mesquite}, a program used for management of cladistical matrices and trees, for further analysis. 

The dynamical characteristics and albedo are ordered as in \citet{Holt2018JovSatSatsClad} and \citet{FraixBurnet2006DwarfGalaxies}, with the colours unordered. The reasoning behind the ordering of dynamical characteristics is related to the stability of the Jovian Trojans. In dynamical space, the Jovian Trojans are relatively stable \citep[e.g.][]{Nesvorny2002SaturnTrojanHypothetical, Robutel2006TrojanRes1, EmeryAsteroidsIVJupTrojan, Holt2020TrojanStability}, and therefore any changes in dynamical properties represent large differences. In contrast to this, the colour ratios represent estimations in compositional structure of the objects. These broad-band colours can be affected by single changes in mineralogy \citep[e.g.][]{DeMeoAsteroidsIVTaxonomy,Reddy2015AstIVCompAst}, and are thus unordered. 

Simulations have suggested that some of the Jovian Trojans are unstable on relatively short timescales \citep[e.g.][]{Levison1997JupTrojanEvol,Tsiganis2005ChaosJupTrojans, DiSisto2014JupTrojanModels,DiSisto2019TrojanEscapes,Holt2020TrojanStability}. In order to account for this, we use only those objects that are present in the {\tt AstDyS} database \citep{Knezevic2017AstDysTrojans}. As the creation of proper elements requires a degree of stability \citep[e.g.][]{Knezevic2003AstDys}, these objects are stable in the swarms for at least $1\e6$ years. In this initial phase, we also only select those Trojans that have available observational data from at least one of the four surveys, \textit{WISE}, SDSS, \textit{Gaia} or {\tt MOVIS}. The result of this is the generation of two distinct matrices, one for each of the two Jovian Swarms. The L$_4$ dataset is smaller with 398 objects, whilst the L$_5$ matrix contains 407 objects. Though these subsets are markedly smaller than the total known populations of the two swarms, they offer a significant advantage over a possible HCM set. For comparison, in the L$_4$ swarm there are only five objects, 4060 Deipylos (1987 YT$_1$), 3793 Leonteus (1985 TE$_3$), 5027 Androgeos (1988 BX$_1$), 5284 Orsilocus (1989 CK$_2$) and 4063 Euforbo (1989 CG$_2$), and one, 7352 (1994 CO), in the L$_5$ that are present in all four surveys. Even if only the largest photometric dataset \citep[SDSS,][]{Szab2007JovTrojanSloneDSS} is considered, our subsets are nearly double those of a restricted HCM-type study (L$_4$:176 objects, L$_5$:232 objects). 
 
\begin{figure}
	\includegraphics[width=\columnwidth]{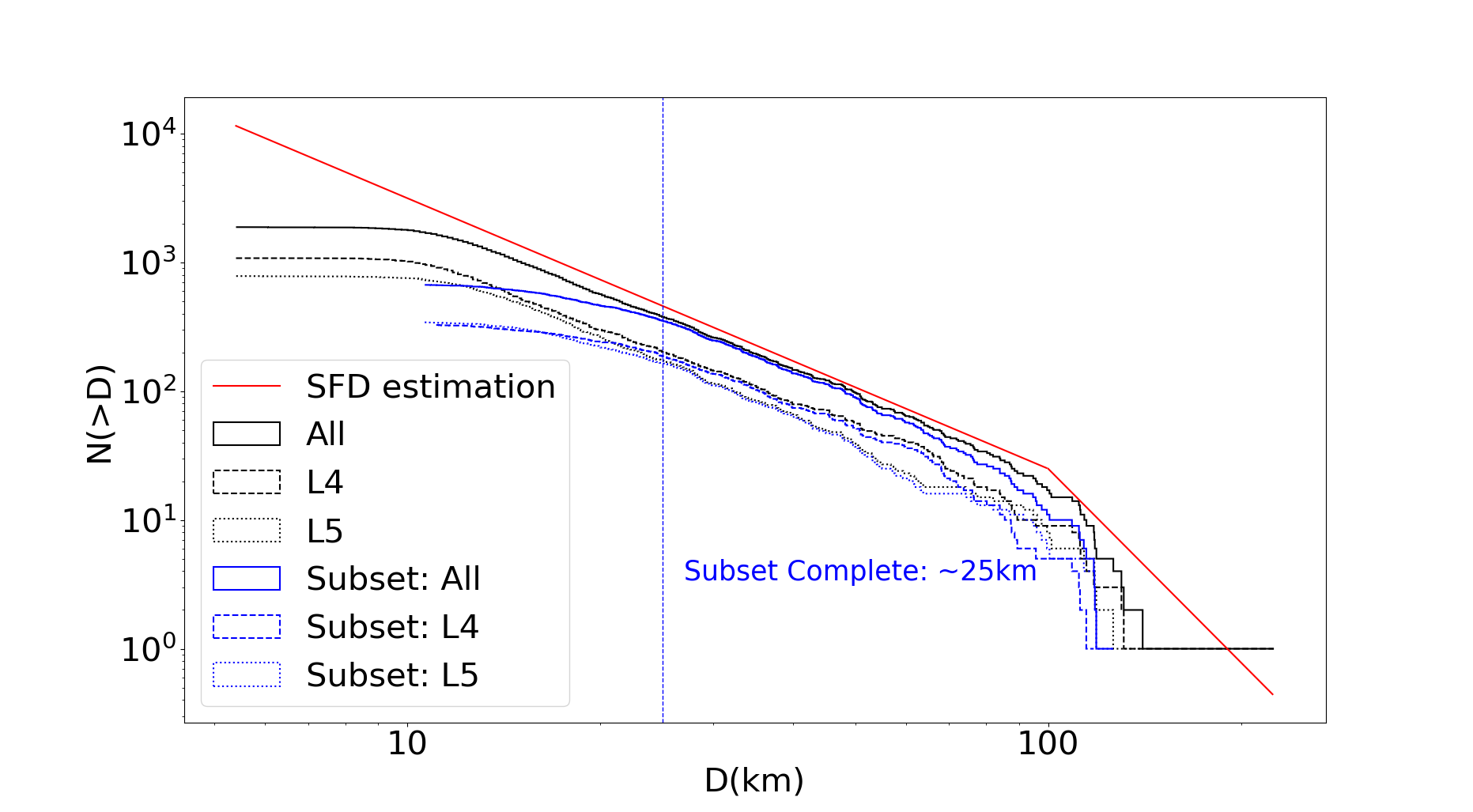}
    \caption{Size-Frequency distribution (SFD) of the Jovian Trojan population (black solid), L$_4$ (black dashed) and L$_5$ (black dot) swarms. We show the SFD of objects used in this work in blue, showing a completeness to 25km. A estimated complete SFD distribution of the Jovian Trojans (red) is also shown \citep{Nesvorny2018SSDynam}.}
    \label{Fig:TrojansSFDSub}
\end{figure}

The objects in our subsets are shown in the context of the swarms in Fig. \ref{Fig:TrojansSub}. In selecting only those objects with observational data available from one or other of the named surveys, we acknowledge that we are introducing a size bias, since larger objects are more likely to have been surveyed. We show the size-frequency distribution of our chosen objects in Fig. \ref{Fig:TrojansSFDSub}. This shows that our subset is complete to approximately 25km diameter. 

In addition to the Jovian Trojans, a fictitious outgroup object is created, with a base 0 for each of the characteristics. The function of this outgroup is to root the trees. In the context of biological cladistics, a related clade, but one that is outside the group of interest, is selected as the outgroup \citep{Farris1982outgroups}. In doing this, the outgroup sets the base character state for each characteristic. For astrocladistics of the Trojans, the dynamics make selection of the outgroup more difficult, as there is no true ancestral state from which ingroup characteristics are derived. For the synthetic outgroup created for this study, the dynamical characteristics are set close to 0 in proper $\Delta$ semi-major axis ($\Delta a_p$), eccentricity ($e_p$) and  sine inclination ($\textrm{sin}i_p$). The calculated mean libration values would be at the closest approach to Jupiter ($56.42\degree$ and $285.72\degree$ for the L$_4$and L$_5$swarms respectively), with low libration amplitudes (L4: $4.044\degree$, L5: $2.73\degree$). These values represent a very stable area of the parameter space. In terms of albedo (L4: 0.024, L5: 0.031) and colours, the object would be very dark, and have a featureless spectrum. Based on these parameters the ougroup served the purpose of rooting each consensus tree without being too close and considered part of the ingroups, or too far away so that the relationship to the populations of interest were lost.

Each matrix is available in the online supplemental material\footnote{\url{https://github.com/TimHoltastro/holt-etal-2021-Jovian-Trojan-astrocladistics.git}} in binned and unbinned form.

\subsection{Trees}
\label{SubSec:Method:Trees}
Each {\tt Mesquite} taxon-character matrix is then used to create a set of phylogenetic {\tt trees using Tree analysis using New Technology} ({\tt TNT}) v1.5 \citep{Goloboff2008TNT, Golboff2016TNT15}, via the Zephyr {\tt Mesquite} package \citep{MesquiteZephyr}. This tree search is based on the concept of maximum parsimony \citep{Maddison1984outgroup}. Each tree generated in the block has a length  and in this case, is a characteristic of the tree itself, and not of the individual branches. This tree length is calculated on the bases of characteristics changing states, for example a change form a 0 to a 1 would constitute a 1 step value. In ordered characteristics, a change from 0 to 2 would be two steps, where as in the unordered, would only be one step. A tree with more changes in character state would have a longer tree length. The {\tt TNT} algorithm \citep{Goloboff2008TNT, Golboff2016TNT15} rearranges the configuration of the trees, attempting to find the set of trees with the lowest tree length, creating a block of the most equally parsimonious trees, those with the same minimum tree length. We use a drift algorithm \citep{Goloboff1996FastPasrimony} search by generating 100 Wagner trees \citep{Farris1970MethodsComp}, with 10 drifting trees per replicate. These starting trees are then checked using a Tree bisection and reconnection (TBR) algorithm \citep{Goloboff1996FastPasrimony} to generate a block of 10000 equality parsimonious dendritic trees. The Nexus files for both matrices, both with and without the tree blocks, are available on the GITHUB repository \footnote{Available from \url{ https://github.com/TimHoltastro/holt-etal-2021-Jovian-Trojan-astrocladistics.git}}. A 0.5 majority-rules consensus tree can be constructed \citep{Margush1981MajorityRules} once the tree block is imported back into {\tt Mesquite} \citep{Mesquite}. This tree is then a hypothesis for the relationships between the Jovian Trojans in the individual swarms. 

As part of the consensus tree, each node (see Fig. \ref{Tree:L4} and Fig. \ref{Tree:L5}) shows the fraction of trees in the block that contain that node ($F_{node}$). This fraction is tabulated in tables \ref{tab:L4} and \ref{tab:L5} for each subclan, clan and superclan. The higher the prevalence of the node, with 1.0000 indicating that the node is in all 10,000 trees, gives higher confidence in the grouping. 

\subsection{Dispersal velocities, Diameter calculations and Escape analysis}
\label{SubSec:Method:DisVelo}
The taxonomic clusters produced by the cladistical methodology can be verified using the established inverse Gauss equations \citep[e.g.][]{Zappala1996EjectionVelocity, Turrini2008IrregularSatsSaturn, Holt2018JovSatSatsClad}, in much the same way that asteroid collisional families are identified and confirmed \citep[e.g.][]{Nesvorny2015AsteroidFamsAIV}. \citet{Holt2018JovSatSatsClad} provide a demonstration of the use of those equations in conjunction with their cladistical analysis of the Jovian and Saturnian satellite systems. In that work, the inverse Gauss equations are used to comment on the relative timing of creation and validity of the clusters in the irregular satellites identified by astrocladistics. The rationale for this is that clusters with low dispersal velocities would most likely indicate families produced by recent breakups, with larger velocities possibly indicating either more energetic disruptions of the family's parent body, or an older family that has had longer to disperse.

We extend the methodology used in \citet{Holt2018JovSatSatsClad} to investigate the mean dispersal velocity in the dynamical parameter space of the clusters we identify in the Jovian Trojan population\footnote{Python 3 program is available from the GITHUB repository: \url{https://github.com/TimHoltastro/holt-etal-2021-Jovian-Trojan-astrocladistics.git}}. In the traditional methodology, the largest object is used as a point of reference for the parameters used in the calculations \citep[specifically $a_r$, $e_r$, $i_r$ and $n_r$ in equations 5-6 in ][]{Holt2018JovSatSatsClad}. In this work we calculate two different dispersal velocities for each Jovian Trojan. As in the original work, we determine the dispersal velocity of each object in a given cluster to the largest object in that cluster ($\Delta V_{\textrm{ref}}$). In addition, we calculate the dispersal velocity from a fictitious centroid at the mean of the cluster proper element space ($\Delta a_{\textrm{prop}}$, $e_p$, $\textrm{sin} i_p$ and period $n$: $\Delta Vm_{\textrm{cent}}$). The inverse Gauss equations also require knowledge of the values of $\omega$ and $\omega+f$ at the initial point of disruption \citep[e.g.][]{Zappala1996EjectionVelocity,Nesvorny2004IrrSatFamilyOrigin,Nesvorny2015AsteroidFamsAIV}. For ease of comparison, we have used $\omega$ as 90$\degree$ and $\omega+f$ as 45$\degree$ . Datasets for each of the clusters are available from the Github repository \footnote{\url{https://github.com/TimHoltastro/holt-etal-2021-Jovian-Trojan-astrocladistics.git}}.

Only 1857 of the 5553 (33.44 per cent) canonical Trojan have measured albedos, and therefore reliable diameters in the NASA {\tt HORIZONS} database. In order to investigate the size distribution of each swarm, we created an estimate of the diameter and volume of each of the 5553 Jovian Trojans in the {\tt AstDyS} database \citep{Knezevic2017AstDysTrojans}. The unknown diameters (D) were calculated from the absolute magnitude ($H$) of the object in the NASA {\tt HORIZONS} database, combined with an estimate of the mean geometric albedo values for the population ($P_v$ = 0.075) using equation \ref{Equ:Dest} \citep{Fowler1992IRASAstData}. 

\begin{equation}
D (km) = \frac{1329}{\sqrt{P_v}} 10^{-0.2H}
\label{Equ:Dest}
\end{equation}

It should be noted that it is highly likely that most Jovian Trojans, particularly the smaller members of the population, are markedly aspherical. Indeed, shapes inferred from occultation observations suggest that several of the targets for the \textit{Lucy} mission are likely irregular in shape  \citep{Buie2015PatroclusOcc, Mottola2020Leucus}. Given the known shapes in this size regime, from the Main belt and Near Earth populations \citep[e.g.,][]{Durech2015AsteroidModels}, it is expected that other Jovian Trojans are also irregular in shape. From this our calculated diameter values should only be taken as estimations, and are available in the Github associated with this study. 

% Given this estimated diameter, we then calculated an approximate volume for each object, under the simplifying assumption of a spherical shape. This is only an approximation, as the actual shape models for Jovian Trojans are only available for  624 Hektor (1907 XM) \cite{Marchis2014Hektor} and 617 Patroclus (1906 VY) \citep{Marchis2006DensityPatroclus}. It should be noted that it is highly likely that most Jovian Trojans, particularly the smaller members of the population, are markedly aspherical. Indeed, shapes inferred from occultation observations suggest that several of the targets for the \textit{Lucy} mission are likely irregular in shape \citep{Buie2015PatroclusOcc, Mottola2020Leucus}. Given the known shapes in this size regime, from the Main belt and Near Earth populations \citep[e.g.,][]{Durech2015AsteroidModels}, it is expected that other Jovian Trojans are also irregular in shape. Even so, we can estimate the volumes as a sphere, gained from the diameter (D) estimation in equation \ref{Equ:Dest} 
%\thcoms{I realised that I don't actually use the volumes. }
 
As part of our analyses, we track the dynamical evolution of the chosen objects, using data presented in \citet{Holt2020TrojanStability} which presented escape fractions of the Trojan swarms and collisional families on a timescale of $4.5\e9$ years. The best fit orbital solution for each Jovian Trojan studied in that work was integrated forwards in time under the gravitational influence of the Sun and four giant planets for $4.5\e9$ years. In addition, eight 'clones' of each object were studied, with initial orbital parameters perturbed from the best fit solution along the Cartesian uncertainties presented in the {\tt HORIZONS} database.  We use this information to comment on the stability of the individual members, and each cluster as a whole. 

\subsection{Full population analysis}
In addition to the subset analysis, we also conducted an analysis of the full L$_4$ swarm (3620 objects) and full L$_5$ swarm (1920 objects), using same techniques presented in section \ref{Sec:Methods}. Since many of the objects in each swarm remain poorly characterised, with many lacking for any information other than an apparent magnitude and orbital solution, there is insufficient information in the matrices for us to place great weight in the results of this additional analysis. We only include this as a computational note for future work, as presented in Table \ref{tab:tree}. 

\begin{table}
\caption{Comparison of times taken to generate each 10,000 tree blocks, as described in section \ref{SubSec:Method:Trees}. (subset) are the matrices used in this work, where as (Pop) are the full known population at their respective Lagrange points. No.: Number of objects in the matrix; $Hrs_{cpu}$: number of CPU hours taken to generate tree block on a single core of Intel Xeon W-2133 CPU at 3.60GHz; $L_{tree}$: Tree length; $I_c$:consensus index \citep{Brooks1986ConsistancyIndex} of the 0.5 consensus tree; $I_r$:retention index \citep{Naylor1995RetentionIndex} of the 0.5 consensus tree. }
\label{tab:tree}
\resizebox{\columnwidth}{!}{
\begin{tabular}{lllllll}
\hline
            & No.  & $Hrs_{cpu}$ & $L_{tree}$ & $I_c$ & $I_r$ \\ 
\hline
L$_4$ (subset) & 398  & 10.72       & 1635.37    & 0.123 & 0.751 \\
L$_5$ (subset) & 407  & 10.69       & 1984.79    & 0.113 & 0.712 \\
L$_4$ (Pop)    & 3620 & 420.15      & 3926       & 0.041 & 0.899 \\
L$_5$ (Pop)    & 1920 & 372.2       & 2794       & 0.054 & 0.883 \\
\hline
\end{tabular}
}
\end{table}

\section{Results and discussion}
\label{Sec:Results}
Here, we present the taxonomic trees resulting from our cladistical analysis of the Jovian Trojan swarms. Each swarm is presented and discussed separately, and we compare our results to the previously identified collisional families \citep{Nesvorny2015AsteroidFamsAIV}. In order to avoid confusion with a specific cluster identified in our cladistics analysis, we use the term \textit{clan} to identify the groups of objects that share a similar heritage. We borrow two conventions from the biological Linnean taxonomy \citep{Linnaeus1758SystemaNaturae}, namely the inclusion of a type object and the use of prefixes. Each clan is named after the member that was first discovered. This object is designated the \textit{type object}. Due to observational bias, in most cases, the type object is the largest member of the clan. The largest member of the group is used as a reference point for the dispersal velocities explained in section \ref{SubSec:Method:DisVelo}, and termed the \textit{reference object}. It is important to note that the type object and the reference object in a clan can be the same object, though this is not always the case. In order to assist with any hierarchical grouping, we use the super and sub prefixes, to denote higher and lower groups. To further improve the clarity of the hierarchical clusters, the superclan's have 'Greater' affixed to the representative name. We choose five members as the minimum number for a clan or subclan. This terminology forms a basis for future expansion of the small Solar system body taxonomic framework.

\begin{figure}
	\includegraphics[width=\columnwidth]{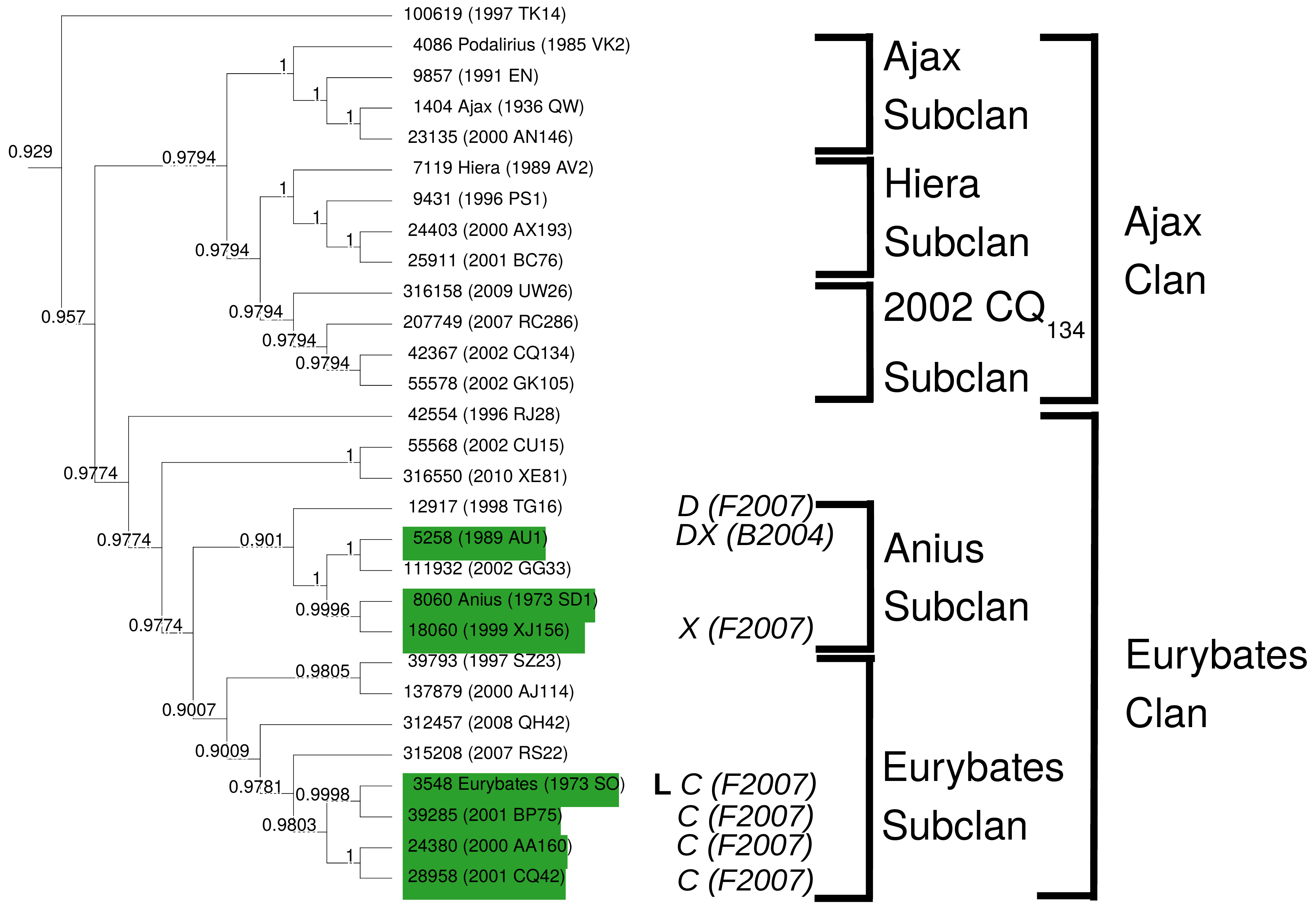}
    \caption{Consensus trees of Greater Ajax superclan, including Ajax and Eurybates clans. An example of trees shown in Appendix \ref{App:clans}. \textit{Letters} associate objects with Bus-Demeo taxonomy \citep{Bus2002AsteroidTax, DeMeo2009AsteroidTax}, classified by associated reference T1989: \citet{Tholen1989Taxonomy}; B2004: \citet{Bendjoya2004JTSpectra}; F2007: \citet{Fornasier2007VisSpecTrojans}; H2012, with associated confidence rating: \citet{Hasselmann2012SDSSTaxonomy}. \textbf{L} indicates objects to be visited by the \textit{Lucy} spacecraft \citep{Levison2017Lucy}. Green highlights are members of the Eurybates collisional family.}
    \label{Fig:greaterAjax}
\end{figure}

The Greater Ajax superclan, shown in Fig. \ref{Fig:greaterAjax}, highlights the hierarchical nature of this new terminology. The superclan is split into two clans, the Ajax and Eurybates clans. The type object of both the Greater Ajax superclan and the Ajax clan is 1404 Ajax (1936 QW), whereas 3548 Eurybates (1973 SO) is the type object of the Eurybates clan. Within both clans, there are two subclans. In the Eurybates clan, there is the Anius subclan with type object 8060 Anius (1973 SD$_1$), and the Eurybates subclans, along with three other objects, namely 42554 (1996 RJ$_{28}$), 55568 (2002 CU$_{15}$), 316550 (2010 XE$_{81}$), not associated with either subclan. In this example set, the Trojan 3548 Eurybates (1973 SO) is therefore the type object of both the Eurybates subclan and Eurybates clan, and is also a member of the Greater Ajax superclan. In this example, 3548 Eurybates (1973 SO) is also the reference object used in dispersal velocity calculations calculations for the Eurybates clan and subclan. 

%T1989: \citet{Tholen1989Taxonomy}; B2004: \citet{Bendjoya2004JTSpectra}; F2007: \citet{Fornasier2007VisSpecTrojans}; H2012, with associated confidence rating: \citet{Hasselmann2012SDSSTaxonomy}.

\subsection{L$_4$ Swarm}
\label{SubSec: L4}
In the L$_4$ Trojan swarm, we analyse a total of 398 objects using the astrocladistical methodology. A total of 10,000 equally parsimonious trees were generated, a process that took 10~hrs, 43~min using a single core of Intel Xeon W-2133 CPU at 3.60GHz. The resulting consensus tree is presented in Fig. \ref{Tree:L4}. The tree has a consistency index of 0.123 \citep{Brooks1986ConsistancyIndex} and a retention index of 0.751 \citep{Naylor1995RetentionIndex}. The consensus tree has a length of 1635.37
\footnote{The tree length, retention index and consistency index are measures of how accuracy a tree represents the true relationships. A smaller tree length implies a more parsimonious, and thus likely tree \citep{Goloboff2015Parsimony}. The two other indices are measures of homoplasy, the independent loss or gain of a characteristic \citep{Brandley2009Homoplasy}. In both indices, a value of 1 indicates no homoplasy, and thus no random events. The consistency index is the ratio of the minimum number of changes in a tree, to the actual number \citep{Givnish1997consistency}. The retention index is similar, but incorporates the maximum number of changes into the index \citep{Farris1989retention}. We direct the interested reader to \citet{Gascuel2005MathEvolPhylogeny} for a more detailed analysis of the mathematics behind these indices.}. 

The superclans, clans and subclans identified in the L$_4$ swarm are listed in Table \ref{tab:L4}. In the L$_4$ swarm, we identify a total of ten unaffiliated clans and eight superclans containing an additional seventeen clans. Each of these trees are shown in detail in Appendix \ref{App:clans}. 

In the L$_4$swarm, there are four canonical collisional families \citep{Nesvorny2015AsteroidFamsAIV}. Here three are represented in the subset, the 1996 RJ, Hektor and Eurybates families. All members of the Eurybates, 1996 RJ and Hektor families in the canonical set used in this study are also in \citet{Rozehnal2016HektorTaxon} and \citet{Vinogradova2015TrjoanFamilies}. There are no representatives of the canonical Arkesilaos family, though \citet{Vinogradova2015TrjoanFamilies} associated this family with their Epeios non-canonical family, of which the largest member, 2148 Epeios (1976 UW) is the type object of the Epeios clan. The only member of the Hektor family, 624 Hektor (1907 XM), is the type object of the Hektor clan, in the Greater Hektor superclan. The Eurybates collisional family provides some place for comment. Seven of the thirteen identified members are clustered the Ajax clan, around 3548 Eurybates (1973 SO), the type object. There are two other clusters of Eurybates family members, three objects in the Philoctetes Clan, and another three that are unassociated with any clan. The fact that these are clustered, but separated in the consensus tree, may indicate that they are victims of `chaining' in HCM, and thus not truly members of the collisional family. 

\begin{figure*}

	\includegraphics[height=0.89\textheight]{Trees/SubMatrixWiseSDSSGaiaMovis-L4-Conserv-app.pdf}
    \caption{Consensus tree of cladistical analysis of 398 L$_4$ Jovian Trojans. Numbers indicate the proportion of the 10000 generated trees where a given branch is present. Colours are indicative of previously identified collisional families: Green: Eurybates; Orange: Hektor; Red; 1996 RJ; after \citet{Nesvorny2015AsteroidFamsAIV}. \textit{Letters} associate objects with Bus-Demeo taxonomy \citep{Bus2002AsteroidTax, DeMeo2009AsteroidTax}, from different sources, T1989: \citet{Tholen1989Taxonomy}; B2004: \citet{Bendjoya2004JTSpectra}; L2004: \citet{Lazzaro2004s3os2asteroids}; F2007: \citet{Fornasier2007VisSpecTrojans}; H2012, with associated confidence rating: \citet{Hasselmann2012SDSSTaxonomy}. \textit{Lucy} Targets are indicated by an \textbf{L}. A high resolution, expanded form of this figure is available in online supplemental material. Blue rectangles correspond to detailed figures in Appendix \ref{App:clans}.}
    \label{Tree:L4}
\end{figure*}

\begin{table}
\caption{Clans, \textbf{superclans} and \textit{subclans} identified in the L$_4$ Trojan swarm. Name: Clan Name; $N$: Number of objects; $D_{\textrm{ref}}$: Reference object diameter; $V_{\textrm{esc}}$: Escape velocity of reference object; $F_{\textrm{esc}}$: fraction of objects that escape the L$_4$ Lagrange point, from \citet{Holt2018JovSatSatsClad};  $\Delta Vm_{\textrm{ref}}$: mean dispersal velocity calculated from inverse Gauss equations, see section \ref{SubSec:Method:DisVelo}, to the reference object, with $1 \sigma$ standard deviation; $\Delta Vm_{\textrm{cen}}$: as $\Delta Vm_{\textrm{ref}}$, with calculations to the fictitious cluster center; $F_{node}$: faction of trees in the block that contain the node.}
\label{tab:L4}
\resizebox{\columnwidth}{!}{%
\begin{tabular}{llllllll}
\hline
Name          & $N$ & $D_{\textrm{ref}}$ & $V_{\textrm{esc}}$ & $F_{\textrm{esc}}$ & $\Delta Vm_{\textrm{ref}}$ & $\Delta Vm_{\textrm{cen}}$  &  $F_{node}$ \\ 
                    &       & km           & ms$^{-1}$ &           & ms$^{-1}$        & ms$^{-1}$      &   \\
\hline
L4-Stentor          & 8      & 71.84     & 24.02      & 0.07               & 11.38$\pm$7.85   & 10.25$\pm$6.51 & 1.0000 \\
L4-1998 WR$_{10}$   & 5      & 34.95     & 11.69      & 0.02               & 26.02$\pm$8.11   & 10.61$\pm$5.33 & 0.7628\\
L4-Periphas         & 5      & 80.17     & 26.81      & 0.44               & 13.02$\pm$6.11   & 10.61$\pm$4.56 & 0.9216\\
L4-Halitherses      & 15     & 37.7      & 12.61      & 0.01               & 21.53$\pm$12.28  & 15.63$\pm$9.9  & 0.9422 \\
L4-Polypoites       & 5      & 68.73     & 22.98      & 0                  & 30.67$\pm$9.95   & 15.49$\pm$8.5  & 0.9799 \\
L4-Ulysses          & 17     & 76.15     & 25.46      & 0.08               & 24.15$\pm$12.43  & 14.25$\pm$8.25 & 0.9694 \\
L4-Idomeneus        & 6      & 112.05    & 37.47      & 0                  & 9.81$\pm$3.51    & 8.37$\pm$3.7   & 1.0000 \\
L4-Halaesus         & 10     & 50.77     & 16.98      & 0.06               & 12.3$\pm$9.36    & 10.56$\pm$6.34 & 1.0000  \\
L4-Agamemnon        & 16     & 131.04    & 43.82      & 0.12               & 28.34$\pm$12.5   & 21.41$\pm$11.01& 1.0000 \\
L4-Thersander       & 10     & 65.92     & 22.04      & 0.14               & 21.82$\pm$10.93  & 17.5$\pm$12.43 & 0.9795 \\
                    &        &           &            &                    &                  &                &  \\
\textbf{L4-Greater Achilles}  & 35     & 130.1     & 43.51      & 0.06               & 16.91$\pm$11.24  & 14.4$\pm$8.6 & 0.9501    \\
L4-Epeios           & 10     & 48.36     & 16.17      & 0                  & 19.55$\pm$7.42   & 8.57$\pm$4.42  & 0.9987 \\
L4-Achilles         & 9      & 130.1     & 43.51      & 0.15               & 9.9$\pm$5.96     & 8.65$\pm$6.54  & 0.9707 \\
L4-1991EL           & 10     & 68.98     & 23.07      & 0.06               & 17.37$\pm$13.76  & 12.69$\pm$8.48 & 0.9799 \\
                    &        &           &            &                    &                  &                &  \\
\textbf{L4-Greater Nestor}    & 27     & 112.32    & 37.56      & 0.63               & 34.04$\pm$20.11  & 29.81$\pm$12.31 & 0.9507 \\
L4-Eurymedon        & 6      & 45.68     & 15.28      & 0.3                & 20.96$\pm$15.31  & 13.48$\pm$6.9  & 0.9013 \\
L4-Nestor           & 7      & 112.32    & 37.56      & 0.4                & 20.71$\pm$13.1   & 14.84$\pm$5.4  & 0.9709 \\
                    &        &           &            &                    &                  &                &  \\
\textbf{L4-Greater Ajax}     & 29     & 85.5      & 28.59      & 0.38               & 40.33$\pm$26.27  & 32.91$\pm$16.75 & 0.9290  \\
L4-Ajax             & 12     & 85.5      & 28.59      & 0.6                & 18.06$\pm$14.39  & 16.82$\pm$10.65 & 0.9794\\
\textit{L4-Ajax Sub} & 4      & 85.5      & 28.59      & 0.64               & 11.29$\pm$0.93   & 7.34$\pm$2.62  & 1.0000 \\
\textit{L4-Hiera Sub}& 4      & 59.15     & 19.78      & 0.36               & 8.31$\pm$1.19    & 4.35$\pm$0.85  & 1.0000 \\
\textit{L4-2002 CQ$_{134}$ Sub} & 4      & 32.16     & 10.75      & 0.81       & 36.08$\pm$17.87  & 20.26$\pm$11.97 & 0.9794 \\
L4-Eurybates        & 16     & 63.88     & 21.36      & 0.23               & 32.58$\pm$28.94  & 24.59$\pm$14.67 & 0.9774\\
\textit{L4-Anius Sub  } & 5      & 53.28     & 17.82      & 0.27         & 12.61$\pm$10.34  & 10.42$\pm$4.62 & 0.9010 \\
\textit{L4-Eurybates Sub} & 8      & 63.88     & 21.36      & 0.24        & 25.23$\pm$20.27  & 19.43$\pm$2.22 & 0.9007\\
                    &        &           &            &                    &                  &                &  \\
\textbf{L4-Greater Hektor} & 28     & 225       & 75.24      & 0.54               & 31.47$\pm$19.35  & 28.99$\pm$21.63 & 0.9593\\
L4-Thersites        & 11     & 89.43     & 29.91      & 0.83               & 34.57$\pm$33.85  & 27.49$\pm$23.04 & 0.9792 \\
L4-Hektor           & 17     & 225       & 75.24      & 0.35               & 31.43$\pm$16.53  & 27.93$\pm$16.05 & 1.0000\\
                    &        &           &            &                    &                  &                &  \\
\textbf{L4-Greater Diomedes}  & 75     & 117.79    & 39.39      & 0.43               & 108.79$\pm$36.64 & 41.85$\pm$23.23 & 0.9782 \\
L4-Philoctetes      & 26     & 33.96     & 11.36      & 0.38               & 25.19$\pm$7.9    & 19.07$\pm$11.21  & 0.9998\\
\textit{L4-Andraimon Sub }  & 10     & 33.96     & 11.36      & 0.77               & 27.48$\pm$6.64   & 23.91$\pm$7.2 & 0.9796  \\
L4-Diomedes         & 12     & 117.79    & 39.39      & 0.76               & 56.9$\pm$28.09   & 40.6$\pm$19.19   & 0.9427\\
L4-Lycomedes        & 20     & 31.74     & 10.61      & 0.45               & 33.18$\pm$16.71  & 29.44$\pm$16.14 & 0.9809\\
\textit{L4-Amphiaraos Sub} & 8      & 26.83     & 8.97       & 0.57               & 13.3$\pm$4.33    & 10.18$\pm$1.9 & 1.0000  \\
                    &        &           &            &                    &                  &                &  \\
\textbf{L4-Greater Telamon }  & 35     & 111.66    & 37.34      & 0.05               & 27.16$\pm$15.83  & 21.68$\pm$11.82 & 0.8646 \\
L4-Telamon          & 5      & 64.9      & 21.7       & 0.27               & 26.96$\pm$18.11  & 20.7$\pm$6.98  & 0.9600 \\
L4-Kalchas          & 6      & 46.46     & 15.54      & 0                  & 16.88$\pm$10.79  & 12.47$\pm$4.59 & 1.0000 \\
L4-Theoklymenos     & 19     & 111.66    & 37.34      & 0.03               & 25.71$\pm$17.83  & 20.08$\pm$12.5 & 0.8390 \\
\textit{L4-Makhaon Sub} & 5      & 111.66    & 37.34      & 0.09               & 13.39$\pm$6.05   & 8.22$\pm$1.17 &0.9691   \\
                    &        &           &            &                    &                  &                &  \\
\textbf{L4-Greater Odysseus}  & 36     & 114.62    & 38.33      & 0.11               & 24.17$\pm$12.28  & 18.41$\pm$9.07 &0.9701  \\
L4-Epistrophos      & 5      & 24        & 8.02       & 0                  & 8.32$\pm$3.94    & 6.99$\pm$2.47  & 1.0000 \\
L4-Odysseus         & 20     & 114.62    & 38.33      & 0                  & 17.6$\pm$11.75   & 12.54$\pm$8.43 & 0.9797 \\
\hline
\end{tabular}%
}
\end{table}

\subsubsection{Unaffiliated L$_4$ clans}
Our results reveal ten clans in the L$_4$ swarm that are unaffiliated with any identified superclan, presented in Fig. \ref{fig:UnassL4clans}. None of the unaffiliated clans can be further split into subclans. Six of the unaffiliated clans, namely the Stentor, 1998WR$_{10}$, Periphas, Halitherses, Polypoites and Ulysses clans, are located at the base of the L$_4$ tree. Each of the ten unaffiliated clans in the L$_4$ swarm contain at least one D-type object. The Agamemnon and Ulysses clans containing five and six D-type members respectively. The Halitherses clan contains one X-type, 13475 Orestes (1973 SX), along with a single D-type, 13362 (1998 UQ$_{16}$), indicating that there may be some heterogeneity to these clans. 

The dynamical stability of all members of the identified unaffiliated clan members was assessed by \citep{Holt2020TrojanStability}. Comparing our list of those clans with the dynamical data from that work, we find that most of the clans exhibit significant dynamic stability, at a level that exceeds the mean stability of the L$_4$ Trojan population as a whole (with the simulations described in \citealt{Holt2020TrojanStability} yielding a mean escape fraction of 0.24 for the L$_4$ cloud over the age of the Solar system). The exception is the Periphas clan, which displays a higher escape fraction (0.44) over the course of those simulations. 

In the following sections, we discuss three of these unaffiliated clans, the Stentor, Idonmeneus and Thersander clans, highlighting several interesting cases. The other seven clans, as shown in Fig. \ref{fig:UnassL4clans}, may contain objects of interest, though we leave further detail discussion for future research. 

\paragraph{Stentor Clan: } The first clan identified in our consensus tree of the L$_4$ Trojans (Figure~\ref{Tree:L4}) is the Stentor clan, shown in more detail in Fig. \ref{fig:Stentor}, after the type object 2146 Stentor (1976 UQ), and consists of a total of 8 objects. The clan includes the two identified members of the 1996 RJ collisional family, 226027 (2002 EK$_{127}$ and 9799 (1996 RJ), \citep{Nesvorny2015AsteroidFamsAIV}, and it seems likely that the other members of the clan represent previously undetected members of the collisional family. The type object of this clan, 2146 Stentor (1976 UQ), is chosen over 9799 (1996 RJ), due to it being discovered nearly 20 years earlier. In this clan, although 2146 Stentor (1976 UQ) (50.76~km) is the type object, 7641 (1986 TT$_6$) is used as the reference frame for our calculations of the clan member's dispersion in  $\Delta V_{\textrm{ref}}$, as the available observational data suggest that it has the largest diameter in the clan (71.84~km). 

Unfortunately, no members of this clan have been classified under the Bus-Demeo system. Almost all members of this clan were found to be dynamically stable in the simulations carried out by \citet{Holt2020TrojanStability}, with the one exception being the clones of the type object, 2146 Stentor (1976 UQ). More than half of the clones of that object (56 per cent) escaped from the Jovian Trojan population over the $4.5\e9$ years of those simulations. The stability of the remainder of the clan is likely the result of most of the members having low $\delta a_{\textrm{prop}}$ ($<0.036~au$), mean libration angles ($<3.5\degree$ from the Lagrange point) and range ($<14\degree$). 

The clan has relatively compact \textit{Gaia} \textit{G} magnitude values (17.56 to 18.11~mag), though there are only two similar sized members, 2146 Stentor (1976 UQ) and 9799 (1996 RJ), in the dataset. Three additional objects, 7641 (1986 TT$_6$),  83983 (2002 GE$_{39}$) and 88225 (2001 BN$_{27}$), have a corresponding SDSS \emph{(g - r)} colour (0.57 to 0.7), indicating that perhaps there is a diagnostic feature for the clan in the visible range. 

\paragraph{Idonmeneus Clan: } The small Idomeneus clan (6 members), Fig. \ref{fig:Idomeneus} contains two D-types, 2759 Idomeneus (1980 GC) and 4063 Euforbo (1989 CG$_2$), along with a small $\Delta Vm$ ($9.81\pm3.51$~ms$^{-1}$), and large reference object, 3793 Leonteus (1985 TE$_3$). This clan also includes 4063 Euforbo (1989 CG$_2$), a 95.62~km object. The clan is entirely stable, with no clones of any member escaping. The members have a relatively low range of reference angle values, fairly close to the $60 \degree$ Lagrange point ($60.44\degree$ to $61.77\degree$), though with a comparatively low libration range ($19.43\degree$ to $29.64\degree$). The clan has a small spread of SDSS colours, particularly in the \emph{(u - g)} colour (1.23-1.51). In the {\tt MOVIS} survey there are narrow \emph{(Y - J)} (0.29 to 0.38) and \emph{(J - Ks)} colour ratios (0.49 - 0.72). The narrow ranges indicate that the colours, along with the dynamics are diagnostic for this clan. 

\paragraph{Thersander Clan: } The Thersander clan, named after 9817 Thersander (6540 P-L) contains 10 objects, and is highlighted in Fig. \ref{fig:Thersander}. This unaffiliated clan, includes 21900 Orus (1999 VQ$_{10}$), a provisionally allocated D-type that is the target of the \textit{Lucy} mission. In the clan, there is also 24341 (2000 AJ$_{87}$), an identified C-type \citep{Fornasier2007VisSpecTrojans}. Close to this clan, there are several members of the Eurybates family, 24341 (2000 AJ$_{87}$), a C-type, 9818 Eurymachos (6591 P-L), a P/X-type \citep{Fornasier2007VisSpecTrojans,Hasselmann2012SDSSTaxonomy} and 65225 (2002 EK$_{44}$). This could have implications for classification of 21900 Orus (1999 VQ$_{10}$), see section \ref{SubSec:lucy} for discussion. The compact SDSS colours are due to only a single object, 53477 (2000 AA$_{54}$), found in the survey. In terms of escapes, a low number of clones escape the swarm, mainly from 14268 (2000 AK$_{156}$) and 24531 (2001 CE$_{21}$).

\subsubsection{Greater Achilles Superclan}
The Greater Achilles superclan contains 35 objects, grouped into three distinct groups, the Epeios (discussed bellow), 1991 El and Achilles clans, as shown in Fig. \ref{Fig:greaterAchilles}. The type object, 588 Achilles (1906 TG), has been classified as a DU-type \citep{Tholen1989Taxonomy}. The majority of the objects in the superclan are classified as D-type, with just two exceptions, both of which are members of the Epeios clan: 12921 (1998 WZ$_5$), a X-type, and 5283 Pyrrhus (1989 BW), which is unclassified, but has an unusual negative spectral slope in \citet{Bendjoya2004JTSpectra}. 

If the more traditional $\delta V_{\textrm{ref}}$  of the superclan is considered, the Eios ($19.55\pm7.42$~ms$^{-1}$) and 1991El ($17.37\pm13.7$~ms$^{-1}$) clans have larger dispersal velocities than the Greater Achilles superclan ($16.91\pm11.2$~ms$^{-1}$), whilst the Epeios ($8.57\pm4.42$~ms$^{-1}$) and Achilles ($8.65\pm6.54$~ms$^{-1}$) clans have smaller $\Delta V{\textrm{cent}}$ than the superclan ($4.4\pm8.6$~ms$^{-1}$). 

The Greater Achilles superclan is relatively stable (0.057 $F_{\textrm{esc}}$). Only 160534 (1996 TA$_{58}$), a member of the Achilles clan, has a high escape fraction (0.78). This is not surprising, as the superclan has a low range of $\delta a_{\textrm{prop}}$ (0.0 to 0.05~au), and is close to the $60 \degree$ Lagrange point ($59.1 \degree$ to $63.1\degree$).

\paragraph{Epeios Clan: }The Epeios clan, named for 2148 Epeios (1976 UW), contains 10 members. The type object was also in the non-canonical Epeios collisional family \citep{Vinogradova2015TrjoanFamilies}. This non-canonical family was associated with the canonical Arkesilaos family \citep{Nesvorny2015AsteroidFamsAIV}, of which we have no members represented. This could indicate with further characterisation in future surveys, members of the Arkesilaos family could form part of this clan. This is supported by the fact that both the Epeios clan and Epeios collisional family \citep{Vinogradova2015TrjoanFamilies} contain X-type objects. 

In this clan, 5283 Pyrrhus (1989 BW) was unclassified, though it has an interesting negative slope in \citet{Bendjoya2004JTSpectra}. Also, within this clan is 12921 (1998 WZ$_5$), an identified X-type \citep{Fornasier2007VisSpecTrojans}. The Epeios clan is entirely stable, with no unstable members. There are a narrow range of SDSS colours, though there are only two members, 37710 (1996 RD$_{12}$) and 168364 (1996 TZ$_{19}$), in the survey. Dynamically, this clan is close to the Lagrange point ($59.1\degree$ to $61.77\degree$), with small libration amplitudes ($4.04\degree$ to $14.33\degree$) and eccentricities (0.01 to 0.1). 

This clan may contain a dynamical pair of objects, 258656 (2002 ES$_{76}$) and 2013~CC$_{41}$ \citep{Holt2020TrojanPair}, the first such objects identified in the Trojan population. Unfortunately neither of these objects are included in this analysis, due to their lack of presence in wide-field surveys. The Epeios clan does not include any D-types, but has a X-type, 12921 (1998 WZ$_5$) \citep{Fornasier2007VisSpecTrojans}, and an object with a potential negative slope, 5283 Pyrrhus (1989 BW) \citep{Bendjoya2004JTSpectra}. These associations are an indication that the 258656-2013 C$_{41}$ pair may have different properties to the majority of the Jovian Trojans.

\subsubsection{Greater Nestor Superclan}
The Great Nestor superclan consists of 37 objects shown in Fig. \ref{Fig:greaterAjax} and includes two distinct clans, Eurymedon and Nestor, as well as several additional members that are not associated with any individual clan. We discuss the Nestor clan in detail below. 
Whilst most of the Trojans are D-types (72.2 per cent), the Greater Nestor superclan contains two large members of other taxonomic types, 659 Nestor (1908 CS), a XC-type \citep{Tholen1989Taxonomy} and 5012 Eurymedon (9507 P-L), a C-type \citep{Hasselmann2012SDSSTaxonomy}, each is the type object of their respective clan. Based on the simulations described in \citet{Holt2020TrojanStability}, the Greater Nestor superclan has the largest escape fraction of any superclan in the L$_4$ swarm, with fully 63 per cent of all test particles generated based on clan members escaping from the Trojan population on a timescale of $4\e9$ years. The more stable members are located in the two clans, but though those clans still exhibit escape fractions higher than the base L$_4$ escape fraction (at 0.3 and 0.4, respectively). The superclan, as a whole, has an average $\delta a$ range (0.03 to 0.11~au), with relatively high eccentricities (0.07 to 0.17). 

\paragraph{Nestor clan: } This clan contains 7 objects, two of which have been taxonomically identified, the XC-type 659 Nestor (1908 CS), and D-type 4060 Deipylos (1987 YT$_1$) \citep{Bendjoya2004JTSpectra}. \citet{Holt2020TrojanStability} noted a slightly larger escape rate amongst the X-types in the Trojans, and this is reflected in this clan. The Nestor Clan has a relatively high escape fraction (0.4), versus that of the L$_4$ swarm (0.23) as a whole. The members of the clan all display centres of libration that are slightly ahead of the $60 \degree$ point ($60.44 \degree$ to $64.43\degree$), though the range of amplitudes is relatively small ($14.33\degree$ to $24.54 \degree$). With the diversity of taxonomic types within the clan, it is not surprising that the members also display a wide range of SDSS colours ,\emph{(b - v)}:0.65 - 0.99, \emph{(u - g)}: 1.62 - 2.29, \emph{(g - r)}: 0.43 - 0.77, \emph{(r - i)}: 0.18 - 0.27. The narrow range of {\tt MOVIS} values are due to only a single representative of the clan, 4060 Deipylos (1987 YT$_1$),\emph{(Y - J)}:0.241, \emph{(J - Ks)}:0.547, \emph{(H - Ks)}:0.137, in the survey. 

\subsubsection{Greater Ajax Superclan}
% This is the example superclan
\label{SS:GreaterAjax}
The 29 objects in this superclan, and the associated Ajax and Eurybates clans, are shown in Fig. \ref{Fig:greaterAjax}. We use this superclan, and the following detailed discussion of both clans, as examples for the rest of the consensus trees, found in Appendix \ref{App:clans}. This superclan includes the many members of the Eurybates collisional family. The cluster is not named the `Eurybates superclan', as 1404 Ajax (1936 QW) was discovered in 1936 \citep{Wyse1938JupiterTrojans}, nearly 40 years before 3548 Eurybates (1973 SO). This superclan is one of the most complex in the L$_4$ swarm, with multiple subclans in each clan. Apart from one unassociated object, 100619 (1997 TK$_{14}$), all objects are in one of the clans. In terms of escapes, the Greater Ajax superclan is has a higher escape fraction (0.38) than the L$_4$ swarm as a whole. The group is dynamically diverse, though they have a compact $\delta a_{\textrm{prop}}$ range ($0.07\degree$ to $0.11\degree$). Relatively compact SDSS values, \emph{(b - v)}:0.65-0.93, \emph{(u - g)}: 1.25-1.72, \emph{(g - r)}: 0.43-0.7, \emph{(r - i)}: 0.15-0.29, may be an actual feature of this superclan, as eight of the 29 superclan objects are represented in the SDSS survey, though the \emph{(i - z)} color has quite a wide range (-0.03-0.26). 	

\paragraph{Ajax clan:} In this clan there are three subclans (Ajax, Hiera and 2002 CQ$_{134}$), each consisting of four objects in a branching format. The Hiera and 2002 CQ$_{134}$ subclans form a sister group to the Ajax subclan. Unfortunately, there are no taxonomically identified members of this clan. With the close association to the Eurybates family, this makes the three largest members of the clan, 1404 Ajax (1936 QW), 4086 Podalirius (1985 VK$_2$) and  7119 Hiera (1989 AV$_2$), all of which have a absolute H-magnitude greater than 9, of particular interest for future telescope observations, see section \ref{Sec: targets}. Most of the escapes in the Greater Ajax superclan come from this clan. The 2002 CQ$_{134}$ subclan has a large escape fraction (0.81), with all members having an escape fraction over 0.65. 

The clan is located well ahead of the $60 \degree$ point, with mean libration angle between $63.1\degree$ and $65.76\degree$. The clan does have a relatively narrow range of dynamical values ($\Delta a_{\textrm{prop}}$:0.09~au-0.11~au, $e_{\textrm{prop}}$:0.03-0.08, $\textrm{sin}i_{\textrm{prop}}$:0.28-0.49), that could be diagnostic. In addition, some of the SDSS values may also be diagnostic, \emph{(b - v)}:0.86-0.93, \emph{(u - g)}:1.44-1.63, \emph{(g - r)}:0.63-0.7, \emph{(r - i)}:0.15-0.24, with three members of the clan represented, 4086 Podalirius (1985 VK$_2$),  24403 (2000 AX$_{193}$) and 42367 (2002 CQ$_{134}$). Two additional members , 207749 (2007 RC$_{286}$) and 316158 (2009 UW$_{26}$), are represented in the {\tt MOVIS} dataset with similar values, \emph{(Y - J)}: 0.469-0.494, \emph{(J - Ks)}: 0.608-0.712. The range of \textit{Gaia} values from six different sized members is broader (17.29-18.93~mag), highlighting the need for further investigations into members of this clan. 

\paragraph{Eurybates clan: } There are two subclans (Anius and Eurybates) in this clan. The Anius subclan has five members, with two duos and a single object, in a 1:2:2 format. The Eurybates subclan (eight members), as expected for the group containing many members of the Eurybates collisional family \citep{Broz2011EurybatesFamily,Nesvorny2015AsteroidFamsAIV}, has a comparatively complex structure (three duos and two singles in 2:1:1:2:2 format). The type object of Eurybates clan, 3548 Eurybates (1973 SO) is a target for the \textit{Lucy} mission. There are three other Eurybates family members, 39285 (2001 BP$_{75}$), 24380 (2000 AA$_{160}$), 28958 (2001 CQ$_{42}$), all C-types \citep{Fornasier2007VisSpecTrojans}, in close association under the Eurybates subclan. The other four members of the Eurybates subclan, 39793 (1997 SZ$_{23}$), 137879 (2000 AJ$_{114}$), 312457 (2008 QH$_{42}$), 315208 (2007 RS$_{22}$), and possibly two in the Anoius subclan, 12917 (1998 TG$_{16}$) and 111932 (2002 GG$_{33}$), are likely previously unidentified members of the collisional family. The age of this collisional family has been identified as approximately,$1.045\pm 0.364\e9$ years \citep{Holt2020TrojanStability}. With that long an age, the possibility for interlopers is quite high, as the true members of the collisional family disperse. 18060 (1999 XJ$_{156}$) is a X-type in \citet{Fornasier2007VisSpecTrojans}, and the corresponding SDSS colours, \emph{(b - v)}:0.70, \emph{(u - g)}:1.69, \emph{(g - r)}:0.48, \emph{(r - i)}:0.21, \emph{(i - z)}:0.06, are different to other members. The Eurybates clan, which includes members of the Eurybates collisional family, has a lower escape fraction (0.23) than the superclan as a whole (0.38). The clan escape fraction (0.23) is similar to the escape fraction of the Eurybates collisional family (0.1881) found by \citet{Holt2020TrojanStability}. If we disregard the X-type (18060 (1999 XJ156), \emph{(g - r)}:0.48), the SDSS \emph{(g - r)} colour is contained within a single bin, \emph{(g - r)}: 0.633-0.7. 

\subsubsection{Greater Hektor superclan}
This superclan contains the only member of the Hektor collisional family \citep{Rozehnal2016HektorTaxon}, 624 Hektor (1907 XM), considered in our analysis. The supercaln also contains many other objects identified as D-type \citep{Roig2008JovTrojanTaxon, Rozehnal2016HektorTaxon}. The exception, 5285 Krethon (1989 EO$_{11}$), is a XD-type \citep{Bendjoya2004JTSpectra} in the Thersites clan, which with further examination could be reidentified as a true D-type. In the superclan, the $\Delta V_{\textrm{ref}}$ and $\Delta V_{\textrm{cent}}$ are similar ($31.47\pm19.35$~ms$^{-1}$ and $28.99\pm21.63$~ms$^{-1}$), as well as in both clans (Thersites clan: $34.57±33.85$~ms$^{-1}$ and $27.49±23.04$~ms$^{-1}$, Hektor clan: $31.43\pm16.53$~ms$^{-1}$ and $27.93\pm16.05$~ms$^{-1}$), with each of mean velocities being smaller than the $V_{\textrm{esc}}$ of 624 Hektor (1907 XM) ($75.24$~ms$^{-1}$). 

Dynamically, the superclan is ahead of the Lagrange point ($63.1\degree$ to $68.4\degree$), with a fairly high libration range ($34.75\degree$ to $60.27\degree$) and $\delta a_{\textrm{prop}}$ (0.09~au to 0.12~au). Some of the compact range of SDSS values, \emph{(b - v)}:0.72-0.86, \emph{(u - g)}:1.16-1.63, \emph{(g - r)}:0.5-0.63, \emph{(i - z)}:0.09-0.2, could be diagnostic, but a wider range of other colours, \emph{(r - i)}:0.18-0.29), {\tt MOVIS}, \emph{(Y - J)}:0.02-0.46, \emph{(J - Ks)}:0.37-1.18, and \textit{Gaia} (15.11-18.38~mag) are indicative of heterogeneity in the superclan. 
\paragraph{Thersites clan: } As with the superclan, almost all members of this clan are identified as D-types \citep[1868 Thersites (2008 P-L), 4946 Askalaphus (1988 BW$_1$), 2797 Teucer (1981 LK), 20995 (1985 VY),][]{Bendjoya2004JTSpectra, Hasselmann2012SDSSTaxonomy}.  Most of the unstable members of the Hektor superclan are in the Thersites clan, with six members of the clan having all nine clones escape, 2797 Teucer (1981 LK), 4946 Askalaphus (1988 BW$_1$), 8317 Eurysaces (4523 P-L), 20995 (1985 VY), 37298 (2001 BU$_{80}$) and 266869 (2009 UZ$_{151}$). This clan has higher eccentricity (0.03 to 0.11) and libration range ($39.85\degree$-$60.27\degree$) compared with the Hektor clan. As with the superclan, the SDSS colours are compact, \emph{(b - v)}:0.79-0.86, \emph{(u - g)}:1.35-1.53, \emph{(g - r)}:0.57-0.63, \emph{(r - i)}:0.23-0.29, \emph{(i - z)}: 0.09-0.2, and with four members in the survey, 2797 Teucer (1981 LK), 4946 Askalaphus (1988 BW$_1$), 20995 (1985 VY) and 38606 (1999 YC$_{13}$), could be diagnostic. There are three members represented in {\tt MOVIS}, 173086 Nireus (2007 RS$_8$), 200023 (2007 OU$_6$), 264155 (2009 VJ$_{109}$) and 266869 (2009 UZ$_{151}$), though 264155 (2009 VJ$_{109}$), has quite different colours, \emph{(Y - J)}:0.294, \emph{(J - Ks)}:1.004, compared to the other three, \emph{(Y - J)}:0.398-0.492, \emph{(J - Ks)}:0.309-0.909. 

\paragraph{Hektor clan: }The type object of this clan, 624 Hektor (1907 XM), is the largest object in the Jovian Trojan population \citep[225km,][]{Marchis2014Hektor}. It is also the largest member of the Hektor collisional family \citep{Rozehnal2016HektorTaxon}. Unfortunately, 624 Hektor (1907 XM) is the only member of the collisional family studied in this analysis, therefore any conclusions about potential family memberships are speculative at best. Most of the instability in this clan in confined to two members, 24275 (1999 XW$_{167}$) and 42230 (2001 DE$_{108}$), both of which have only a single clone remaining at the end of the \citet{Holt2020TrojanStability} simulations. The clan as a reasonably high $\Delta a_{\textrm{prop}}$ values (0.09~au - 0.12~au). As with the Hektor superclan, most of the SDSS colours, \emph{(b - v)}:0.72-0.86, \emph{(u - g)}:1.16-1.63, \emph{(g - r)}:0.5-0.63, \emph{(i - z)}:0.09-0.2, are compact, with a range of \emph{(r - i)} (0.18-0.29), {\tt MOVIS}, \emph{(Y - J)}:0.02-0.46, \emph{(J - Ks)}:0.49-1.18, and \textit{Gaia} (15.11-18.38~mag) values. The type object, 624 Hektor (1907 XM), shows a level of heterogeneity in the spectra \citep{Perna2018HektorAgamemnonSpec}, agreeing with the compact values for the clan. Interestingly, 1583 Antilochus (1950 SA) and 3801 Thrasymedes (1985 VS), were identified as potential asteroid pair \citep{Milani1993JovianTrojFamilies}, though this was not confirmed by \citet{Holt2020TrojanPair}. In our analysis these two objects are next to one another in the dendritic tree (Fig. \ref{Fig:greaterHektor}), lending strength to our analysis. 

\subsubsection{Greater Diomedes superclan}
\label{SS:GreaterDiomedes}
This is the largest superclan in the L$_4$ swarm, with 71 members. It also has the largest $\Delta V_{\textrm{ref}}$ of any superclan ($108.79\pm36.64$~ms$^{-1}$). The $\Delta V_{\textrm{cent}}$ is more reasonable ($41.85\pm23.23$~ms$^{-1}$), closer to the $V_{\textrm{esc}}$ of 1437 Diomedes (1937 PB) ($39.3$~ms$^{-1}$), the type object of the superclan. The superclan includes two \textit{Lucy} targets, 11351 Leucus (1997 TS$_{25}$) and 15094 Polymele (1999 WB$_2$). They are both provisionally classified differently, with 15094 Polymele (1999 WB$_2$) being a X-type \citep{Buie2018LeucusPolymele,SouzaFeliciano2020LucyTargVis} and 11351 Leucus (1997 TS$_{25}$) a D-type \citep{Buie2018LeucusPolymele}. They are in two separate clusters, with 5094 Polymele (1999 WB$_2$) not in any clan, and 11351 Leucus (1997 TS$_{25}$) in the Diomedes clan with another DX-type, 1437 Diomedes (1937 PB). The dynamical stability of the different clans within the superclan is markedly variable, with some significantly less stable than others (e.g. Diomedes clan which has an escape fraction of 0.76, compared to the Philoctetes clan, with an escape fraction of 0.38). The escape rates within each clan, however, are relatively consistent - so all objects within an unstable clan are similarly unstable, whilst those in the stable clans are all relatively stable, and each of theses clans has a larger escape fraction than that of the overall L$_4$ swarm (0.2335). 

\paragraph{Philoctetes clan: } This clan with 26 members, displays a high diversity of taxonomic types, three X-types \citep[19725 (1999 WT$_4$), 24233 (1999 XD$_{94}$) and 23963 (1998 WY$_8$);][]{Hasselmann2012SDSSTaxonomy}, a C-type \citep[24420 (2000 BU$_{22}$),][]{Fornasier2007VisSpecTrojans}, and a D-type \citep[9590 (1991 DK$_1$);][]{Hasselmann2012SDSSTaxonomy} in the Andraimon subclan. This clan also contains three members of the Eurybates family \citep[24420 (2000 BU$_{22}$), 111805 (2002 CZ$_{256}$) and 24426 (2000 CR$_{12}$),][]{Nesvorny2015AsteroidFamsAIV}, and a fourth non-canonical member \citep[63291 (2001 DU$_{87}$),][]{Rozehnal2016HektorTaxon}. A large fraction of this clan is represented in the SDSS database (0.6923), with relatively compact colours, \emph{(b - v)}:0.58-0.93, \emph{(u - g)}:1.16-1.63, \emph{(g - r)}:0.37-0.7, \emph{(i - z)}:-0.03-0.2, though there is a wide \emph{(r - i)} range (0.1-0.24). There is only a single representative of the clan in the \textit{Gaia} survey (19725 (1999 WT$_4$), 18.67~mag), so the value range here is only indicative. As the largest object in the clan, 1869 Philoctetes is relatively small (33.96 km), the $V_{\textrm{esc}}$ ($11.36$~ms$^{-1}$) is lower than the $\Delta V_{\textrm{ref}}$ ($25.1\pm7.9$~ms$^{-1}$).

\paragraph{Diomedes clan: } This mid-sized (12 members) clan, contains 11351 Leucus (1997 TS$_{25}$), a D-type \citep{Fornasier2007VisSpecTrojans} \textit{Lucy} target. The type object of the clan, 1437 Diomedes (1937 PB) is also classified as a DX-type \citep{Tholen1989Taxonomy}. The $\Delta V_{\textrm{cent}}$ for the clan is relatively high ($56.9\pm28.09$~ms$^{-1}$), though close to the $V_{\textrm{esc}}$ of the large type object ($39.3$~ms$^{-1}$). With relatively high $\Delta a_{\textrm{prop}}$ values (0.11 to 0.16~au) and mean center of libration values ($67.09\degree$ to $73.74\degree$;	Amplitude: $50.06\degree$ to $75.59\degree$), it is unsurprising that this clan has a high escape rate (0.76). In the SDSS dataset, there are only three members represented, 5209 (1989 CW$_1$), 43706 Iphiklos (1416 T-2) and 83977 (2002 CE$_{89}$), and with only two in the MOVIS database, 11397 (1998 XX$_{93}$) and 65228 (2002 EH$_{58}$), it is difficult to make any conclusions regarding colour distribution. The wide range of WISE (W1:0.08-0.26, W2:0.06-0.28) albedos indicate that there is a variety of compositions in this clan. 

\subsubsection{Greater Telamon superclan}
The Greater Telamon superclan which has 35 members, including three separated clans, Telmon, Kalchas and Theoklymenos. The Telmon and Kalchas clans are relativly small, with 5 and 6 members respectively.   The Theoklymenos clan is larger, at 19 members, and contains a X-type \citep[5023 Agapenor (1985 TG$_3$)][]{Hasselmann2012SDSSTaxonomy}, and two D-types \citep[24390 (2000 AD$_{177}$ and 3063 Makhaon (1983 PV)][]{Fornasier2007VisSpecTrojans, Lazzaro2004s3os2asteroids}.  We discuss the Kalachas clan in more detail below. 

This is one of the most stable superclans ($F_{\textrm{esc}}$: 0.05) in the Trojan population. Most of the escape values in the superclan originate with the type object, 1749 Telamon (1949 SB), where all nine particles escape \citep{Holt2020TrojanStability}. Within this only supercaln 3063 Makhaon (1983 PV) has a higher escape fraction (0.33) higher than the L$_4$ swarm (0.23).

Other superclan members have all nine clones stay in the L$_4$ Trojan region.  With moderate $\Delta a$ values (0.04-0.09~au) and a location near the Lagrange point ($59.61\degree$ to $64.43\degree$), this stability is not surprising. In general, the clan has low \textit{WISE} albedos (W1:0.102-0.239 , W2:0.102-0.251). The exception is 24225 (1999 XV80) (W1:0.378, W2:0.378), which extends the ranges of the superclan as well as the Theoklymenos clan. The SDSS values are relatively diverse, \emph{(b - v)}:0.65-0.93, \emph{(u - g)}:1.35-1.72, \emph{(g - r)}:0.43-0.7, \emph{(i - z)}:-0.03-0.26), particularly the \emph{(r - i)} color (0.16-0.34).

\paragraph{Kalachas clan: } The Kalachas clan contains two X-type objects, 4138 Kalchas (1973 SM) and 7152 Euneus (1973 SH$_1$) \citep{Bendjoya2004JTSpectra}, both of similar size (46.46~km and 45.52~km respectively). The smaller of the two, 7152 Euneus (1973 SH$_1$) has a low $\Delta V_{\textrm{ref}}$ (5.4ms$^{-1}$) to 138 Kalchas (1973 SM), which is the reference object for the clan. Even though they were not identified in \citet{Holt2020TrojanPair}, their $\Delta V_{\textrm{ref}}$, similar properties and sizes, indicate that these two large objects could be an ancient disrupted binary pair \citep{Vokroulicky2008AsteroidPairs,Pravec2019AsteroidPairs}. 

All members of this clan are stable over the life of the Solar system \citep{Holt2020TrojanStability}. The clan has very low proper eccentricities (0.0161-0.0532) and sin$i$ (0.0102-0.119) values, and with mid-range $\delta a_{\textrm{prop}}$ values, places the clan within the stable parameter space \citep{Nesvorny2002SaturnTrojanHypothetical, DiSisto2014JupTrojanModels, Hellmich2019TrojanYarkovski, Holt2020TrojanStability}. The $\Delta V_{\textrm{ref}}$ of the clan is relatively small ($16.88\pm10.7$~ms$^{-1}$), and close to the $V_{\textrm{esc}}$ of 4138 Kalchas (1973 SM) ($15.5$~ms$^{-1}$).  With a relatively high fraction of objects (50 per cent) represented in the SDSS catalogue, the \emph{(b - v)}, \emph{(u - g)}, \emph{(g - r)} and \emph{(i - z)} colours are possibly diagnostic, \emph{(b - v)}:0.72-0.86, \emph{(u - g)}:1.44-1.72, \emph{(g - r)}:0.5-0.7, \emph{(i - z)}:0.09-0.15. The range of \emph{(r - i)} SDSS colours (0.16-0.23) are mainly due to 89924 (2002 ED$_{51}$), \emph{(r - i)}: 0.225 being a possible outlier. 

\subsubsection{Greater Odysseus superclan}
The Odysseus superclan (36 members) contains two clans, Epistrophs (5 members) and Odysseus (20 members), neither of which is discussed here in detail. There is a diversity of taxonomic types in this superclan. The type object, 1143 Odysseus (1930 BH) is classified as a D-type \citep{Tholen1989Taxonomy}, though there are two other objects with taxonomic classifications, namely 24882 (1996 RK$_{30}$) which is an X-type, and 21372 (1997 TM$_{28}$) classified as a C-type \citep{Hasselmann2012SDSSTaxonomy}. The Epistrophos clan contains two D-types \citep[39293 (2001 DQ$_{10}$) and 23382 Epistrophos (4536 T-2)][]{Hasselmann2012SDSSTaxonomy}. There is a X-type \citep[13463 Antiphos (5159 T-2)][]{Fornasier2007VisSpecTrojans}, another D-type \citep[15535 (2000 AT$_{177}$)][]{Fornasier2007VisSpecTrojans}, and a X-type \citep[24485 (2000 YL$_{102}$)][]{Hasselmann2012SDSSTaxonomy}, that are not associated with any clan. 

The range of albedos and colours reflect the diversity in the superclan. Much of this can, however, be explained by several outliers, for example, 9713 Oceax (1973 SP$_1$)in the Odysseus clan has high \textit{WISE} (W1:0.336 , W2:0.336) and geometric (0.168) albedos compared with the rest of the objects. A particularly interesting object is 128383 (2004 JW$_{52}$), in terms of its colours. The SDSS colours for 128383 (2004 JW$_{52}$) are high for \emph{(b - v)} (1.55) and \emph{(g - r)} (1.3), but low for \emph{(i - z)}, (-0.55),  the opposite of the rest of the superclan, \emph{(b - v)}:0.649-0.857, \emph{(g - r)}:0.433-0.7, \emph{(i - z)}:-0.0167-0.25. This one outlier accounts for much of the SDSS variation. 

The superclan ($\Delta V_{\textrm{cent}}: 18.41\pm9.07$~ms$^{-1}$) and clans (Odysseus: $\Delta V_{\textrm{cent}}: 12.54\pm8.43$~ms$^{-1}$) are fairly compact, particularly the Epistrophos clan ($\Delta V_{\textrm{cent}}: 6.99\pm2.47$~ms$^{-1}$). This superclan is also quite stable ($F_{\textrm{esc}}$: 0.11), with the majority of the instability coming form the unaffiliated superclan members, such as 22404 (1995 ME$_4$), where all the clones escape. The Epistrophos and Odyssesus clan members are all completely stable, due to both sets being close to the Lagrange point ($60.44\degree$ to $61.77\degree$ and $59.1\degree$ to $61.77\degree$ respectively). 

\subsection{L$_5$ Swarm}
\label{SubSec: L5}
In our analysis of the L$_5$ swarm, we present a consensus tree of 407 objects in Fig. \ref{Tree:L5}.  A total of 10,000 equally parsimonious trees took approximately 10 hrs 26 minutes to find using a single core of Intel Xeon W-2133 CPU at 3.60GHz. The consensus tree has a length of 1984.79, with a consensus index of 0.113 \citep{Brooks1986ConsistancyIndex} and retention index of 0.712 \citep{Naylor1995RetentionIndex}. The superclans, clans and sublclans identified in the L$_5$ swarm are listed in Table\ref{tab:L5}. In the L$_5$ swarm, there are seven clans unaffiliated with any superclan with six subclans within them. There is a small number of large superclans (three), compared with the L$_4$ swarm, and each superclan contains a larger number of clans and subclans. In total there are 14 clans containing a total of 14 subclans. Overall, the L$_5$ swarm contains more hierarchical structure than the L$_4$ swarm, shown in Fig. \ref{Tree:L5}.

In the L$_5$ swarm, there are two canonical collisional families, 2001 UV$_{209}$ and the larger Ennomos family \citep{Nesvorny2015AsteroidFamsAIV}. \citet{Vinogradova2015TrjoanFamilies} questioned the existence of any collisional families in the L$_5$swarm, thought they did note some clustering around 247341 2001 UV$_{209}$, 11487 (1988 RG$_{10}$), and 4709 Ennomos (1988 TU$_2$). \citet{Rozehnal2016HektorTaxon} has a similar dataset to the canonical one, with a few extra objects. The non-canonical 2001 UV$_{209}$ and several Ennomos family members are in the Cebriones and Troilus clans of the Greater Patroclus superclan, along with the two canonical Ennomos family members. The Ennomos family is more problematic. In our subset, there are nine members, spread throughout the L$_5$swarm. There is a small cluster of three members in the Aneas clan, though the largest member of the collisional family, 4709 Ennomos (1988 TU$_2$), is located in the Cebriones clan, Greater Pratoclus superclan,  with two other non-canonical members. The hierarchical structure seen in the L$_5$ swarm through astrocladistics could indicated that the dynamical history of the swarm is more complex than can be reliably identified by HCM, and as indicated by the lack of confident clusters in \citet{Vinogradova2015TrjoanFamilies}. 

%Tree
\begin{figure*}
	\includegraphics[height=0.90\textheight]{Trees/SubMatrixWiseSDSSGaiaMovis-L5-Conserv-app.pdf}
    \caption{Consensus tree of cladistical analysis of 407 L$_5$ Jovian Trojans. Numbers indicate fraction of 10000 trees where branch is present. Colours are indicative of previously identified collisional families: Brown: Ennomos; Purple: 2001 UV$_{209}$ after \citet{Nesvorny2015AsteroidFamsAIV}. \textit{Letters} associate objects with Bus-Demeo taxonomy \citep{Bus2002AsteroidTax, DeMeo2009AsteroidTax} from different sources, T1989: \citet{Tholen1989Taxonomy}; B2004: \citet{Bendjoya2004JTSpectra}; F2004: \citep{Fornasier2004L5TrojanSpec}; F2007: \citet{Fornasier2007VisSpecTrojans}; H2012, with associated confidence rating: \citet{Hasselmann2012SDSSTaxonomy}. \textit{Lucy} Targets are indicated by an \textbf{L}. A high resolution, expanded form of this figure is available in online supplemental material. Blue rectangles correspond to detailed figures in Appendix \ref{App:clans}.}
    \label{Tree:L5}
\end{figure*}

\begin{table}
\caption{Clans, \textbf{superclans} and \textit{subclans} identified in the L$_5$ Trojan swarm. $Name$: Family Name; $N$: Number of members; $D_{\textrm{ref}}$: Reference object diameter; $V_{\textrm{esc}}$: Escape velocity of reference object; $F_{\textrm{esc}}$: fraction of objects that escape the L$_5$ Lagrange point, from \citet{Holt2018JovSatSatsClad}; $\Delta V_{\textrm{ref}}$: dispersal velocity relative to the reference object (calculated using the inverse Gauss equations; see section~\ref{SubSec:Method:DisVelo}), with $1 \sigma$ standard deviation; $\Delta V_{\textrm{cent}}$: as $\Delta V_{\textrm{ref}}$, with calculations to the fictitious cluster center;  $F_{node}$: faction of trees in the block that contain the node.}
\label{tab:L5}
\resizebox{\columnwidth}{!}{%
\begin{tabular}{llllllll}
\hline
$Name$              & $No.$ & $D_{refobj}$ & $V_{esc}$ & $F_{esc}$ & $\Delta V_{ref}$ & $\Delta V_{cent.}$ & $F_{node}$\\
                    &       & km           & $ms^{-1}$ &           & $ms^{-1}$        & $ms^{-1}$         & \\
\hline
L5-Asteropaios      & 17     & 57.65     & 19.28      & 0.06               & 25.94$\pm$11.96  & 15.48$\pm$6.14 & 0.9304 \\
\textit{L5-Lykaon Sub} & 5      & 50.87     & 17.01      & 0.07               & 10.22$\pm$2.14   & 7.06$\pm$1.84  & 0.9588 \\
\textit{L5-1988 RS$_{10}$ Sub}& 7      & 32.14     & 10.75      & 0.02              & 9.97$\pm$6.12    & 8.31$\pm$5.99  & 0.8942 \\
                    &        &           &            &                    &                  &                &  \\
L5-Dolon            & 20     & 42.52     & 14.22      & 0.31               & 46.28$\pm$27.85  & 34.71$\pm$19.81 & 0.9999\\
\textit{L5-Erichthonios Sub } & 5      & 27.53     & 9.21       & 0.04    & 13.36$\pm$10.86  & 9.95$\pm$6.73  & 0.9999 \\
\textit{L5-Dolon Sub} & 11     & 42.52     & 14.22      & 0.35               & 25.77$\pm$13.42  & 23.52$\pm$14.58 & 1.0000 \\
                    &        &           &            &                    &                  &                &  \\
L5-Apisaon          & 9      & 40.67     & 13.6       & 0.67               & 26.29$\pm$7.75   & 20.31$\pm$7.29 & 0.9794 \\
                    &        &           &            &                    &                  &                &  \\
L5-Khryses          & 8      & 53.2      & 17.79      & 0.53               & 14.41$\pm$5.08   & 8.02$\pm$3.15 & 0.9784  \\
                    &        &           &            &                    &                  &                &  \\
L5-1999 RU$_{12}$   & 5      & 24.01     & 8.03       & 0.84               & 29.24$\pm$11.18  & 17.87$\pm$4.84 & 1.0000 \\
                    &        &           &            &                    &                  &                &  \\
L5-1990 VU$_1$          & 21     & 63.19     & 21.13      & 0.52               & 28.35$\pm$28.23  & 24.87$\pm$18.96 & 0.9391 \\
\textit{L5-1990 VU$_1$ Sub}&7      & 59.3      & 19.83      & 0.73               & 61.38$\pm$30.9   & 31.94$\pm$9.74 & 0.9788 \\
\textit{L5-Idaios Sub} & 8      & 44.55     & 14.9       & 0.38               & 13.72$\pm$5.31   & 8.93$\pm$3.59  & 0.9395 \\
                    &        &           &            &                    &                  &                &  \\
L5-Anchises         & 8      & 99.55     & 33.29      & 0.88               & 32.88$\pm$10.82  & 22.51$\pm$13.02 & 0.9612\\
                    &        &           &            &                    &                  &                &  \\
\textbf{L5-Greater Patroclus} & 133    & 140.36    & 46.94      & 0.1                & 31.57$\pm$20.49  & 31.62$\pm$15.09 & 0.8377\\
L5-Memnon           & 23     & 118.79    & 39.72      & 0.11               & 15.31$\pm$7.57   & 15.05$\pm$6.59  & 0.9534\\
\textit{L5-Memnon Sub}        & 9      & 118.79    & 39.72      & 0.15               & 16.02$\pm$7.97   & 12.24$\pm$7.9 & 0.9332   \\
\textit{L5-Amphios Sub}       & 9      & 38.36     & 12.83      & 0.14               & 14.06$\pm$8.13   & 13.04$\pm$7.89 & 0.9727  \\
L5-1971 FV$_1$          & 18     & 75.66     & 25.3       & 0.07               & 12.76$\pm$7.58   & 9.43$\pm$5.5  & 0.9065  \\
\textit{L5-Lampos Sub}       & 6      & 35.39     & 11.83      & 0.22               & 21.47$\pm$4.71   & 11.11$\pm$1.99 & 1.0000 \\
\textit{L5-1971 FV$_1$ Sub}      & 8      & 75.66     & 25.3       & 0                  & 6.35$\pm$0.93    & 5.28$\pm$2.55 & 0.9801  \\
L5-1989 TX$_{11}$         & 5      & 28.26     & 9.45       & 0                  & 10.8$\pm$1.35    & 4.33$\pm$2.38 & 0.9248\\
L5-Phereclos        & 17     & 94.62     & 31.64      & 0.01               & 9.97$\pm$5.41    & 8.14$\pm$3.84 & 0.8886  \\
\textit{L5-Pandarus Sub}      & 5      & 82.03     & 27.43      & 0.04               & 18.61$\pm$0.94   & 7.48$\pm$3.44 & 0.9996  \\
\textit{L5-Phereclos Sub}     & 12     & 94.62     & 31.64      & 0                  & 10.61$\pm$5.66   & 7.29$\pm$3.49 & 0.889  \\
L5-Troilus          & 13     & 100.48    & 33.6       & 0.09               & 18.95$\pm$7.41   & 12.88$\pm$5.81 & 1.0000 \\
\textit{L5-Troilus Sub}      & 5      & 100.48    & 33.6       & 0                  & 15.95$\pm$5.0    & 8.2$\pm$4.36 & 1.0000   \\
\textit{L5-1988 RY$_{11}$ Sub}     & 5      & 39.75     & 13.29      & 0.2                & 22.58$\pm$6.65   & 10.33$\pm$4.42 & 1.0000 \\
L5-Cebriones        & 17     & 95.98     & 32.09      & 0.23               & 24.9$\pm$10.97   & 19.04$\pm$7.42 & 1.0000 \\
\textit{L5-Bitias Sub}        & 7      & 47.99     & 16.05      & 0.35               & 23.47$\pm$9.83   & 13.74$\pm$4.09 & 0.9798 \\
                    &        &           &            &                    &                  &                &  \\
\textbf{L5-Greater Aneas }    & 64     & 118.22    & 39.53      & 0.2                & 65.85$\pm$34.23  & 38.7$\pm$20.27 & 0.8884 \\
L5-1988 RH$_{13}$         & 6      & 53.1      & 17.76      & 0.65               & 36.74$\pm$22.51  & 23.44$\pm$15.43 & 0.9350 \\
L5-1994 CO           & 5      & 47.73     & 15.96      & 0.13               & 28.95$\pm$5.47   & 21.4$\pm$11.93 & 0.9997 \\
L5-1989 UQ$_5$          & 5      & 25.91     & 8.66       & 0                  & 25.34$\pm$4.92   & 9.72$\pm$7.17  & 1.0000 \\
L5-Sarpedon         & 17     & 77.48     & 25.91      & 0.04               & 23.15$\pm$13.77  & 18.98$\pm$6.16 & 0.9596 \\
\textit{L5-Hippokoon Sub}     & 5      & 18.43     & 6.16       & 0                  & 12.03$\pm$4.65   & 9.9$\pm$5.74 & 1.0000    \\
\textit{L5-Sarpedon Sub}     & 11     & 77.48     & 25.91      & 0.04               & 16.65$\pm$13.65  & 15.97$\pm$9.07 & 0.9596 \\
L5-Aneas            & 26     & 118.22    & 39.53      & 0.2                & 34.55$\pm$18.97  & 22.42$\pm$13.03 & 0.9544\\
\textit{L5-Helicaon Sub}      & 5      & 32.54     & 10.88      & 0.27               & 8.95$\pm$3.27    & 7.18$\pm$3.51 & 1.0000  \\
\textit{L5-Iphidamas Sub}    & 5      & 49.53     & 16.56      & 0.33               & 15.8$\pm$6.57    & 13.08$\pm$5.2  & 0.9795 \\
\textit{L5-Aneas Sub}        & 8      & 118.02    & 39.47      & 0.07               & 30.7$\pm$20.66   & 23.17$\pm$13.27 & 0.9744 \\
                    &        &           &            &                    &                  &                &  \\
\textbf{L5-Greater Astyanax}  & 80     & 126.29    & 42.23      & 0.45               & 92.28$\pm$43.46  & 50.89$\pm$28.51 & 0.9647 \\
L5-Mentor           & 20     & 126.29    & 42.23      & 0.41               & 48.51$\pm$24.27  & 32.58$\pm$18.62 & 0.9732  \\
\textit{L5-1988 RR$_{10}$ Sub}     & 5      & 29.08     & 9.72       & 0                  & 19.21$\pm$3.39   & 9.84$\pm$4.86 & 1.0000   \\
\textit{L5-Mentor Sub}       & 10     & 126.29    & 42.23      & 0.58               & 36.72$\pm$8.02   & 24.68$\pm$11.69 & 0.9732 \\
L5-Helenos          & 11     & 34.05     & 11.39      & 0.12               & 40.81$\pm$19.19  & 21.03$\pm$11.21 & 1.0000 \\
L5-Astyanax         & 41     & 53.98     & 18.05      & 0.55               & 141.79$\pm$42.59 & 42.8$\pm$28.01 & 0.9735  \\
\textit{L5-Ophelestes Sub}   & 5      & 32.39     & 10.83      & 0.18               & 20.79$\pm$12.53  & 13.7$\pm$6.25 & 1.0000  \\
\textit{L5-Astyanax Sub}      & 6      & 53.98     & 18.05      & 0.54               & 99.92$\pm$57.95  & 55.04$\pm$28.89 & 0.9796 \\
\textit{L5-Acamas Sub}       & 6      & 43.86     & 14.67      & 0.74               & 102.44$\pm$11.67 & 31.58$\pm$31.34 & 1.0000\\
\textit{L5-1989 UX$_5$ Sub}      & 9      & 32.19     & 10.76      & 0.56               & 15.28$\pm$8.38   & 12.97$\pm$7.93 & 0.9735\\
\hline
\end{tabular}%
}
\end{table}

\subsubsection{Unaffiliated L$_5$ clans}
There are seven clans that are unaffiliated with any superclan in the L$_5$ swarm. In this section we discuss the Dolan (20 members), 1990 VU$_1$ (21 members) and Anchises (8 members) clans. The values for the other four clans, Asteropaios (17 members), Apisaon (9 members), Khryses (8 members) and 1990 VU$_1$ are available in the Github repository\footnote{\url{https://github.com/TimHoltastro/holt-etal-2021-Jovian-Trojan-astrocladistics.git}}. In the superclan, the Asteropaios, Dolan and 1990 VU$_1$ each have have two subclans, shown in Fig. \ref{fig:UnassL5clans}, unlike the L$_4$ clans. All unaffiliated clans in the L$_5$ swarm are located between the Patroclus and Aneas superclans in the tree. Except for the 1999 RU$_2$ clan, which does not contain any taxonomically identified objects, each clan contains at least one D-type object. Most of the unaffiliated clans have escape fractions higher than that of the L$_5$ swarm \citep[0.2489][]{Holt2020TrojanStability}. The exception is the stable Asteropaios clan ($F_{\textrm{esc}}$:0.06). The 1992 RU$_2$ ($F_{\textrm{esc}}$:0.84) and Anchises clans ($F_{\textrm{esc}}$: 0.88) are particularly unstable.

\paragraph{Dolon Clan: } This clan contains 55419 (2001 TF$_{19}$), which is a member of the Ennomos collisional family \citep{Nesvorny2015AsteroidFamsAIV}. In this clan there are three X-type objects \citep[11554 Asios (1993 BZ$_{12}$), 32482 (2000 ST$_{354}$) and 29314 Eurydamas (1994 CR$_{18}$);][]{Hasselmann2012SDSSTaxonomy}, along with two D-types \citep[9430 Erichthonios (1996 HU$_{10}$) and 11488 (1988 RM$_{11}$);][]{Fornasier2004L5TrojanSpec,Hasselmann2012SDSSTaxonomy}, both located in the Erichthonios subclan. This diversity in types is reflected in the geometric (0.03-0.14) and \textit{WISE} (W1:0.07-0.29, W2:0.08-0.3) albedo ranges of the clan. The clan is close ($295.12\degree$ to $297.46\degree$) to the L$_5$ Lagrange point ($300\degree$) resulting in an overall escape fraction ($F_{\textrm{esc}}$:0.31) similar to the overall L$_5$ swarm \citep[0.2489][]{Holt2020TrojanStability}, though the Erichthonios subclan is much more stable ($F_{\textrm{esc}}$:0.04). The overall $\Delta V_{\textrm{ref}}$ and $\Delta V_{\textrm{cent}}$ of the clan are relatively high ($46.28\pm27.85$~ms$^{-1}$ and $34.71\pm19.81$~ms$^{-1}$ respectively), in comparison to the small reference object ($V_{\textrm{esc}}: 14.22$~ms$^{-1}$). The SDSS \emph{(u - g)} (1.51-1.62), \emph{(g - r)} (0.48-0.64) and \emph{(i - z)} (0.01-0.18) as well as the {\tt MOVIS} \emph{(J - Ks)} (0.53-0.9) colours are compact and fairly diagnostic for the clan. 

\paragraph{1990 VU$_1$ clan: } There are two identified subclans (1990 VU$_1$ and Idaios subclans) in this clan. The type object, 1990 VU$_1$ has been identified as a XD-type \citep{Bendjoya2004JTSpectra}, with five other D-types \citep[16070 (1999 RB$_{101}$), 58008 (2002 TW$_{240}$), 15977 (1998 MA$_{11}$), 30705 Idaios (3365 T-3) and 47969 (2000 TG$_{64}$)][]{Fornasier2004L5TrojanSpec, Hasselmann2012SDSSTaxonomy} present in the clan. In terms of stability, this clan has an escape fraction (0.52), nearly double that of the L$_5$ swarm as a whole \citep[0.2489]{Holt2020TrojanStability}. There is a wide variety of escape fraction of members this clan, with all 9 particles of the type object 1990 VU$_1$ escaping, but two other members, 30705 Idaios (3365 T-3) and 301760 (2010 JP$_{42}$), being completely stable. This range of stability is not unexpected, as the clan has a wide variance in $\Delta V_{\textrm{cent}}$ ($24.87\pm18.96$~ms$^{-1}$), $e_{\textrm{prop}}$ (0.04-0.14) and $\textrm{sin}i_{\textrm{prop}}$ (0.01-0.43). The \textit{Gaia} G magnitude is constrained (17.67-18.19~mag), with only four similar sized objects represented, more analysis is needed. The \emph{(i - z)} SDSS colour (0.05-0.317) has a narrow range in this clan.

\paragraph{Anchises clan: } This clan is unstable ($F_{\textrm{esc}}$: 0.88), including all clones of the type object 1173 Anchises (1930 UB). This particular object was studied by \citet{Horner2012AnchisesThermDynam}, who found that it will most likely escape the Trojan population and evolve to become either a Centaur or Jupiter family comet on hundred-million year timescales. The clan is located a few degrees from the $300\degree$ Lagrange point ($291.6\degree$ to $296.29\degree$) with a decent range ($32.05\degree$ to $51.55\degree$) and $\delta a_{\textrm{prop}}$ (0.08 to 0.12). The type object 1173 Anchises (1930 UB), along with 11089 (1994 CS$_8$) are X-types \citep{Tholen1989Taxonomy, Fornasier2004L5TrojanSpec}. There is also a D-type, \citep[11552 Boucolion (1993 BD$_4$)][]{Hasselmann2012SDSSTaxonomy} in this clan. In the \textit{Gaia} G band, 1173 Anchises (1930 UB) shows a different  value (16.75~mag) to the other two smaller, measured objects, 11089 (1994 CS$_8$) and 11552 Boucolion (1993 BD$_4$) (18.47 and 18.32~mag respectively). Unfortunately, 1173 Anchises (1930 UB) was not observed in either of SDSS or {\tt MOVIS} surveys. This is of note, as the SDSS \emph{(b - v)} (0.69-0.93), \emph{(g - r)} (0.48-0.7), \emph{(r - i)} (0.2-0.27), as well as the {\tt MOVIS} \emph{(Y - J)} (0.23-0.32) and \emph{(J - Ks)} (0.44-0.53) colours are plausibly diagnostic of the clan.  

\subsubsection{Greater Patroclus Superclan}
This large (133 members) superclan contains six clans, as shown in Fig. \ref{Fig:greaterPatroclus}. Of these, we discuss the Memnon (23 members), Troilus (13 members) and Cebriones (17 members) clans in the following sections. The details of the other clans, 1971 FV$_1$ (18 members), 1989 TX$_{11}$ (5 members) and Phereclos (17 members), are available on the Github repository\footnote{\url{https://github.com/TimHoltastro/holt-etal-2021-Jovian-Trojan-astrocladistics.git}}. There is a diversity of taxonomic types represented in this superclan, though as with other superclans, the members are predominantly D-types. 
Overall, the superclan is more stable ($F_{\textrm{esc}}$: 0.1) than the L$_5$ swarm as a whole. Each of the clans has a lower escape rate than the L$_5$ swarm, with several having no escapees, see Table\ref{tab:L5}, for details. The clan is clustered close to the $300\degree$ Lagrange point ($296.29\degree$-$303.31\degree$), with relatively low $\delta a_{\textrm{prop}}$ (0.0-0.11~au).

The largest member of the Ennomos collisional family \citep[4709 Ennomos (1988 TU$_2$)][]{Nesvorny2015AsteroidFamsAIV}, is within this superclan, though it is not used as the type object. The actual type object, 617 Patroclus (1906 VY) was discovered over 80 years earlier and thus is considered the type for the superclan, though it is not associated with any clan in this analysis. The binary 617 Patroclus (1906 VY) \citep{Merline2001PatroclusBinary} is currently the only \textit{Lucy} target in the L$_5$ swarm. The $\Delta V_{\textrm{ref}}$ to 617 Patroclus (1906 VY) ($31.57\pm20.49$~ms$^{-1}$) is smaller than the $V_{\textrm{esc}}$ ($46.94$~ms$^{-1}$) and similar to the $\Delta V_{\textrm{cent}}$ ($31.62\pm15.09$~ms$^{-1}$). 

\paragraph{Memnon clan: }The Memnon clan has several D-types \citep[30505 (2000 RW$_{82}$), 3317 Paris (1984 KF), 80119 (1999 RY$_{138}$) and 105808 (2000 SZ$_{135}$)][]{Bendjoya2004JTSpectra, Hasselmann2012SDSSTaxonomy}, though the type object, 2895 Memnon (1981 AE$_1$) is a C-type \citep{Bendjoya2004JTSpectra}. As with the superclan, this clan is stable ($F_{\textrm{esc}}$: 0.11) and close to the L$_5$ Lagrange point ($296.29\degree$-$303.31\degree$). 

A representative of the 2001 UV$_{209}$ collisional family \citep[37519 Amphios (3040 T-3)][]{Nesvorny2015AsteroidFamsAIV}, is within this clan, and is the the type object of the Amphios subclan, which has a small $\Delta V_{\textrm{ref}}$ ($14.06\pm8.13$~ms$^{-1}$) and $\Delta V_{\textrm{cent}}$ ($13.04\pm7.89$~ms$^{-1}$), close to the $V_{\textrm{esc}}$ ($12.83$~ms$^{-1}$). The objects in this subclan may represent unidentified members of the 2001 UV$_{209}$ collisional family, or at least closely associated objects. 

The clan has mid range \emph{(b - v)} (0.74-0.93), \emph{(u - g)} (1.18-1.73) and \emph{(g - r)} (0.52-0.7) SDSS colours, with high \emph{(i - z)} values (0.12-0.34). The two {\tt MOVIS} objects 295336 (2008 HY$_8$), \emph{(Y - J)}:0.559373, \emph{(J - Ks)}:0.973755, \emph{(H - Ks)}:0.407764, and 369886 (2012 RM$_6$), \emph{(Y - J)}:0.318022, \emph{(J - Ks)}:0.585282, show quite different colours. Further characterisation of the large objects in the clan, 2895 Memnon (1981 AE$_1$), 3317 Paris (1984 KF) and 37519 Amphios (3040 T-3), would be required to resolve this dichotomy in the colours.

\paragraph{Troilus clan: } Within the clan there are two small subclans, the Troilus and 1988 RY$_{11}$ subclans. The Troilus subclan, which in includes the type object, 1208 Troilus (1931 YA), of the clan, is entirely stable. The members of the 1988 RY$_{11}$ subclan have a higher escape fraction ($F_{\textrm{esc}}$: 0.2), though even this is lower than the over all L$_5$ escape fraction (0.2489).

The type object, 1208 Troilus (1931 YA), is an interesting case. It is the type object of the Troilus clan, which also contains a single member of the Ennomos collisional family \citep[76867 (2000 YM$_5$)][]{Nesvorny2015AsteroidFamsAIV}. It is classified as FCU-type \citep{Tholen1989Taxonomy}, designating it as an unusual object. It is the only 'F-type' in the Trojan swarm. This type was degenerated under the modern Bus-Demeo system \citep{Bus2002AsteroidTax} into the B-types, closely associated with the other C-types in the Trojans. As the type object is relatively large, the $\Delta V_{\textrm{ref}}$ ($18.95\pm7.41$~ms$^{-1}$) and $\Delta V_{\textrm{cent}}$ ($12.88\pm5.81$~ms$^{-1}$) of the clan is lower than the $V_{\textrm{esc}}$ ($33.6$~ms$^{-1}$). The clan is clustered centrally around the L$_5$ Lagrange point ($298.63\degree$ to $302.14\degree$), which likely indicates that it dates back to the time the Jovian Trojans were captured. The SDSS \emph{(b - v)} (0.72-0.91), \emph{(g - r)} (0.5-0.7) and \emph{(i - z)} (0.07-0.23) colours are relatively constrained. An initial tight {\tt MOVIS} bin is due to only a single object \citep[299491 (2006 BY$_{198}$)][]{Popescu2018MOVISTaxa}.

\paragraph{Cebriones clan: } 4709 Ennomos (1988 TU$_2$), the largest member of the Ennomos collisional family, is in the Cebriones clan \citep{Nesvorny2015AsteroidFamsAIV}. Again, 4709 Ennomos (1988 TU$_2$) is not used as the type object, as the chosen type object, 2363 Cebriones (1977 TJ$_3$) was discovered earlier. A non-canonical family member, 32496 (2000 WX182) \citep[32496 (2000 WX$_{182}$)][]{Rozehnal2016HektorTaxon}, is also in the clan. This is complicated by two members of the non-canonical 2001 UV$_{209}$ family \citep[17171 (1999 NB$_{38}$) and 24470 (2000 SJ$_{310}$), ][]{Rozehnal2016HektorTaxon} that are also present in the clan. 

2363 Cebriones (1977 TJ$_3$) is a D-type object \citep{Tholen1989Taxonomy}, and the only classified member of the clan. This clan has the highest escape rate in the Greater Patroclus superclan ($F_{\textrm{esc}}$: 0.23), and even this is lower than that of the overall L$_5$ swarm \citep[0.2489][]{Holt2020TrojanStability}. In terms of colours, there are an insufficient number of multi-spectral observations to ascertain any trends, with only two members represented in the SDSS data, 17415 (1988 RO$_{10}$) and 129135 (2005 AD$_{21}$), and two different objects in {\tt MOVIS}, 51969 (2001 QZ${292}$) and 53419 (1999 PJ$_4$).

\subsubsection{Greater Aneas Superclan}
This superclan (64 members) contains five clans, 1988 RH$_{13}$ (6 members), 1994 CO (5 members), 1989 UQ$_5$ (5 members), Sarpedon (17 members) and Aneas (26 members) clans, with subclans in the Aneas clan (Hippokoon and Sarpendon subclans) and Aneas clans (Helicaon, Iphidamas and Aneas subclans). The only clan discussed in detail here is the Aneas clan. Almost all taxonomically identified members of this superclan are D-types \citep{Tholen1989Taxonomy,Bendjoya2004JTSpectra,Fornasier2004L5TrojanSpec, Hasselmann2012SDSSTaxonomy}. The only exception is 17419 (1988 RH$_{13}$), the type object of the 1988 RH$_{13}$ Clan, a C-type, though with a comparatively low confidence score \citep[62, ][]{Hasselmann2012SDSSTaxonomy}. Over-all the superclan has a relatively low escape rate ($F_{\textrm{esc}}$: 0.2), when compared with the L$_5$ swarm \citep[0.2489][]{Holt2020TrojanStability}. Within the Greater Aneas superclan, the majority of unstable members are in the 1988 RH$_{13}$ Clan, which has an escape rate of 0.65. Other clans have a similar or lower escape rate than the superclan. Though 1172 Anease (1930 UA) is a large object (118.02km), the reference object for the dispersal velocities in the superclan is 1867 Deiphobus (1971 EA) (118.22km). The $\Delta V_{\textrm{ref}}$ ($65.85\pm34.23$~ms$^{-1}$) is high. The $\Delta V_{\textrm{cent}}$ ($38.7\pm20.27$~ms$^{-1}$), though still quite high, is closer to the $V_{\textrm{esc}}$ ($39.53$~ms$^{-1}$).

\paragraph{Aneas clan: } The Aneas clan contains several D-type objects, including the type object 1172 Aneas (1930 UA) \citep{Tholen1989Taxonomy}. The three members of the Ennomos collisional family present in the clan \citep[36624 (2000 QA$_{157}$), 1867 Deiphobus (1971 EA) and 247967 (2003 YD$_{149}$)][]{Nesvorny2015AsteroidFamsAIV} form a cluster with 34746 2001 QE$_{91}$, however this does not fulfill the minimum requirements for a subclan (five objects). There are three other subclans Helicaon, Iphidamas and Aneas subclans, each containing at least one D-type. As in the Greater Anease superclan, 1172 Aneas (1930 UA) is the dynamical reference object for $\Delta V_{\textrm{ref}}$ calculations. The over all escape fraction of the clan ($F_{\textrm{esc}}$:0.2) is similar to the Greater Aneas superclan ($F_{\textrm{esc}}$:0.2), though the Helicaon (0.27 $F_{\textrm{esc}}$) and Iphidamas (0.33 $F_{\textrm{esc}}$) subclans have a slightly higher rates. In the SDSS colours, \emph{(b - v)} (0.649-0.857), \emph{(g - r)} (0.5-0.633)and \emph{(i - z)} (-0.0167-0.25) are relatively constrained. The \emph{(u - g)} (1.294-1.847) and \emph{(r - i)} (0.0682-0.267) values would also be relatively compact, except for the outlier 129147 (2005 CY70), which has comparatively high values, \emph{(u - g)}:2.28, \emph{(r - i)}:0.37.  

\subsubsection{Greater Astyanax Superclan} This is the terminal superclan in the L$_5$ tree. It contains 809 members, of which 41 are in the Astyanax clan, discussed in detail below. The Mentor clan (20 members) is also discussed. The remaining Helenos clan contains 11 members, and the values are presented in the Github repository\footnote{\url{https://github.com/TimHoltastro/holt-etal-2021-Jovian-Trojan-astrocladistics.git}}. 

This superclan has a diversity of taxonomic types. The majority of the superclan is D-types, but the type object of the Mentor clan, 3451 Mentor (1984 HA$_1$) is a well recognised X-type \citep{Bus2002AsteroidTax, Hasselmann2012SDSSTaxonomy}. There are also two CX-types in the Astyanax clan \citep[24454 (2000 QF$_{198}$) and 16560 Daitor (1991 VZ$_5$)][]{Hasselmann2012SDSSTaxonomy}. The Helenos clan contains one taxonomic identified member, 4829 Sergestus (1988 RM1), an XD-type \citep{Fornasier2004L5TrojanSpec}. This diversity of taxonomic types is reflected in the wide range of all colour values (W1:0.07-0.4, W2:0.03-0.4, G-mag:15.86-18.7~mag, \emph{(b - v)}:0.65-0.91, \emph{(u - g)}:1.18-1.95, \emph{(g - r)}:0.44-0.68, \emph{(r - i)}:0.07-0.4, \emph{(i - z)}:-0.26-0.29, \emph{(Y - J)}:0.05-0.46, \emph{(J - Ks)}:0.07-1.18, \emph{(H - Ks)}:0.04-0.81). The superclan has a large $\Delta V_{\textrm{ref}}$ ($92.28\pm43.46$~ms$^{-1}$) compared to the $V_{\textrm{esc}}$ of the largest member, 3451 Mentor ($42.2$~ms$^{-1}$), though the $\Delta V_{\textrm{cent}}$ ($50.89\pm28.51$~ms$^{-1}$) is more reasonable. The escape fraction of the supergroup (0.42) is higher than the L$_5$ swarm. The superclan has a large range of high $\delta a_{\textrm{prop}}$ values (0.07~au - 0.15~au), though the smaller values are limited to the 1988RR$_{10}$ subclan ($\delta a_{\textrm{prop}}$:0.07-0.11) within the Mentor clan. 

\paragraph{Mentor Clan: } 3451 Mentor (1984 HA$_1$), the type object of the Mentor clan, is a large (126.29km) X-type \citep{Bus2002AsteroidTax, Hasselmann2012SDSSTaxonomy}. There is also a X-type \citep[34785 (2001 RG$_{87}$)][]{Fornasier2004L5TrojanSpec}, and two D-types \citep[5130 Ilioneus (1989 SC$_7$) and 17416 (1988 RR$_{10}$)][]{Fornasier2004L5TrojanSpec}. The $\Delta V_{\textrm{ref}}$ ($48.51\pm24.27$~ms$^{-1}$) is close to the $F_{\textrm{esc}}$ of 3451 Mentor (1984 HA$_1$) ($42.23$~ms$^{-1}$), and the $\Delta V_{\textrm{cent}}$ ($32.58\pm18.62$~ms$^{-1}$). Even amongst the Trojans, which are some of the darkest objects in the Solar system \citep{Grav2012JupTrojanWISE}, the Mentor clan has a range of low geometric (0.0367-0.107) and \textit{WISE} \citep[W1:0.0557-0.171, W2:0.0276-0.177][]{Grav2011JupTrojanWISEPrelim, Grav2012JupTrojanWISE} albedos. Unfortunately, there are only two representatives in the SDSS dataset: 3451 Mentor (1984 HA$_1$) and 133862 (2004 BR$_{38}$), and only a single representative in the {\tt MOVIS} database, 289501 (2005 EJ$_{133}$), and therefore any comments on colours are preliminary. 

\paragraph{Astyanax Clan: } This is one of the largest clans in our analysis and at 41 members is larger than some superclans. Consequently, it does have a large $\Delta V_{\textrm{ref}}$ ($141.79\pm42.59$~ms$^{-1}$) and  $\Delta V_{\textrm{cent}}$ ($42.8\pm28.01$~ms$^{-1}$) relative to the $V_{\textrm{esc}}$ ($18.05$~ms$^{-1}$) of the small type object (1871 Astyanax  (1971 FF), 53.98km). Two of the subclans, Ophelestes ($\Delta V_{\textrm{ref}}:20.79\pm12.5$~ms$^{-1}$ , $\Delta V_{\textrm{cent}}:3.7\pm6.25$~ms$^{-1}$) and 1989 UX ($\Delta V_{\textrm{ref}}:15.28\pm8.38$, $\Delta V_{\textrm{cent}}:12.97\pm7.93$~ms$^{-1}$), have low dispersal velocities, though these are higher than the $V_{\textrm{esc}}$ of the respective type objects (52767 Ophelestes  (1998 MW$_{41}$):$10.83$~ms$^{-1}$ and 9030 (1989 UX$_5$):$10.76$~ms$^{-1}$ respectively). Within this clan, there are two members of the Enominos collisional family in this clan \citep[17492 Hippasos (1991 XG$_1$) and 98362 (2000 SA$_{363}$)][]{Nesvorny2015AsteroidFamsAIV} clustered close together in the Astyanax subclan. The small \textit{Gaia} range (17.836-18.381~mag) is due to only two objects being represented, 16560 Daitor (1991 VZ$_5$) and 17492 Hippasos(1991 XG$_1$). The majority of the objects (60.09 per cent) are in the SDSS colour set. The \emph{(b - v)} (0.649-0.926) and \emph{(g - r)} (1.183-1.958) values are low and constrained, where as the \emph{(u - g)} (0.5-0.633) and \emph{(i - z)}(-0.15-0.317) are on the high end and broad.

\section{Identified priority targets}
\label{Sec: targets}
One of the outcomes of this work is to identify priority targets for future observations. Here, we collate these objects and describe the rationale for their selection. A summary of these objects is presented in Table \ref{lightcurves}. 
\paragraph*{1404 Ajax (1936 QW),  4086 Podalirius (1985 VK$_2$) and 7119 Hiera (1989 AV$_2$): } These three objects are located in the Ajax clan. All three are fairly large, with \textit{H} magnitudes brighter than 9. They are of interest due to a lack of taxonomically identified objects in the Ajax clan. This clan is close to the Eurybates clan, which contains multiple members of the Eurybates collisional family, along with 3548 Eurybates (1973 SO), a \textit{Lucy} target. 
\paragraph*{2456 Palamedes (1966 BA$_1$): } The largest object (\textit{H} magnitude of 9.3) for the Thersites clan, which contains 21900 Orus (1999 VQ$_{10}$), a \textit{Lucy} target. Only a single member of the clan, 53477 (2000 AA$_{54}$), has SDSS colour values. Further classification and observations of 2456 Palamedes (1966 BA$_1$) would help to provide context for the smaller \textit{Lucy} target, 21900 Orus (1999 VQ$_{10}$), and the clan as a whole.
\paragraph*{5283 Pyrrhus (1989 BW): } This object is the largest in the Epeios clan. In \citet{Bendjoya2004JTSpectra}, it is reported as having a negative spectral slope. Unfortunately, it not represented in either of the multi-band surveys. This clan is of interest, as the only taxonomically identified object, 12921 (1998 WZ$_5$), a X-type amongst the prominently D-types of the Greater Achilles superclan. The 258656 (2002 ES$_{76}$)-(2013 CC$_{41}$) pair identified by \citet{Holt2020TrojanPair} is also potentially in the Epeios clan, close to 5283 Pyrrhus (1989 BW).
\paragraph*{659 Nestor (1908 CS): } An XC-type amongst the mostly D-types of the L$_4$ Trojan swarm. It is also one of the largest members of the population (with a \textit{H} magnitude of 8.99), and is the type member of the Greater Nestor superclan, which has a variety of taxonomic types. Additional observations of this object would help to understand the diversity of objects in the Trojan population.   
\paragraph*{1437 Diomedes (1937 PB): } This is the type object of the Diomedes clan, which includes 11351 Leucus (1997 TS$_{25}$), a small \textit{Lucy} target. Further observations of this object could provide more details on 11351 Leucus (1997 TS25) (\textit{H} mag: 10.7), and being a brighter object (with an absolute magnitude of 8.3), is able to be observed more easily. Like 2456 Palamedes (1966 BA$_1$), 1437 Diomedes (1937 PB) offers an opportunity to provide some context, prior to visitation of a related object by \textit{Lucy}.
\paragraph*{4138 Kalchas (1973 SM) and 7152 Euneus  (1973 SH$_1$): } These objects are identified X-types in a very stable clan, with absolute magnitudes of 10.1 and 10.2, respectively. Another large X-type in the population, 617 Patroclus (1906 VY), is part of a binary, and a \textit{Lucy} target. Though not in the same clan, further investigations on 4138 Kalchas (1973 SM) and 7152 Euneus (1973 SH$_1$) could provide some details on other X-types in a stable configuration.
\paragraph*{1173 Anchises (1930 UB): } The subject of dynamical and thermophysical studies by \citet{Horner2012AnchisesThermDynam} and the type object of the unaffiliated L$_5$ Anchises clan. This object is one of the darkest objects (0.05 albedo) in the Trojan population, though it is quite large, over 100km, and has an H-magnitude of 8.89. We echo the call of \citet{Horner2012AnchisesThermDynam} for further investigation into this object, particularly in broadband colours, as the object is not represented in SDSS or {\tt MOVIS} databases. 
\paragraph*{2895 Memnon (1981 AE$_1$) and 37519 Amphios (3040 T-3): } Both of these objects are located in the stable L$_5$ Memnon clan, part of the Greater Patroclus superclan. One of only two members of the 2001 UV$_{209}$ collisional family included in this analysis, is 37519 Amphios (3040 T-3) \citep{Nesvorny2015AsteroidFamsAIV}, also in the Memnon clan. The objects are the type of their respective subclans. The Memnon clan is also the closest clan to 617 Patroclus (1906 VY), a \textit{Lucy} target not affiliated with any clan. Both of these objects could provide additional information about the context of 617 Patroclus (1906 VY), though 2895 Memnon (1981 AE$_1$) is the cladistically closer object. 37519 Amphios (3040 T-3) is an interesting object in its own right, due to it's affiliation with the 2001 UV$_{209}$ collisional family, and may be the largest remnant of the collision that created that family. 
\paragraph*{1208 Troilus (1931 YA): } A relatively large object (\textit{H} mag 8.99), 1208 Troilus (1931 YA) is the only F/B-type object identified in the Trojan swarm \citep{Tholen1989Taxonomy, Bus2002AsteroidTax}. Though this taxonomic type is associated with the C-types, there are none identified in the Troilus clan. This could indicate that the object is unique in the Trojan population. Further detailed observations could help to place this object in a wider small Solar system body context, and possibly identify previously unknown associations between the Jovian Trojans and other populations.
\paragraph*{4709 Ennomos (1988 TU$_2$): } The largest member of the Ennomos collisional family \citep{Broz2011EurybatesFamily}. The object is a member of the Cebriones clan, which has limited colour information. Further characterisation of this object would help to understand the diversity of collisional family members in the Jovian Trojans.  
\paragraph*{128383 (2004 JW$_{52}$): }This relatively small object (\textit{H} mag 13.1) was removed at the binning stage from the analysis, due to due to its anomalous colour. If the object was included, the SDSS colours would consist of two bins, this object and everything else. The object has high \emph{(b - v)} and \emph{(g - r)} colours (1.55 and 1.3 respectively) in comparison to the rest of the Jovian Trojan population ( 0.10-1.275 and 0.300-1.045), as well as low \emph{(i - z)} values (-0.55, compared with -0.37-0.45). These anomalous values could be explained if the object was an interloper in the Trojan population, but this is contradicted by the stability. The object has an approximately 0.55 fractional escape rate, though only after spending an average of $3.7\e{9}$ in the L$_4$ Trojan swarm \citep{Holt2020TrojanStability}. Further characterisation and investigations into this object could help to resolve this discrepancy and discover the history of the object. 

\begin{table}
\centering
\caption{Physical and observational parameters for the priority targets identified in this work, taken from the Asteroid Lightcurve Database \citep[\url{http://www.minorplanet.info/lightcurvedatabase.html}, retrieved 2020 October 22, ][]{Warner2009ALCDB}. Here, P denotes the rotation period of the asteroid, and $A_{\textrm{min}}$ and $A_{\textrm{max}}$ are the minimum and maximum amplitudes of the asteroid's lightcurve. $H$ is the absolute magnitude of the asteroid, and $p_V$ the geometric albedo.}
%} 

%\url{http://www.minorplanet.info/lightcurvedatabase.html}
\label{lightcurves}
\begin{tabular}{l l  l l c c c c c}
\hline

Ast$_{no.}$& P & $A_{\textrm{min}}$ & $A_{\textrm{max}}$ & $H$ & $p_V$  \\
& hrs & mag    & mag  & mag &  &  \\ \hline
659   & 15.98  & 0.22 & 0.31 & 8.71  & 0.040$\pm$0.004 \\
1173  & 11.60  & 0.16 & 0.73 & 8.91  & 0.035$\pm$0.002 \\
1208  & 56.17  & -    & 0.20 & 9.00  & 0.037$\pm$0.002 \\
1404  & 29.38  & -    & 0.30 & 9.41  & 0.050$\pm$0.003 \\
1437  & 24.49  & 0.34 & 0.70 & 8.21  & 0.028$\pm$0.001 \\
2456  & 7.24   & 0.05 & 0.27 & 9.37  & 0.026$\pm$0.002 \\
2895  & 7.52   & 0.08 & 0.48 & 10.14 &  -               \\
4086  & 10.43  & 0.08 & 0.16 & 9.29  & 0.056$\pm$0.004 \\
4138  & 29.20  & 0.10 & 0.40 & 10.12 & 0.057$\pm$0.007 \\
4709  & 12.28  & 0.31 & 0.47 & 8.77  & 0.078$\pm$0.005 \\
5283  & 7.32   & -    & 0.11 & 9.76  & 0.072$\pm$0.007 \\
7119  & 400.00 & -    & 0.10 & 9.85  & 0.036$\pm$0.005 \\
7152  & 9.73   & -    & 0.09 & 10.34 &  -        \\
37519 & 50.93  & -    & 0.30 & 11.10 &  -        \\
\hline
\hline
\end{tabular}    
\end{table}

\section{\textit{Lucy} context}
\label{SubSec:lucy}
At the time of writing, five of the Jovian Trojans have been selected as targets to be visited by the \textit{Lucy} spacecraft in the late 2020's to early 2030's \citep{Levison2017Lucy}. Each of these objects are included in our astrocladistical analysis, which allows us to provide additional information on the context of those targets, in advance of the mission.
\paragraph*{3548 Eurybates (1973 SO)} is the largest fragment of the Eurybates collisional family \citep{Broz2011EurybatesFamily}, and a member of the Greater Ajax superclan, as described in section \ref{SS:GreaterAjax}. Six other members of the preciously identified Eurybates collisional family, are also located within the clan. The majority of the objects that are thought to be closely associated with 3548 Eurybates (1973 SO) can be found in the Eurybates subclan, and are all classified as as C-types \citep{Fornasier2007VisSpecTrojans}. The C-types are relatively rare in the Trojan population, comprising only approximately 12.79 per cent by number, compared with over 60, by mass in the Main Belt \citep{DeMeo2013SDSSTaxonomy}. Other members of the Eurybates clan include two D-types, 12917 (1998 TG$_{16}$) \citep{Fornasier2007VisSpecTrojans} and 5258 (1989 AU$_1$) \citep{Bendjoya2004JTSpectra}, and a X-type, 18060 (1999 XJ$_{156}$) \citep{Fornasier2007VisSpecTrojans}, with all three in the Anius subclan, a sister subclan to the Eurybates subclan. This complexity of closely associate subclans, may indicates that 3548 Eurybates (1973 SO) may be different to other C-types.
\paragraph*{15094 Polymele (1999 WB$_2$)} is a member of the Greater Diomedes superclan, as described in section \ref{SS:GreaterDiomedes}, along with 11351 Leuchus (1997 TS$_{25}$). It is not associated with any clan, though it is worth noting that if falls relatively close to the Philoctetes clan, which contains several X-types, a C-type, a D-type, and three members of the Eurybates collisional family. The diversity in this superclan, and the associated Philoctetes clan, means that it is hard to anticipate the physical nature of 15094 Polymele. It may have a shared heritage with any of the other members of the clan, and observations by \textit{Lucy} may well shed new light on its true nature and affiliation.
\paragraph*{11351 Leucus (1997 TS$_{25}$)}, like 15094 Polymele (1999 WB2), is a member of the Greater Diomedes superclan. Specifically, 11351 Leucus (1997 TS$_{25}$) is located well within the Diomedes clan, and the type object 1437 Diomedes (1937 PB) \citep{Tholen1989Taxonomy} is a well recognised DX-type. This suggests that 11351 Leucus (1997 TS$_{25}$) is representative of the majority of D-type Jovian Trojans \citep{Fornasier2007VisSpecTrojans}. This close association could imply that 11351 Leucus (1997 TS$_{25}$) has a common origin and physical composition to that larger object, and as such, that \textit{Lucy}'s visit will provide valuable data on an object that could be representative of the majority of the Trojan population that is associated with the D-types.
\paragraph*{21900 Orus (1999 VQ$_{10}$)}, located in the unaffiliated Thersander Clan, is another provisional D-type. The only other classified object in the clan, 24341 (2000 AJ$_{87}$), is a C-type \citep{Fornasier2007VisSpecTrojans}. In addition, there are several other closely associated C-types. This could suggest that 21900 Orus (1999 VQ$_{10}$) has a different composition to 11351 Leucus (1997 TS$_{25}$), despite both being designated D-types. This further highlights the diversity of taxonomic types in the Trojan swarms, and could be confirmed with analysis of the \textit{Lucy} data, as it becomes available. Indications of the differences between 21900 Orus (1999 VQ$_{10}$) and 121351 Leucus (1997 TS$_{25}$) could be investigated using observations of 2456 Palamedes (1966 BA$_1$), the largest object in the Thersander clan, of which 21900 Orus (1999 VQ$_{10}$) is a member. 
\paragraph*{The 617 Patroclus (1906 VY)/Menoetius} binary system is, so far the only \textit{Lucy} target in the L$_5$ swarm. Being a large object, it is very well studied \citep{Merline2001PatroclusBinary, Marchis2006DensityPatroclus}, and has a well established taxonomy as a X-type \citep{Tholen1989Taxonomy}, though we note that in the original classification, as well as the \textit{Lucy} documentation \citep{Levison2017Lucy}, it is a 'P-type'. In our analysis, 617 Patroculus (1906 VY) is the type object for the Greater Patroclus superclan. The binary is not, itself, associated with any of the clans, although it is close to the Memnon clan. Part of the issue is that in our analysis 617 Patroclus (1906 VY) is not represented in the SDSS catalogue. The inclusion of these data could potentially bring the object into the Memnon clan. Being close to the Memnon clan may associate it with other large members, 2895 Memnon (1981 AE$_1$) and 37519 Amphios (3040 T-3), though neither of these have any colour values, beyond the size dependent \textit{Gaia} G magnitudes. The relatively large 37519 Amphios (3040 T-3) is interesting due to it's inclusion in the 2001 UV$_{209}$ collisional family. While inclusion of 617 Patroclus (1906 VY) in the family would be unreasonable, as the family creation event would have disrupted the binary \citep{Nesvorny2019Binaries}, this may indicate a link between the family and the binary. Further analysis of several of these objects, as discussed in section \ref{Sec: targets}, could help further classify these objects, and place 617 Patroclus (1906 VY) in context prior to \textit{Lucy}'s arrival, in 2033.

\section{Future surveys}
\label{Sec:FutureSurvey}
In this work, we use astrocladistics to investigate the Jovian Trojan population, drawing upon observational data obtained by the latest generation of wide-field surveys. In the coming decade, several new surveys will come online, providing a wealth of new data that could be incorporated in future studies. Here, we comment on the potential for the use of the astrocladistical methodology in the analysis of that data, and discuss how those surveys will improve our understanding of the Jovian Trojan population.  

\paragraph*{\textit{Gaia} DR3:} In this work we use single \textit{G}-band (330 to 1050~nm) data taken from \textit{Gaia} DR2 \citep{Spoto2018GaiaDR2}. Whilst this single band data can provide some information about the objects, The \textit{Gaia} \textit{G}-band magnitudes are clearly linked to size, to first approximation, but also to some extent albedo and distance. Albedos within the Jovian Trojans are low, and relatively consistent \citep{Romanishin2018AlbedoCenJTH}. Distance is also normalised somewhat, due to the librations of the population around the Lagrange points. In the \textit{Gaia} DR2 dataset, there are two additional two bands, \textit{G$_{BP}$}-band (330-680~nm) and \textit{G$_{RP}$}-band (630-1050~nm) \citep{Evans2018GaiaDR2} for stellar objects, but data in these bands is not available for Solar system objects. These data are expected to be included in the full \textit{Gaia} DR3 release, which is currently scheduled for release in early 2022, and once available, could be incorporated into future astrocladistical surveys in a similar way to the SDSS and {\tt MOVIS} colours.

\paragraph*{The Vera Rubin Observatory,} with the Legacy Survey of Space and Time (Rubin Obs. LSST), is expected to receive first light in 2023. During the first few years that Vera Rubin is active, estimates suggest that more than 280,000 Jovian Trojans are expected to be discovered \citep{LSST2009ScienceBook}. Of those objects, it is likely that more than 150 observations will be made of at least 50,000, which will be sufficient for those objects to be characterised in five broadband colours \citep{LSST2009ScienceBook}. This will provide a much larger context for taxonomy in the Jovian Trojan population, and small Solar system bodies in general. Astrocladistics is a tool that could be used to further analyse these data, and that is ideally suited to the analysis of such vast and sprawling datasets. Assuming that the currently observed L$_4$/L$_5$ numerical asymmetry holds \citep{Jewitt2000JovTrojan, Nakamura2008TrojanSurfaceDenSizeEst,Yoshida2008JovTrojanSize, Vinogradova2015JupTrojanMass}, it is expected that those observations would yield results for approximately 33,000 objects in the vicinity of L$_4$, and 17,000 around L$_5$. Given that the computational requirements for cladistical analysis increases approximately with a trend of $n^{3/2}$ \citep{Goloboff2008TNT,Golboff2016TNT15}, we estimate that, using current computational architecture, the analysis of such large datasets would require approximately 2700 CPU-hours for the L$_5$ analysis and 7500 CPU-hours for the population around L$_4$. In order for this to be feasible, further testing into the {\tt TNT} 1.5 parallelisation \citep{Golboff2016TNT15} will be required. 

\paragraph*{The \textit{James Web Space Telescope} (\textit{JWST})} is currently scheduled for launch in 2021. The telescope will provide detailed analysis of many Solar system objects \citep{Rivkin2020TrojanJWSTGTO}. In contrast to the work of \textit{Gaia} and the Vera Rubin observatory, which are undertaking wide ranging surveys, the \textit{JSWT} is instead a targeted mission, providing detailed IR spectra on specific objects, rather than broadband colours on many objects. Whilst the time required for such observations will doubtless be incredibly highly sought after, two members of the Jovian Trojan population, 617 Patroclus (1906 VY) and 624 Hektor (1907 XM), have already been approved for study under the Guaranteed Time Observations program \citep{Rivkin2020TrojanJWSTGTO}. Once those observations are complete, the results can be placed in a wider context due to this work. As JWST is a limited time mission, we recommend the prioritisation of those targets identified in section \ref{Sec: targets} to provide the most benefit. 

\paragraph*{\textit{Twinkle}} is a low-cost, community funded, space telescope, scheduled for launch in 2023 or 2024 \citep{Savini2018TWINKLE}. The mission will provide spectral analysis in three bands in the visible and near-IR (0.4-1~$\mu$m, 1.3-2.42~$\mu$m, and 2.42-4.5~$\mu$m). In terms of the Jovian Trojans, the mission will be able to provide detailed observations down to approximately 15th magnitude. Over the seven year initial lifetime, Twinkle is expected to observe 50 or so of the largest Trojans \citep{Edwards2019TwinkleSB,Edwards2019RemoteTwinkleSSB}, all of which are included in this work. This will provide further characterisation of these bodies, particularly in the IR range. Astrocladistics can offer added value to analysis of Twinkle observations, through associations within clans. 

\paragraph*{The \textit{Nancy Grace Roman Space Telescope} (\textit{RST}, formally \textit{WFIRST})} is currently in development, with an expected launch date in 2025. Once launched, there will a number of opportunities for small body Solar system science using \textit{RST}, including the the ability to obtain a wealth of data for the Jovian Trojans \citet{Holler2018SolarSysWFIRST}. Using the wide-filed imaging system, in the near-IR (0.6-2.0~$\mu$m), \textit{RST} will be able to observe the majority of the currently known Jovian Trojans. In conjunction with the broadband Rubin Observatory LSST colours, those observations will yield a large database of Jovian Trojan characteristics. As computational capabilities and algorithm optimisations increases prior to launch, astrocladistics will provide a tool capable of analysing such large datasets. 

\section{Conclusion}
\label{Sec: Conclusion}

In this work we apply the new astrocladistical technique to the Jovian Trojans. We combine dynamical characteristics with colour information from the Sloan Digital Sky Survey (SDSS), \textit{WISE}, \textit{Gaia} DR2 and {\tt MOVIS}, into a holistic taxonomic analysis. We create two matrices, one for the L$_4$ and one the L$_5$ Trojans, comprised of 398 and 407 objects respectively. As part of this analysis, we find clustering beyond the previously identified collisional families \citep{Nesvorny2015AsteroidFamsAIV}. These clusters we term `clans', which provide the beginnings of a taxonomic framework, the results of which are presented visually using a consensus dendritic tree. Our results yield a hierarchical structure, with individual clans often congregating within a larger `superclan', and with other clans being further broken down into one or more `subclans'. These subclans, clans and super clans form clusters of objects with a possible common origin. With the next generation wide-field surveys and the \textit{Lucy} mission, these clusters will be able to be placed in a wider context under the new paradigm. 

In our analysis of the members of the L$_4$ swarm, we identify a total of ten unaffiliated clans and eight superclans that, in turn, contain an additional seventeen clans. Within our analysis, we include 13 members of the Eurybates collisional family \citep{Nesvorny2015AsteroidFamsAIV}, the largest in the Trojan population. Seven of these, including 3548 Eurybates (1973 SO), a \textit{Lucy} target, cluster into the Eurybates clan, a part of the Greater Ajax superclan. Other canonical family members cluster together, though are separated, possibly indicating that they are not true collisional family members, but suffer from one of the inherent issues with the methodology used to identify families. 

The L$_5$ swarm shows more hierarchical structure: seven unaffiliated clans, with six subclans within them. The L$_5$ swarm is found to contain at least three large superclans, with each superclan containing a larger number of clans and subclans. In total, there are 14 clans containing 14 subclans in the L$_5$ swarm. The only \textit{Lucy} target in the L$_5$ swarm, 617 Patroclus (1906 VY), is the type object of the Greater Patroculus superclan, thought it is not specifically part of any clan, it is close to the Memnon clan, which includes  2001 UV$_{209}$ collisional family member, 37519 Amphios (3040 T-3). The other members of the larger Ennominos collisional family \citep{Nesvorny2015AsteroidFamsAIV} are distributed throughout the dendritic tree, indicating that perhaps the original HCM \citep{Zappala1990HierarchicalClustering1} is inappropriate for describing the history of the swarm.   

A key outcome of our astrocladistical analysis is that we identify 15 high priority targets for follow-up observations. These are all comparatively large and bright objects that should be observed to provide further context for the Jovian Trojan swarms as a whole. Several are closely related to \textit{Lucy} targets that could provide additional information in preparation for \textit{in-situ} observations. 

All of the future \textit{Lucy} targets \citep{Levison2017Lucy} are included in our analysis. Our results therefore provide a taxonomic context for the mission, and extend the value of discoveries made. By associating the \textit{Lucy} targets with other clan members, inferences can be made about their nearest relatives, and the swarms as a whole. 

Whilst the focus of this work is on the current generation of wide-field surveys, several new observatories will be coming on line in the next few decades. The Vera Rubin Observatory, with the Legacy Survey of Space and Time (Rubin Obs. LSST), the \textit{James Web Space Telescope} (\textit{JWST}), \textit{Twinkle} and the \textit{Nancy Grace Roman Space Telescope} (\textit{RST}, formerly \textit{WFIRST}) will all be able to observe the Jovian Trojan population and further characterise these objects. Astrocladistics offers a method of analysis that will allow a timely and detailed analysis of the relationships between the Jovian Trojans, based on the observations made by these next generation telescopes, and helps to identify high priority targets for competitive observational time. The Jovian Trojans are the remnants of the early Solar system, held dynamically stable for the past $4.5\e{9}$ years. They are vital clues to this early period in the story of the Solar system. Astrocladistical analysis of these objects provides us with insights into their history and how they are related to one another.  

\section*{Data availability}
A GitHub repository has been created for this study \url{https://github.com/TimHoltastro/holt-etal-2021-Jovian-Trojan-astrocladistics.git}, and is publicly available. Matrices, trees and clan dispersal velocity datasets are stored in this repository. The Python 3 binning and dispersal velocity programs, as well as the tree-search parameters and nexis (.nex) files containing the tree-banks are also available. 

\section*{Acknowledgements}

%The Acknowledgements section is not numbered. Here you can thank helpful colleagues, acknowledge funding agencies, telescopes and facilities used etc. Try to keep it short.

This research was in part supported by the University of Southern Queensland's Strategic Research Initiative program. TRH was supported by the Australian Government Research Training Program Scholarship. Dr. Pablo Goloboff provided assistance with {\tt TNT}, which is subsidized by the Willi Hennig Society, as well as additional comments on the methodology. We thank Dr. Didier Fraix-Burnet and an anonymous reviewer for their comments in improving this manuscript. Some analysis was conducted in Python 3.7 under the Anaconda software environment \citep{Anaconda240}. 

%%%%%%%%%%%%%%%%%%%%%%%%%%%%%%%%%%%%%%%%%%%%%%%%%%

%%%%%%%%%%%%%%%%%%%% REFERENCES %%%%%%%%%%%%%%%%%%

\bibliographystyle{mnras}
\bibliography{library}

%%%%%%%%%%%%%%%%%%%%%%%%%%%%%%%%%%%%%%%%%%%%%%%%%%

%%%%%%%%%%%%%%%%% APPENDICES %%%%%%%%%%%%%%%%%%%%%

\appendix

%each of the characteristics with bins, R^2 values, fraction completeness. 
\section{Characteristics used in the Matrix} \label{App:char}

This appendix details the characteristics used in the analysis. In total there are 17 values that are binned using the Python 3 \citep{Anaconda240} program, available at the associated Github (\url{https://github.com/TimHoltastro/holt-etal-2021-Jovian-Trojan-astrocladistics.git}). This binning program is based on one developed in \citet{Holt2018JovSatSatsClad}. $R^2$ values are the correlation between the binned values and the original data. The binning program sets the number of bins once an $R^2$ value greater than 0.99 is reached, or the maximum number of bins, 15 is reached. Each characteristic is binned independently for the L4 and L5 Trojan matrices.

\subsection{$\Delta a_p$}
Proper $\Delta$ semi-major axis of the object. From AsyDys database \url{https://newton.spacedys.com/astdys/}\\
Reference: \citet{Knezevic2017AstDysTrojans}\\
Units: au\\
L4 Bin Number: 13\\
L4 $R^2$ value: 0.9902\\
L4 Bin deliminators: [0.0004417  0.01277692 0.02495385 0.03713077 0.04930769 0.06148462
 0.07366154 0.08583846 0.09801538 0.11019231 0.12236923 0.13454615
 0.14672308 0.1589    ]\\
L5 Bin Number: 13\\
L5 $R^2$ value: 0.9902\\
L5 Bin deliminators: [0.0041526  0.01563846 0.02697692 0.03831538 0.04965385 0.06099231
 0.07233077 0.08366923 0.09500769 0.10634615 0.11768462 0.12902308
 0.14036154 0.1517    ]\\

\subsection{$e_p$}
Proper eccentricity of the object. From AsyDys database \url{https://newton.spacedys.com/astdys/}\\
Units: n/a\\
Reference: \citet{Knezevic2017AstDysTrojans}\\
L4 Bin Number: 15\\
L4 $R^2$ value: 0.9900\\
L4 Bin deliminators: [0.0035364  0.01460667 0.02551333 0.03642    0.04732667 0.05823333
 0.06914    0.08004667 0.09095333 0.10186    0.11276667 0.12367333
 0.13458    0.14548667 0.15639333 0.1673    ]\\
L5 Bin Number: 15\\
L5 $R^2$ value: 0.9876\\
L5Bin deliminators: [0.0041151  0.01662667 0.02895333 0.04128    0.05360667 0.06593333
 0.07826    0.09058667 0.10291333 0.11524    0.12756667 0.13989333
 0.15222    0.16454667 0.17687333 0.1892    ]\\

\subsection{$\textrm{sin}i_p$}
Sine of the proper inclination of the object. From AsyDys database \url{https://newton.spacedys.com/astdys/}\\
Units: n/a\\
Reference: \citet{Knezevic2017AstDysTrojans}\\
L4 Bin Number: 15\\
L4$R^2$ value: 0.9870\\
L4 Bin deliminators: [0.0101936 0.06476   0.11852   0.17228   0.22604   0.2798    0.33356
 0.38732   0.44108   0.49484   0.5486    0.60236   0.65612   0.70988
 0.76364   0.8174   ]\\
L5 Bin Number: 13\\
L5 $R^2$ value: 0.9901\\
L5 Bin deliminators: [0.012521   0.06543077 0.11766154 0.16989231 0.22212308 0.27435385
 0.32658462 0.37881538 0.43104615 0.48327692 0.53550769 0.58773846
 0.63996923 0.6922    ]\\

\subsection{MeanLib}
Mean libration value, relative to Jupiter. Calculated using {\tt REBOUND} \citep{Rein2012REBOUND, Rein2015WHFAST} as outlined in section \ref{SubSec:Method:Matrix} of the text.\\
Units: degree\\
Reference: n/a\\
L4 Bin Number: 15\\
L4 $R^2$ value: 0.9838\\
L4 Bin deliminators: [56.4248396  57.77509172 59.10538938 60.43568704 61.7659847  63.09628236
 64.42658001 65.75687767 67.08717533 68.41747299 69.74777065 71.07806831
 72.40836597 73.73866362 75.06896128 76.39925894]\\
L5 Bin Number: 14\\
L5 $R^2$ value: 0.9908\
L5 Bin deliminators: [285.72582824 286.91482596 288.0862523  289.25767863 290.42910496
 291.60053129 292.77195762 293.94338395 295.11481029 296.28623662
 297.45766295 298.62908928 299.80051561 300.97194195 302.14336828
 303.31479461]\\

\subsection{LibRange}
Range of the objects libration, relative to Jupiter. Calculated using {\tt REBOUND} \citep{Rein2012REBOUND, Rein2015WHFAST} as outlined in section \ref{SubSec:Method:Matrix} of the text.\\
Units: degree\\
Reference: n/a\\
L4 Bin Number: 14\\
L4 $R^2$ value: 0.9904\\
L4 Bin deliminators: [ 4.04450175  9.22096281 14.325954   19.43094519 24.53593638 29.64092757
 34.74591876 39.85090995 44.95590114 50.06089233 55.16588352 60.27087471
 65.3758659  70.48085709 75.58584828]\\
L5 Bin Number: 14\\
L5 $R^2$ value: 0.9908\\
L5 Bin deliminators: [ 2.7354308   7.67859255 12.55350552 17.42841848 22.30333145 27.17824441
 32.05315738 36.92807035 41.80298331 46.67789628 51.55280924 56.42772221
 61.30263518 66.17754814 71.05246111]\\

\subsection{albedo}
Geometric albedo of the object. From NASA-JPL {\tt HORIZONS} Solar System Dynamics Database \url{https://ssd.jpl.nasa.gov/} \citet{Giorgini1996JPLSSdatabase}.\\
Units: n/a\\
Reference: \citet{Giorgini1996JPLSSdatabase}\\
L4 Bin Number: 15\\
L4 $R^2$ value: 0.9830\\
L4 Bin deliminators: [0.024827   0.03653333 0.04806667 0.0596     0.07113333 0.08266667
 0.0942     0.10573333 0.11726667 0.1288     0.14033333 0.15186667
 0.1634     0.17493333 0.18646667 0.198     ]\\
L5 Bin Number: 15\\
L5 $R^2$ value: 0.9817\\
L5 Bin deliminators: [0.030831   0.04226667 0.05353333 0.0648     0.07606667 0.08733333
 0.0986     0.10986667 0.12113333 0.1324     0.14366667 0.15493333
 0.1662     0.17746667 0.18873333 0.2       ]\\
 
\subsection{W1Alb}
Near Infrared values from the \textit{WISE} survey using the W1 filter ($3.4\mu m$).\\ 
Units: magnitude\\
Reference: \citet{Grav2011JupTrojanWISEPrelim, Grav2012JupTrojanWISE}\\
L4 Bin Number: 15\\
L4 $R^2$ value: 0.9824\\
L4 Bin deliminators: [0.055661 0.0786   0.1012   0.1238   0.1464   0.169    0.1916   0.2142
 0.2368   0.2594   0.282    0.3046   0.3272   0.3498   0.3724   0.395   ]\\
L5 Bin Number: 15\\
L5 $R^2$ value: 0.9794\\
L5 Bin deliminators: [0.065666   0.08826667 0.11053333 0.1328     0.15506667 0.17733333
 0.1996     0.22186667 0.24413333 0.2664     0.28866667 0.31093333
 0.3332     0.35546667 0.37773333 0.4       ]\\
 
\subsection{W2Alb}
Near Infrared values from the \textit{WISE} survey using the W2 filter ($4.6\mu m$).\\ 
Units: magnitude\\
Reference: \citet{Grav2011JupTrojanWISEPrelim, Grav2012JupTrojanWISE}\\
L4 Bin Number: 15\\
L4 $R^2$ value: 0.9838\\
L4 Bin deliminators: [0.035641   0.05993333 0.08386667 0.1078     0.13173333 0.15566667
 0.1796     0.20353333 0.22746667 0.2514     0.27533333 0.29926667
 0.3232     0.34713333 0.37106667 0.395     ]\\
L5 Bin Number: 15\\
L5 $R^2$ value: 0.9773\\
L5 Bin deliminators: [0.027628 0.0528   0.0776   0.1024   0.1272   0.152    0.1768   0.2016
 0.2264   0.2512   0.276    0.3008   0.3256   0.3504   0.3752   0.4     ]\\

\subsection{g$_{mag}$-mean}
Mean \textit{G}-band magnitude from the \textit{GAIA} survey. Filter passband from 330~nm to 1050~nm \citet{Evans2018GaiaDR2}.\\
Units: magnitude\\
Reference: \citet{Spoto2018GaiaDR2}
L4 Bin Number: 15\\
L4 $R^2$ value: 0.9894\\
L4 Bin deliminators: [15.10926146 15.38560874 15.65787207 15.9301354  16.20239873 16.47466206
 16.74692539 17.01918872 17.29145205 17.56371538 17.83597871 18.10824204
 18.38050537 18.65276871 18.92503204 19.19729537]\\
L5 Bin Number: 12\\
L5 $R^2$ value: 0.9904\\
L5 Bin deliminators: [15.85627031 16.11791172 16.37645066 16.6349896  16.89352854 17.15206747
 17.41060641 17.66914535 17.92768429 18.18622323 18.44476217 18.7033011
 18.96184004]\\

\subsection{\emph{(b - v)}}
Index of Johnson \textit{B} (442nm) and Johnson \textit{V} (540nm) band magnitudes, calculated from SDSS photometry \citep{Fukugita1996SDSSphoto}.\\
Units: magnitude\\
Reference: \citet{Szab2007JovTrojanSloneDSS}\\
L4 Bin Number: 15\\
L4 $R^2$ value: 0.9591\\
L4 Bin deliminators: [0.50896    0.57933333 0.64866667 0.718      0.78733333 0.85666667
 0.926      0.99533333 1.06466667 1.134      1.20333333 1.27266667
 1.342      1.41133333 1.48066667 1.55      ]\\
L5 Bin Number: 15 \\
L5 $R^2$ value: 0.9878\\
L5 Bin deliminators: [0.60968    0.63133333 0.65266667 0.674      0.69533333 0.71666667
 0.738      0.75933333 0.78066667 0.802      0.82333333 0.84466667
 0.866      0.88733333 0.90866667 0.93      ]\\

\subsection{\emph{(u - g)}}
Index of \textit{U} (354.3~nm) and \textit{G} (477~nm) band magnitudes taken from the SDSS \citep{Fukugita1996SDSSphoto}.\\
Units: magnitude\\
Reference: \citet{Szab2007JovTrojanSloneDSS}\\
L4 Bin Number: 15\\
L4 $R^2$ value: 0.9656\\
L4 Bin deliminators: [0.873585   0.96933333 1.06366667 1.158      1.25233333 1.34666667
 1.441      1.53533333 1.62966667 1.724      1.81833333 1.91266667
 2.007      2.10133333 2.19566667 2.29      ]\\
L5 Bin Number: 15\\
L5 $R^2$ value: 0.9724\\
L5 Bin deliminators: [0.62835 0.74    0.85    0.96    1.07    1.18    1.29    1.4     1.51
 1.62    1.73    1.84    1.95    2.06    2.17    2.28   ] \\

\subsection{\emph{(g - r)}}
Index of \textit{G} (477~nm)  and \textit{R } (623.1~nm) band magnitudes taken from the SDSS \citep{Fukugita1996SDSSphoto}.\\
Units: magnitude\\
Reference: \citet{Szab2007JovTrojanSloneDSS}\\
L4 Bin Number: 15\\
L4 $R^2$ value: 0.9560\\
L4 Bin deliminators: [0.299      0.36666667 0.43333333 0.5        0.56666667 0.63333333
 0.7        0.76666667 0.83333333 0.9        0.96666667 1.03333333
 1.1        1.16666667 1.23333333 1.3       ]\\
L5 Bin Number: 15\\
L5 $R^2$ value: 0.9851\\
L5 Bin deliminators: [0.4197 0.44   0.46   0.48   0.5    0.52   0.54   0.56   0.58   0.6
 0.62   0.64   0.66   0.68   0.7    0.72  ]\\

\subsection{\emph{(r - i)}}
Index of \textit{R} (623.1~nm) and \textit{I} (762.5~nm) band magnitudes taken from the SDSS \citep{Fukugita1996SDSSphoto}.\\ 
Units: magnitude\\
Reference: \citet{Szab2007JovTrojanSloneDSS}\\
L4 Bin Number: 15\\
L4 $R^2$ value: 0.9890\\
L4 Bin deliminators: [0.09976 0.116   0.132   0.148   0.164   0.18    0.196   0.212   0.228
 0.244   0.26    0.276   0.292   0.308   0.324   0.34   ]\\
L5 Bin Number: 15\\
L5 $R^2$ value: 0.9841\\
L5 Bin deliminators: [0.06824    0.09066667 0.1127619  0.13485714 0.15695238 0.17904762
 0.20114286 0.2232381  0.24533333 0.26742857 0.28952381 0.31161905
 0.33371429 0.35580952 0.37790476 0.4       ]\\

\subsection{\emph{(i - z)}}
Index of \textit{I} (762.5~nm) and \textit{Z} (913.4~nm) band magnitudes taken from the SDSS \citep{Fukugita1996SDSSphoto}.\\ 
Units: magnitude\\
Reference: \citet{Szab2007JovTrojanSloneDSS}\\
L4 Bin Number: 15\\
L4 $R^2$ value: 0.9614\\
L4 Bin deliminators: [-0.55087 -0.492   -0.434   -0.376   -0.318   -0.26    -0.202   -0.144
 -0.086   -0.028    0.03     0.088    0.146    0.204    0.262    0.32   ]\\
L5 Bin Number: 15\\
L5 $R^2$ value: 0.9656\\
L5 Bin deliminators: [-0.37082    -0.31533333 -0.26066667 -0.206      -0.15133333 -0.09666667
 -0.042       0.01266667  0.06733333  0.122       0.17666667  0.23133333
  0.286       0.34066667  0.39533333  0.45      ]\\

\subsection{\emph{(Y - J)}}
Index of \textit{Y} ($1.02~\mu m$) and \textit{J} ($1.25~\mu m$) band magnitudes from the VISTA survey \citep{Sutherland2015VISTA}, in the MOVIS database \citep{Popescu2016MPMOVIS}.\\
Units: magnitude\\
Reference: \citet{Popescu2018MOVISTaxa}\\
L4 Bin Number: 15\\
L4 $R^2$ value: 0.9875\\
L4 Bin deliminators: [0.02060934 0.0655506  0.1098277  0.1541048  0.1983819  0.242659
 0.2869361  0.3312132  0.3754903  0.4197674  0.4640445  0.5083216
 0.5525987  0.5968758  0.6411529  0.68543   ]\\
L5 Bin Number: 15\\
L5 $R^2$ value: 0.9886\\
L5 Bin deliminators: [0.05425359 0.09975333 0.14458067 0.189408   0.23423533 0.27906267
 0.32389    0.36871733 0.41354467 0.458372   0.50319933 0.54802667
 0.592854   0.63768133 0.68250867 0.727336  ]\\

\subsection{\emph{(J - Ks)}}
Index of \textit{J} ($1.25~\mu m$) and \textit{K} ($2.15~\mu m$) band magnitudes from the VISTA survey \citep{Sutherland2015VISTA}, in the MOVIS database \citep{Popescu2016MPMOVIS}.\\
Units: magnitude\\
Reference: \citet{Popescu2018MOVISTaxa}\\
L4 Bin Number: 15\\
L4 $R^2$ value: 0.9846\\
L4 Bin deliminators: [0.14045928 0.25723273 0.37228047 0.4873282  0.60237593 0.71742367
 0.8324714  0.94751913 1.06256687 1.1776146  1.29266233 1.40771007
 1.5227578  1.63780553 1.75285327 1.867901  ]\\
L5 Bin Number: 15\\
L5 $R^2$ value: 0.9890\\
L5 Bin deliminators: [0.06778045 0.16160333 0.25403967 0.346476   0.43891233 0.53134867
 0.623785   0.71622133 0.80865767 0.901094   0.99353033 1.08596667
 1.178403   1.27083933 1.36327567 1.455712  ]\\

\subsection{\emph{(H - Ks)}}
Index of \textit{H} ($1.65~\mu m$) and \textit{K} ($2.15~\mu m$) band magnitudes from the VISTA survey \citep{Sutherland2015VISTA}, in the MOVIS database \citep{Popescu2016MPMOVIS}.\\
Units: magnitude\\
Reference: \citet{Popescu2018MOVISTaxa}\\
L4 Bin Number: 8\\
L4 $R^2$ value: 0.9991\\
L4 Bin deliminators: [-0.33295512 -0.2505985  -0.1688955  -0.0871925  -0.0054895   0.0762135
  0.1579165   0.2396195   0.3213225 ]\\
L5 Bin Number: 14\\
L5 $R^2$ value: 0.9906\\
L5 Bin deliminators: [-0.1558507  -0.05802146  0.03845707  0.13493561  0.23141414  0.32789268
  0.42437121  0.52084975  0.61732829  0.71380682  0.81028536  0.90676389
  1.00324243  1.09972096  1.1961995 ]\\

\subsection{$\textrm{tax}_{c}$}
Canonical taxonomic designation, based on the \citep{DeMeo2009AsteroidTax}. Note: any 'P-type' have been modernised into the X-types. Reference used is in $\textrm{tax}_{ref}$.

\subsection{$\textrm{tax}_{ref}$}
Source of canonical taxonomic classification ($\textrm{tax}_{c}$)  Tholen1989: \citet{Tholen1989Taxonomy}; Bendjoya2004: \citet{Bendjoya2004JTSpectra}; Fornasier2004 \citep{Fornasier2004L5TrojanSpec}; Lazzaro2004: \citet{Lazzaro2004s3os2asteroids}; Fornasier2007: \citet{Fornasier2007VisSpecTrojans}; H2012: \citet{Hasselmann2012SDSSTaxonomy}.

%Table of each family, showing individual members
\section{Individual Superclans, clans and subclans} \label{App:clans}

\begin{figure*}
\centering
\subfloat[1998WR10 clan.\label{fig:1998WR10}]{\includegraphics[width=0.43\textwidth]{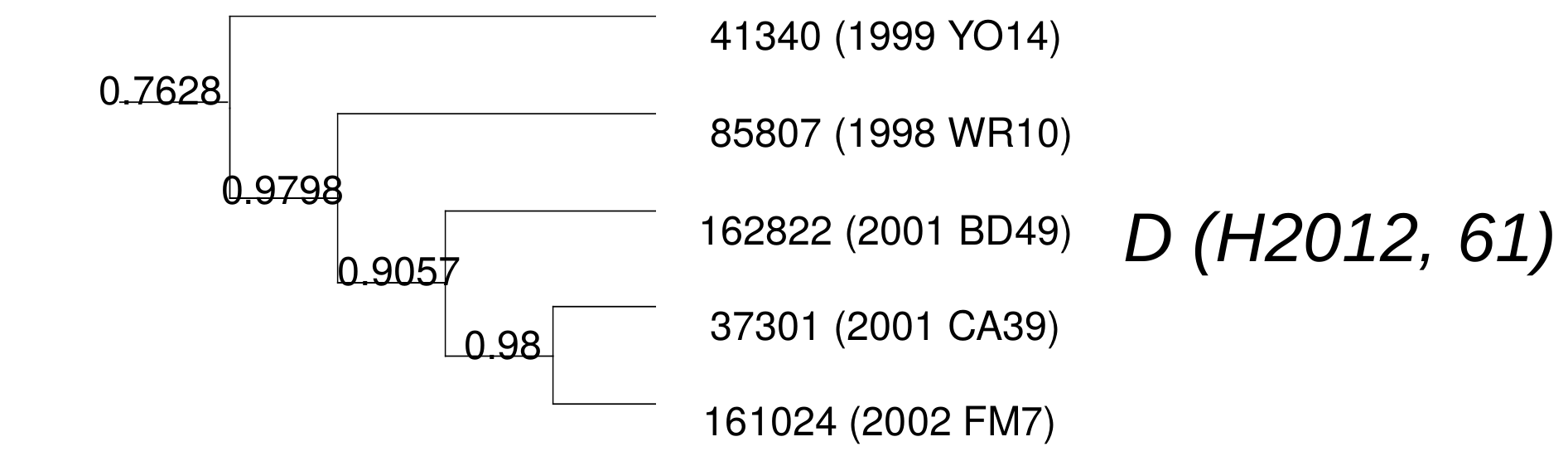}}\hfill
\subfloat[Agamemnon clan.\label{fig:Agamemnon}] {\includegraphics[width=0.43\textwidth]{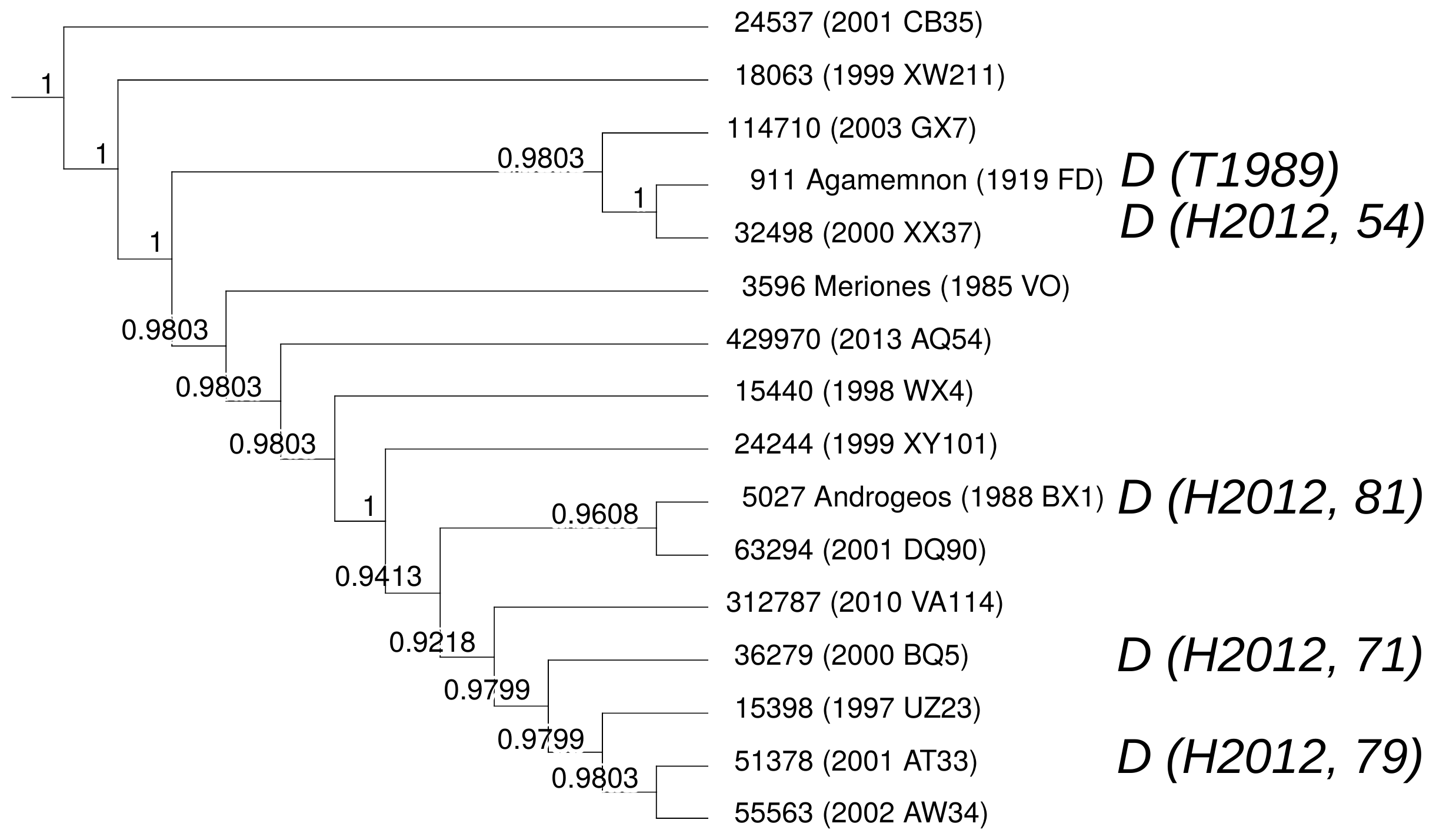}}\hfill
\subfloat[Halaeusus clan.\label{fig:Halaeusus}]{\includegraphics[width=0.43\textwidth]{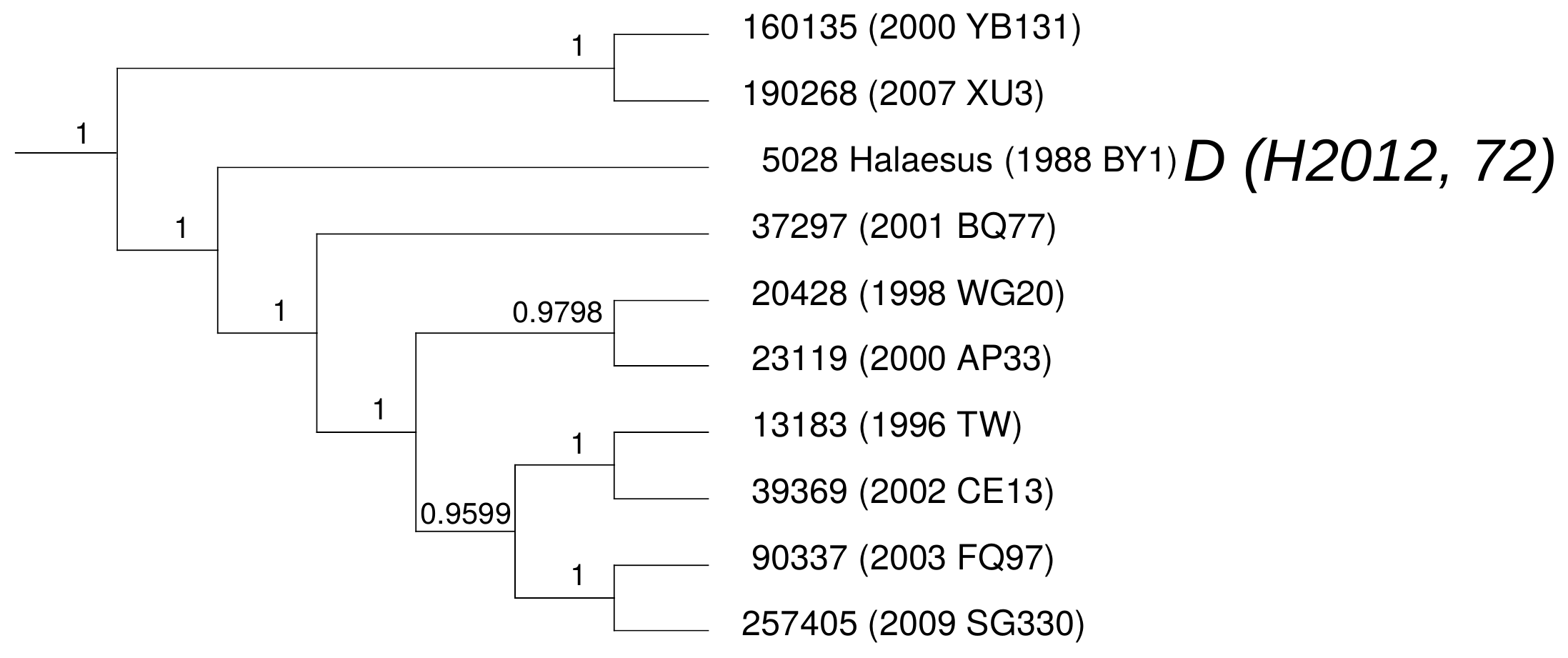}}\hfill
\subfloat[Halitherses clan.\label{fig:Halitherses}]{\includegraphics[width=0.43\textwidth]{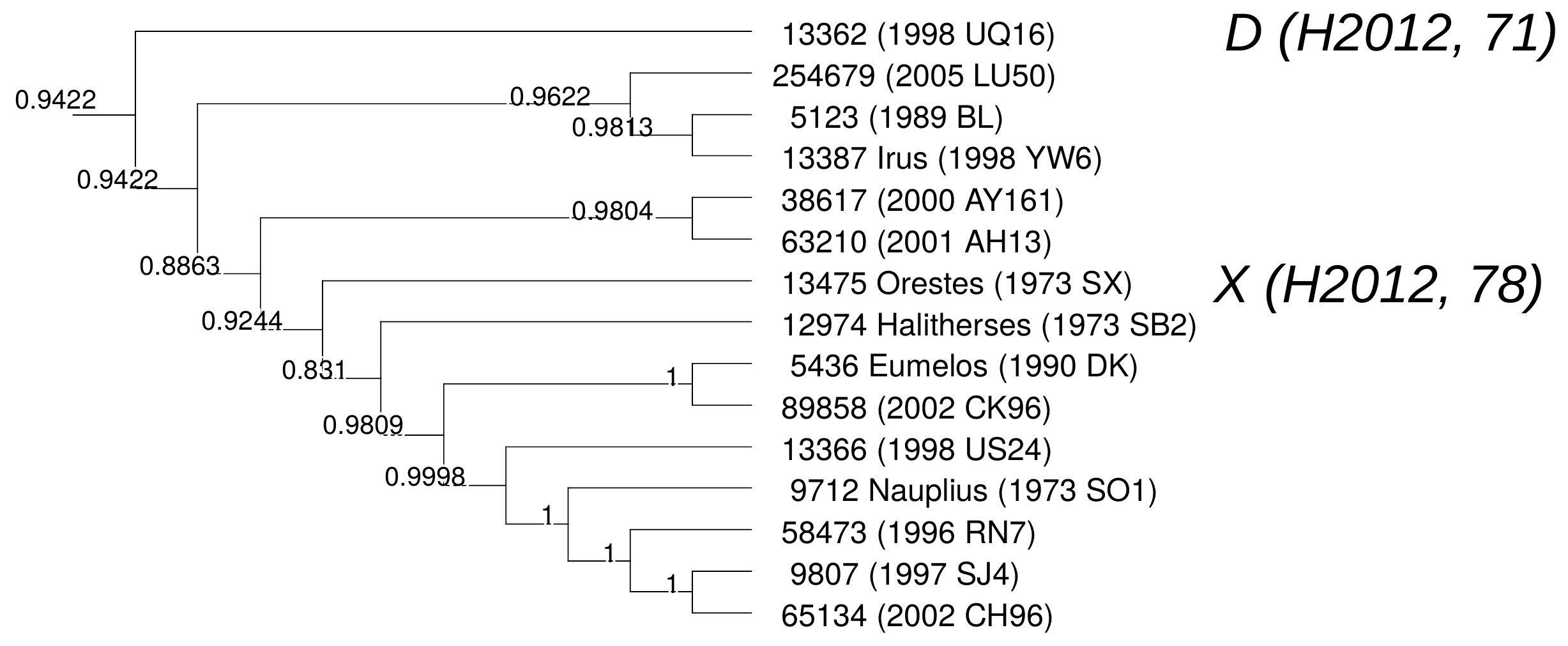}}\hfill
\subfloat[Idomeneus clan.\label{fig:Idomeneus}]{\includegraphics[width=0.43\textwidth]{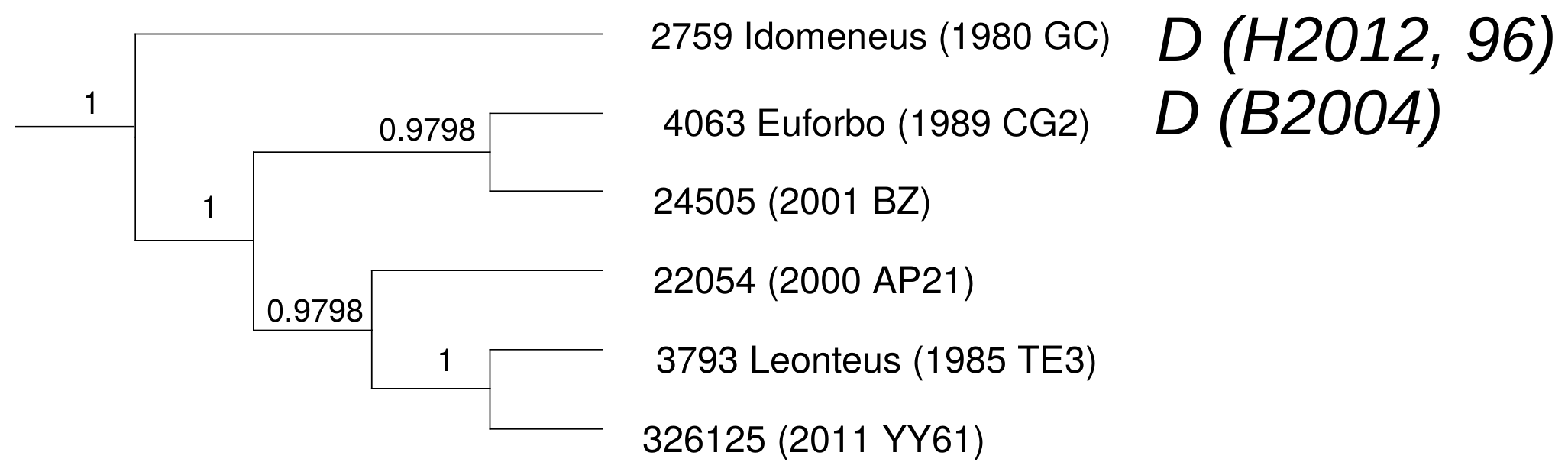}}\hfill
\subfloat[Periphas clan.\label{fig:Periphas}] {\includegraphics[width=0.43\textwidth]{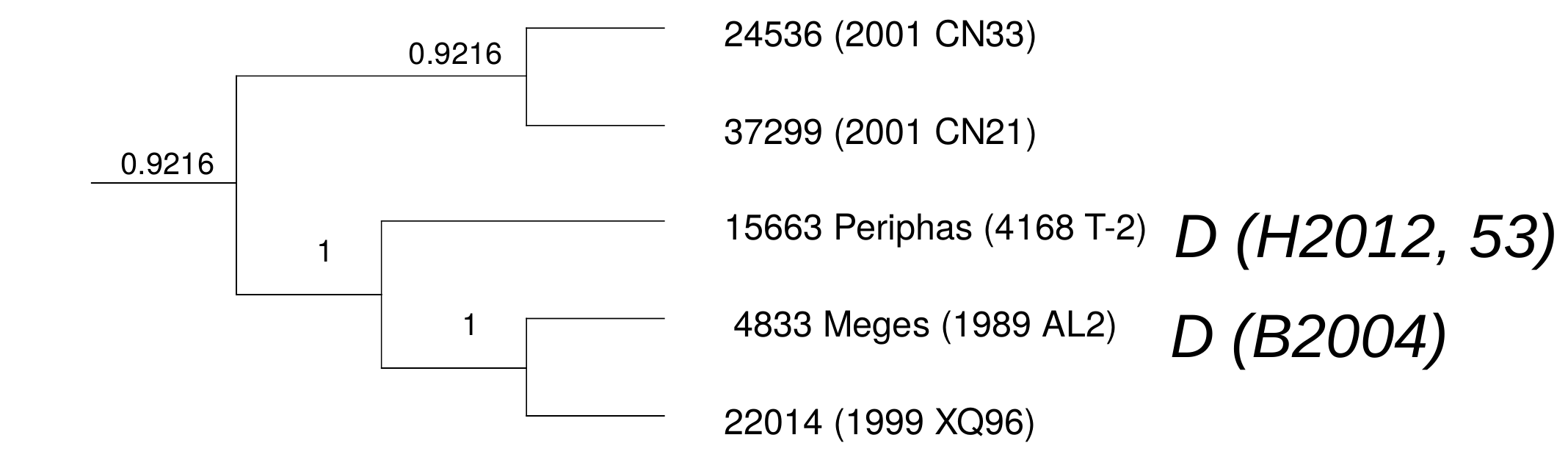}}\hfill
\subfloat[Polypoites clan.\label{fig:Polypoites}]{\includegraphics[width=0.43\textwidth]{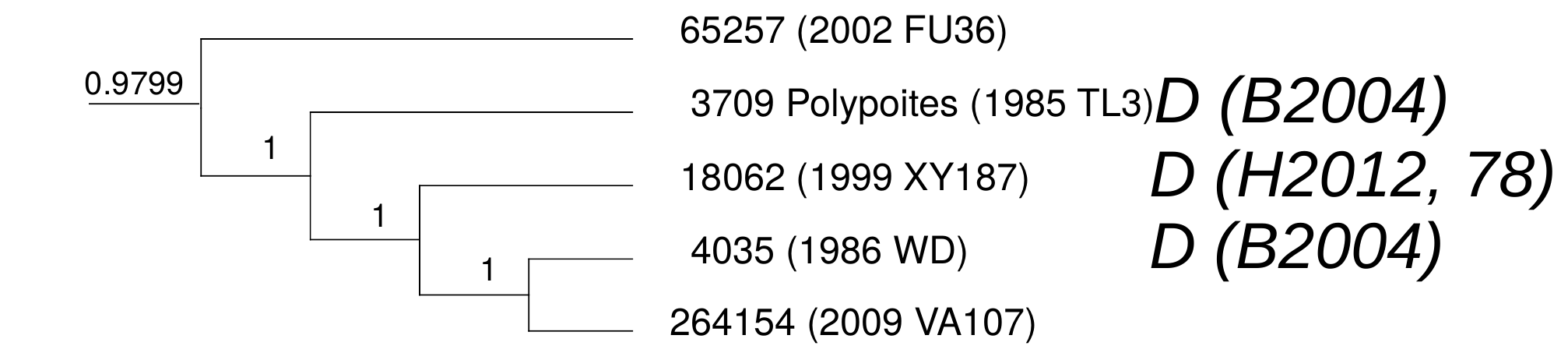}}\hfill
\subfloat[Stentor clan.\label{fig:Stentor}]{\includegraphics[width=0.43\textwidth]{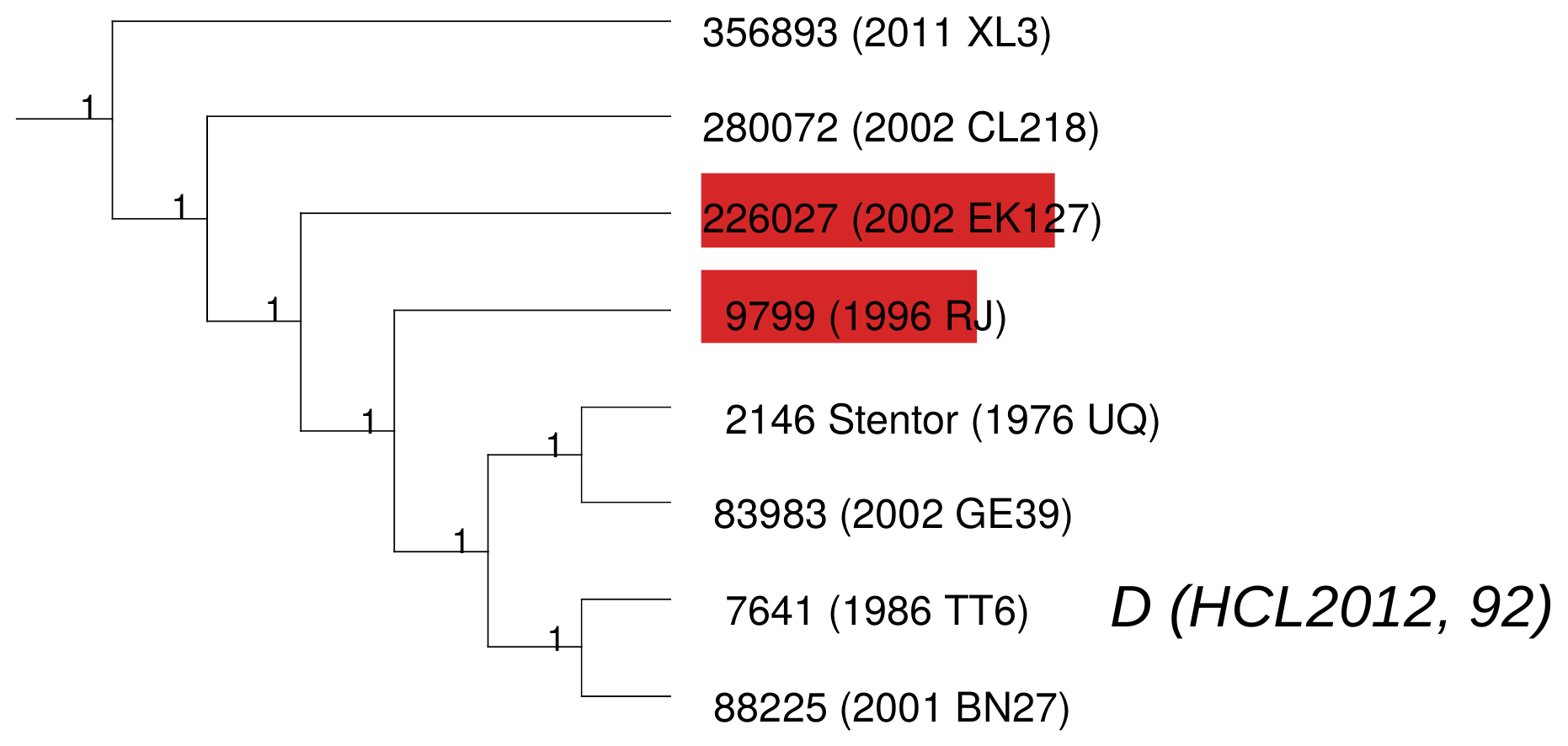}}\hfill
\subfloat[Thersander clan.\label{fig:Thersander}] {\includegraphics[width=0.43\textwidth]{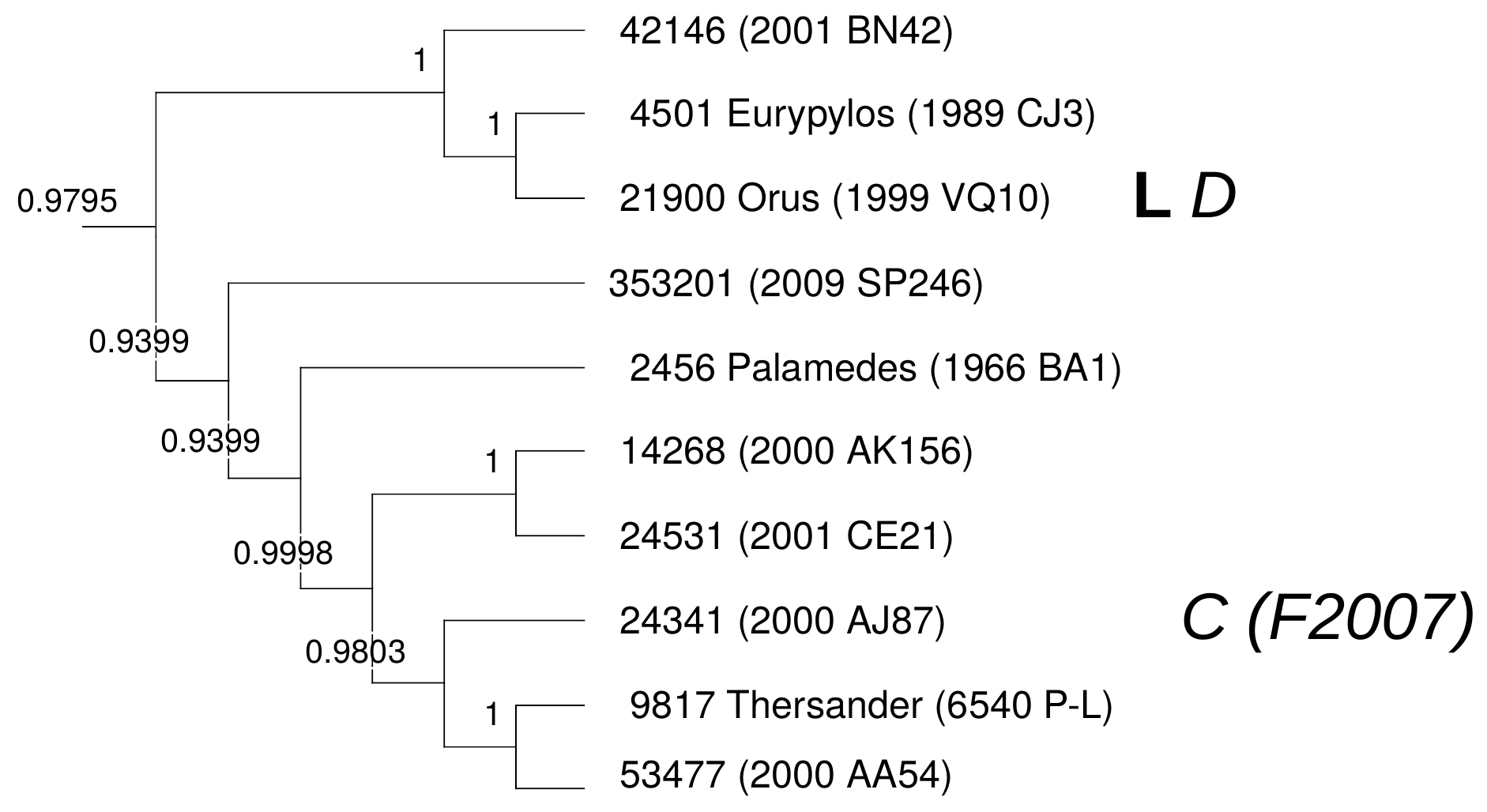}}\hfill
\subfloat[Ulysses clan.\label{fig:Ulysses}]{\includegraphics[width=0.43\textwidth]{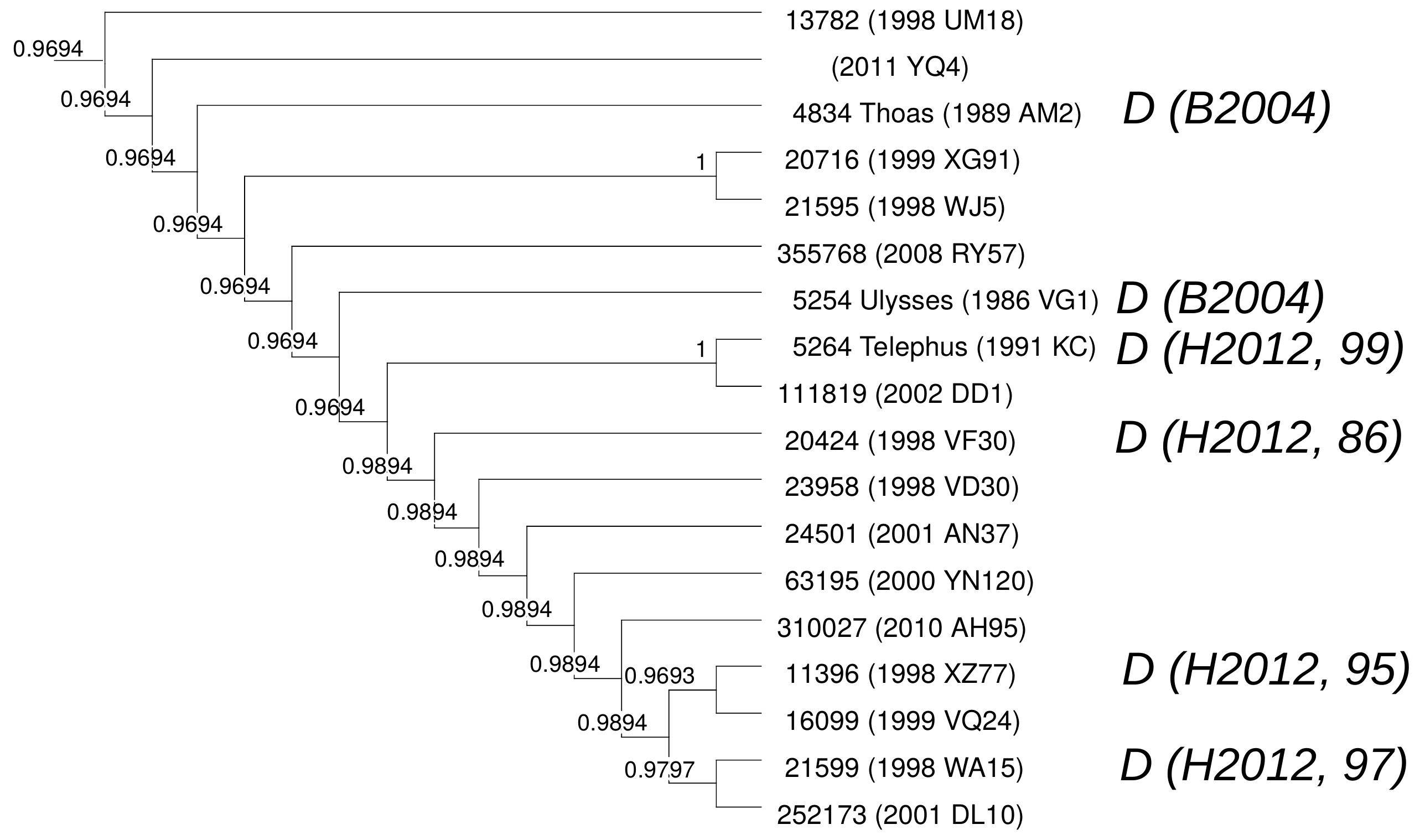}}
\caption{Consensus trees of L4 Trojans that are not associated with any superclan. Numbers indicate fraction of 10000 trees where branch is present. \textit{Letters} associate objects with Bus-Demeo taxonomy \citep{Bus2002AsteroidTax, DeMeo2009AsteroidTax}, classified by associated reference T1989: \citet{Tholen1989Taxonomy}; B2004: \citet{Bendjoya2004JTSpectra}; F2007: \citet{Fornasier2007VisSpecTrojans}; H2012, with associated confidence rating: \citet{Hasselmann2012SDSSTaxonomy}. \textbf{L} indicates objects to be visited by the \textit{Lucy} spacecraft \citep{Levison2017Lucy}. Red highlights are members of the 1996 RJ collisional family.} \label{fig:UnassL4clans}
\end{figure*}

\begin{figure*}
	\includegraphics[width=0.95\textwidth]{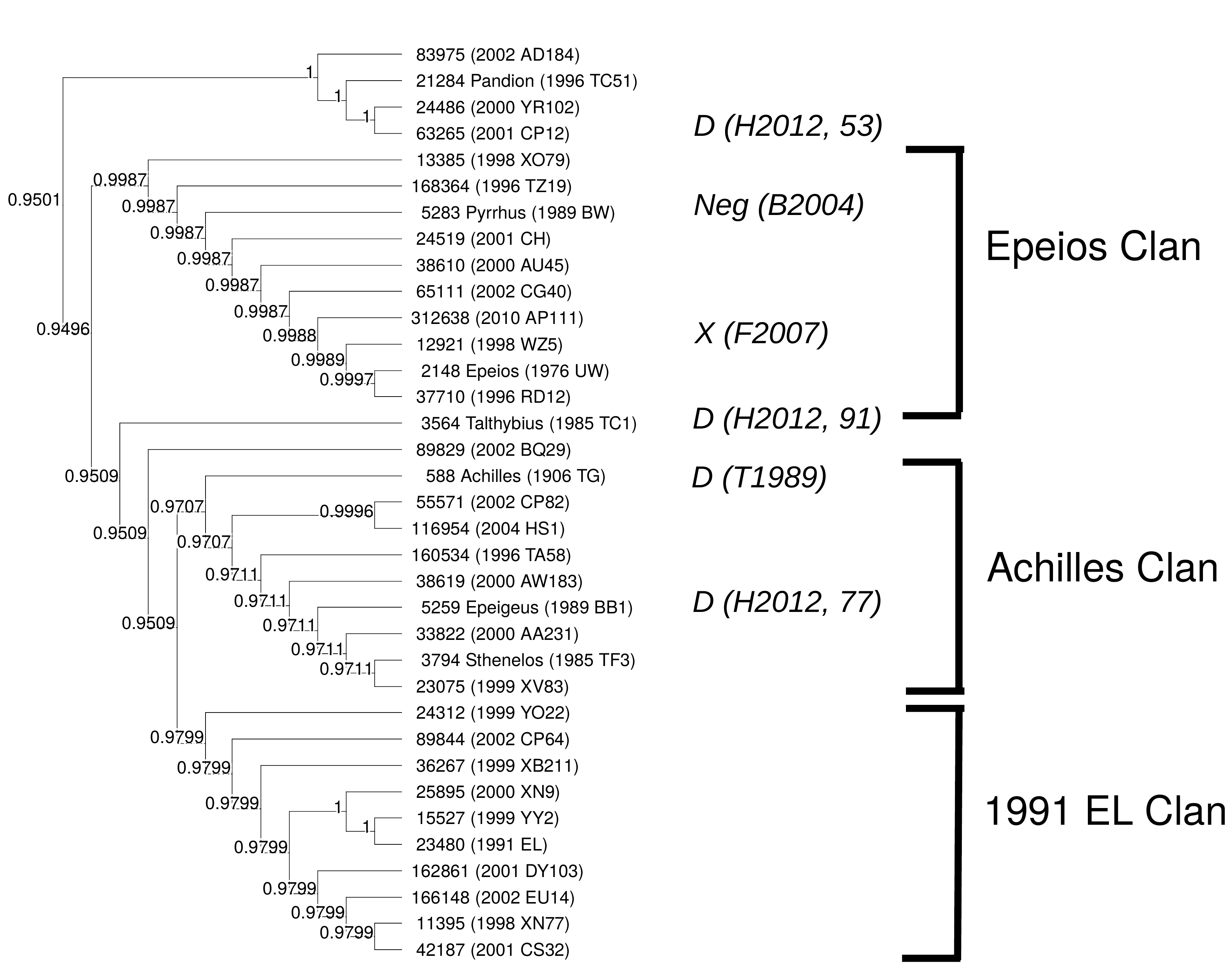}
    \caption{Consensus tree of the L4 Greater Achilles superclan, including Epeios, achilles and 1991 EL clans. Numbers indicate fraction of 10000 trees where branch is present. \textit{Letters} associate objects with Bus-Demeo taxonomy \citep{Bus2002AsteroidTax, DeMeo2009AsteroidTax}, classified by associated reference T1989: \citet{Tholen1989Taxonomy}; B2004: \citet{Bendjoya2004JTSpectra}; F2007: \citet{Fornasier2007VisSpecTrojans}; H2012, with associated confidence rating: \citet{Hasselmann2012SDSSTaxonomy}.}
    \label{Fig:greaterAchilles}
\end{figure*}

\begin{figure*}
	\includegraphics[width=0.95\textwidth]{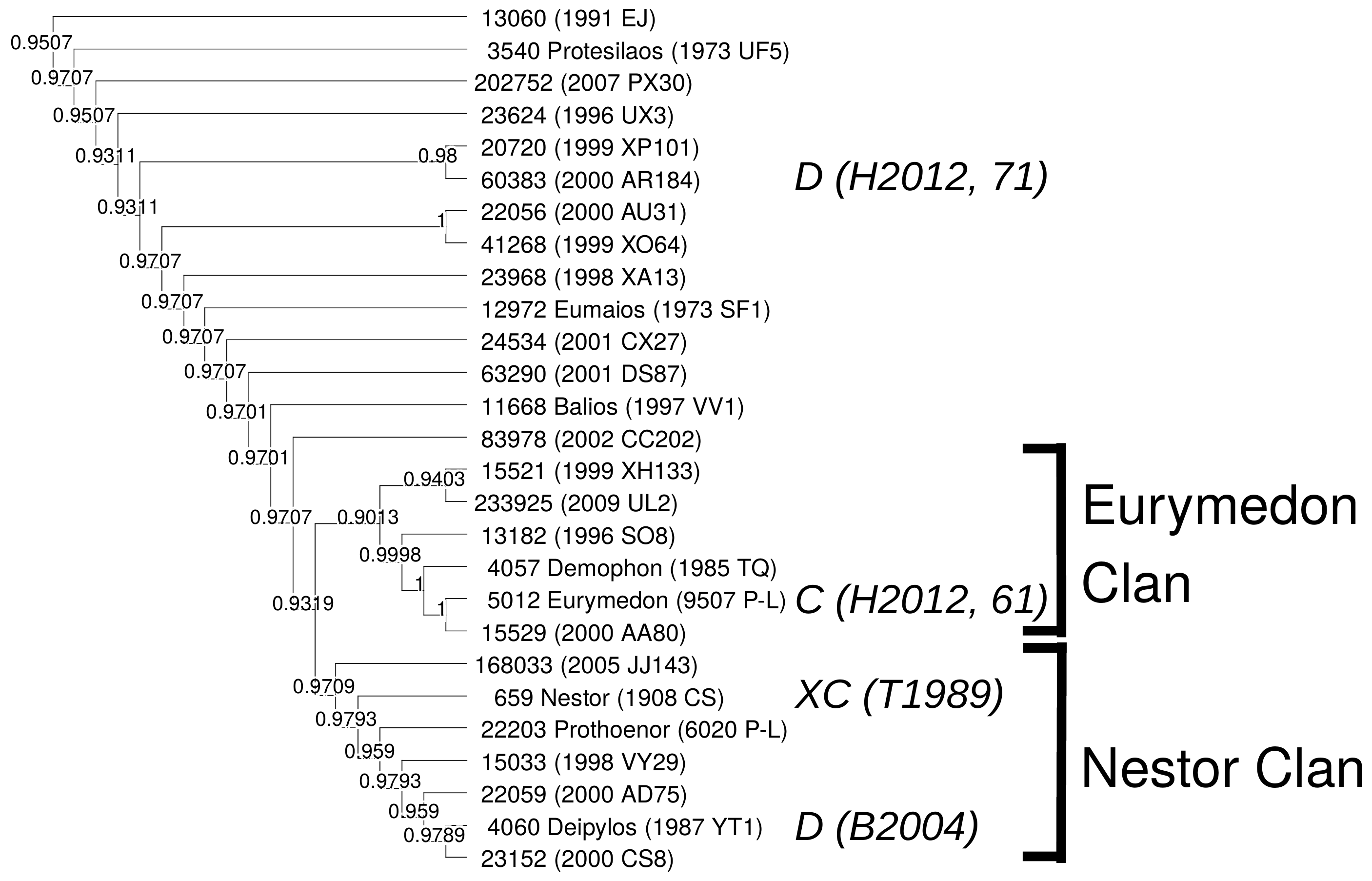}
    \caption{Consensus tree of the L4 Greater Nestor superclan, including Eurymedon and Nestor clans. Numbers indicate fraction of 10000 trees where branch is present. \textit{Letters} associate objects with Bus-Demeo taxonomy \citep{Bus2002AsteroidTax, DeMeo2009AsteroidTax}, classified by associated reference T1989: \citet{Tholen1989Taxonomy}; B2004: \citet{Bendjoya2004JTSpectra}; F2007: \citet{Fornasier2007VisSpecTrojans}; H2012, with associated confidence rating: \citet{Hasselmann2012SDSSTaxonomy}.}
    \label{Fig:greaterNestor}
\end{figure*}

\begin{figure*}
	\includegraphics[width=0.95\textwidth]{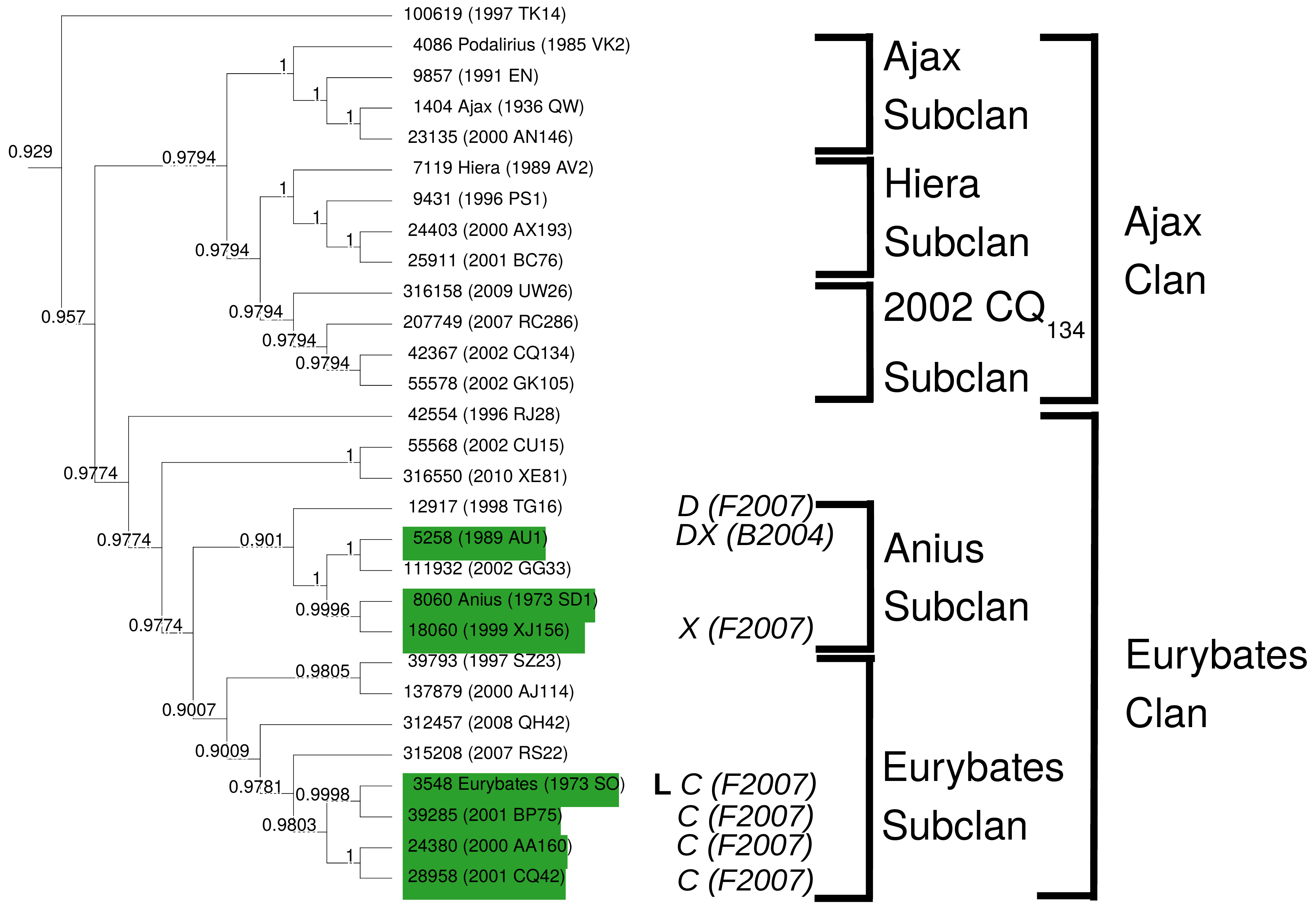}
    \caption{Consensus tree ofthe L4 Greater Ajax superclan, including Ajax and Eurybates clans. This is a duplicate of Fig. \ref{Fig:greaterAjax}, and is included here for completeness. Numbers indicate fraction of 10000 trees where branch is present. \textit{Letters} associate objects with Bus-Demeo taxonomy \citep{Bus2002AsteroidTax, DeMeo2009AsteroidTax}, classified by associated reference T1989: \citet{Tholen1989Taxonomy}; B2004: \citet{Bendjoya2004JTSpectra}; F2007: \citet{Fornasier2007VisSpecTrojans}; H2012, with associated confidence rating: \citet{Hasselmann2012SDSSTaxonomy}.. \textbf{L} indicates objects to be visited by the \textit{Lucy} spacecraft \citep{Levison2017Lucy}. Green highlights are members of the Eurybates collisional family.}
    \label{Fig:greaterAjaxApx}
\end{figure*}

\begin{figure*}
	\includegraphics[width=0.95\textwidth]{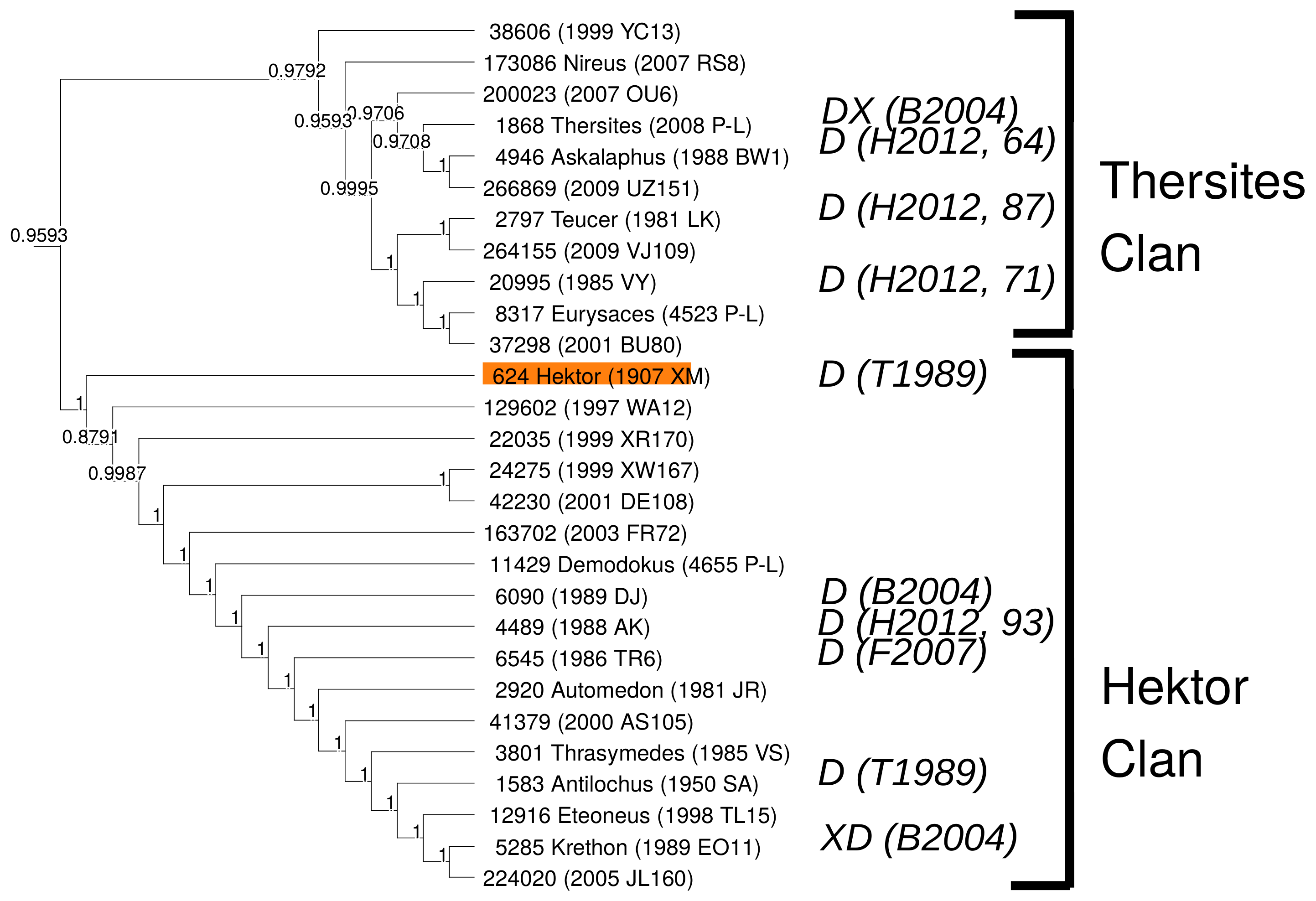}
    \caption{Consensus tree of the L4 Greater Hektor superclan, including Thersites and Hektor clans. Numbers indicate fraction of 10000 trees where branch is present. \textit{Letters} associate objects with Bus-Demeo taxonomy \citep{Bus2002AsteroidTax, DeMeo2009AsteroidTax}, classified by associated reference T1989: \citet{Tholen1989Taxonomy}; B2004: \citet{Bendjoya2004JTSpectra}; F2007: \citet{Fornasier2007VisSpecTrojans}; H2012, with associated confidence rating: \citet{Hasselmann2012SDSSTaxonomy}.. Orange highlights are members of the Hektor collisional family.}
    \label{Fig:greaterHektor}
\end{figure*}

\begin{figure*}
	\includegraphics[width=0.95\textwidth]{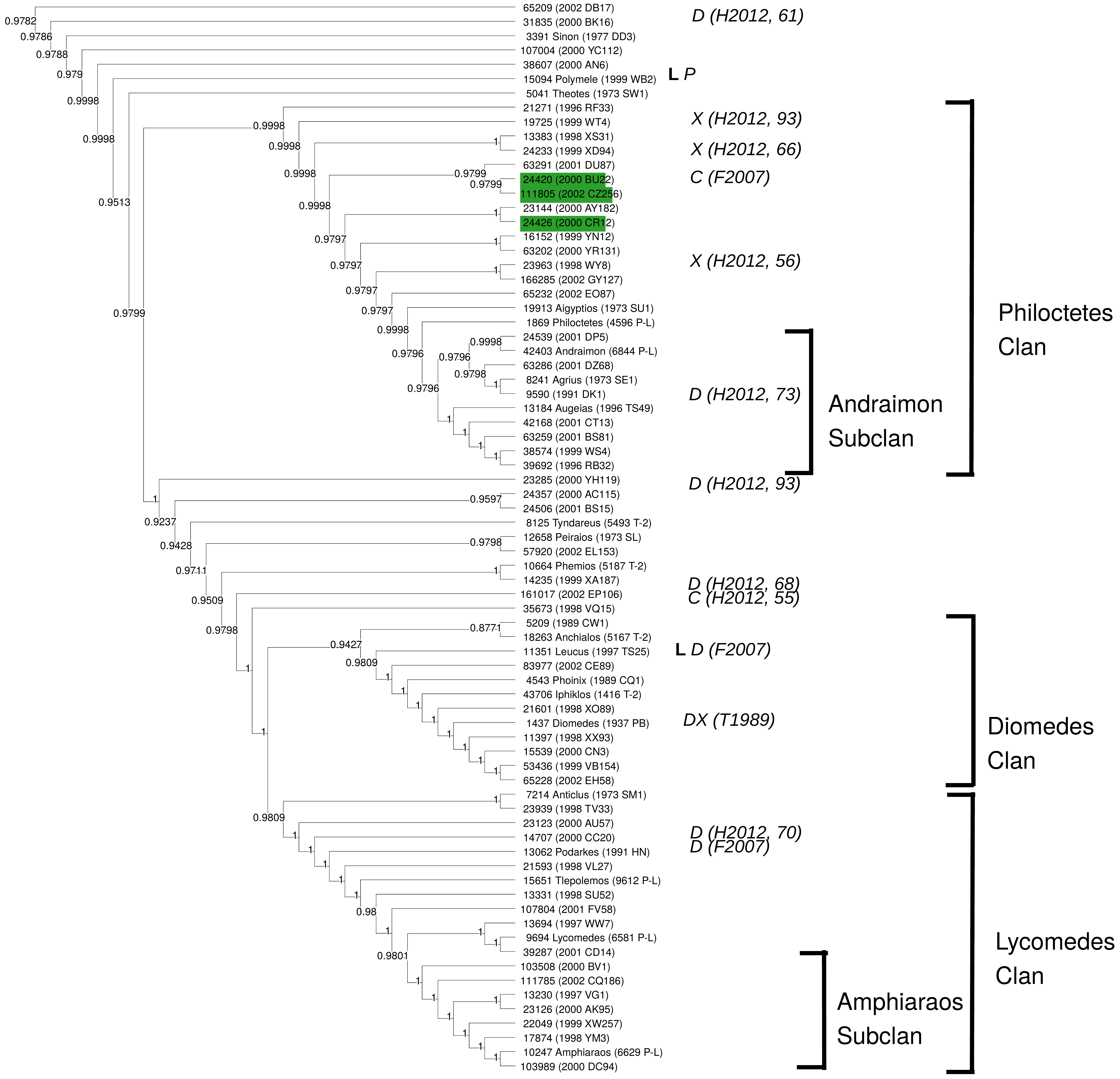}
    \caption{Consensus tree of the L4 Greater Diomedes superclan, including Philoctetes, Diomedes, and Lycomedes clans. Numbers indicate fraction of 10000 trees where branch is present. \textit{Letters} associate objects with Bus-Demeo taxonomy \citep{Bus2002AsteroidTax, DeMeo2009AsteroidTax}, classified by associated reference T1989: \citet{Tholen1989Taxonomy}; B2004: \citet{Bendjoya2004JTSpectra}; F2007: \citet{Fornasier2007VisSpecTrojans}; H2012, with associated confidence rating: \citet{Hasselmann2012SDSSTaxonomy}.. \textbf{L} indicates objects to be visited by the \textit{Lucy} spacecraft \citep{Levison2017Lucy}. Green highlights are members of the Eurybates collisional family.}
    \label{Fig:greaterDiomedes}
\end{figure*}

\begin{figure*}
	\includegraphics[width=0.95\textwidth]{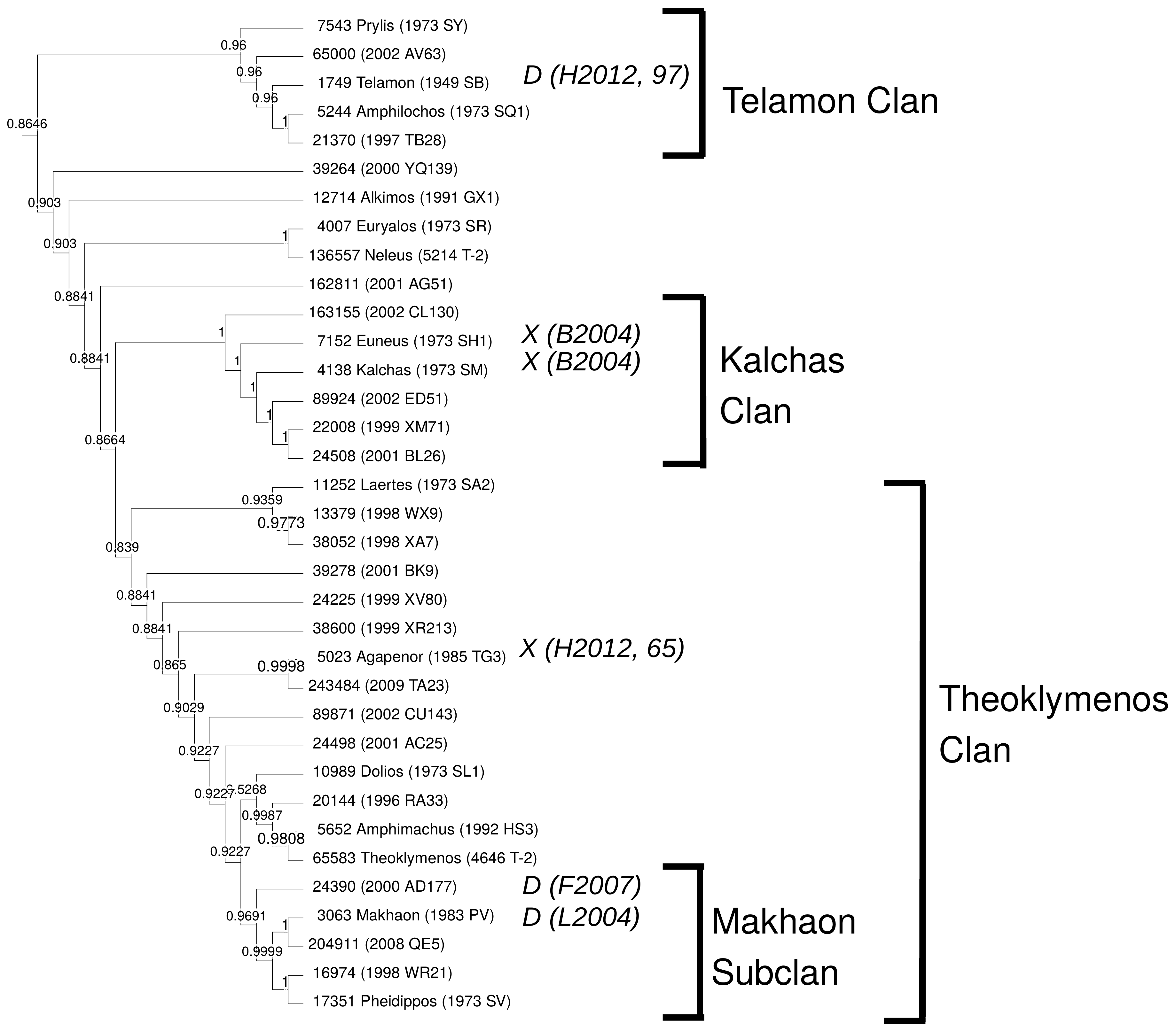}
    \caption{Consensus tree of the L4 Greater Telamon superclan, including Telamon, Kalchas, and Theoklymenos clans. Numbers indicate fraction of 10000 trees where branch is present. \textit{Letters} associate objects with Bus-Demeo taxonomy \citep{Bus2002AsteroidTax, DeMeo2009AsteroidTax}, classified by associated reference T1989: \citet{Tholen1989Taxonomy}; B2004: \citet{Bendjoya2004JTSpectra}; L2004: \citet{Lazzaro2004s3os2asteroids}; F2007: \citet{Fornasier2007VisSpecTrojans}; H2012, with associated confidence rating: \citet{Hasselmann2012SDSSTaxonomy}.}
    \label{Fig:greaterTelamon}
\end{figure*}

\begin{figure*}
	\includegraphics[width=0.95\textwidth]{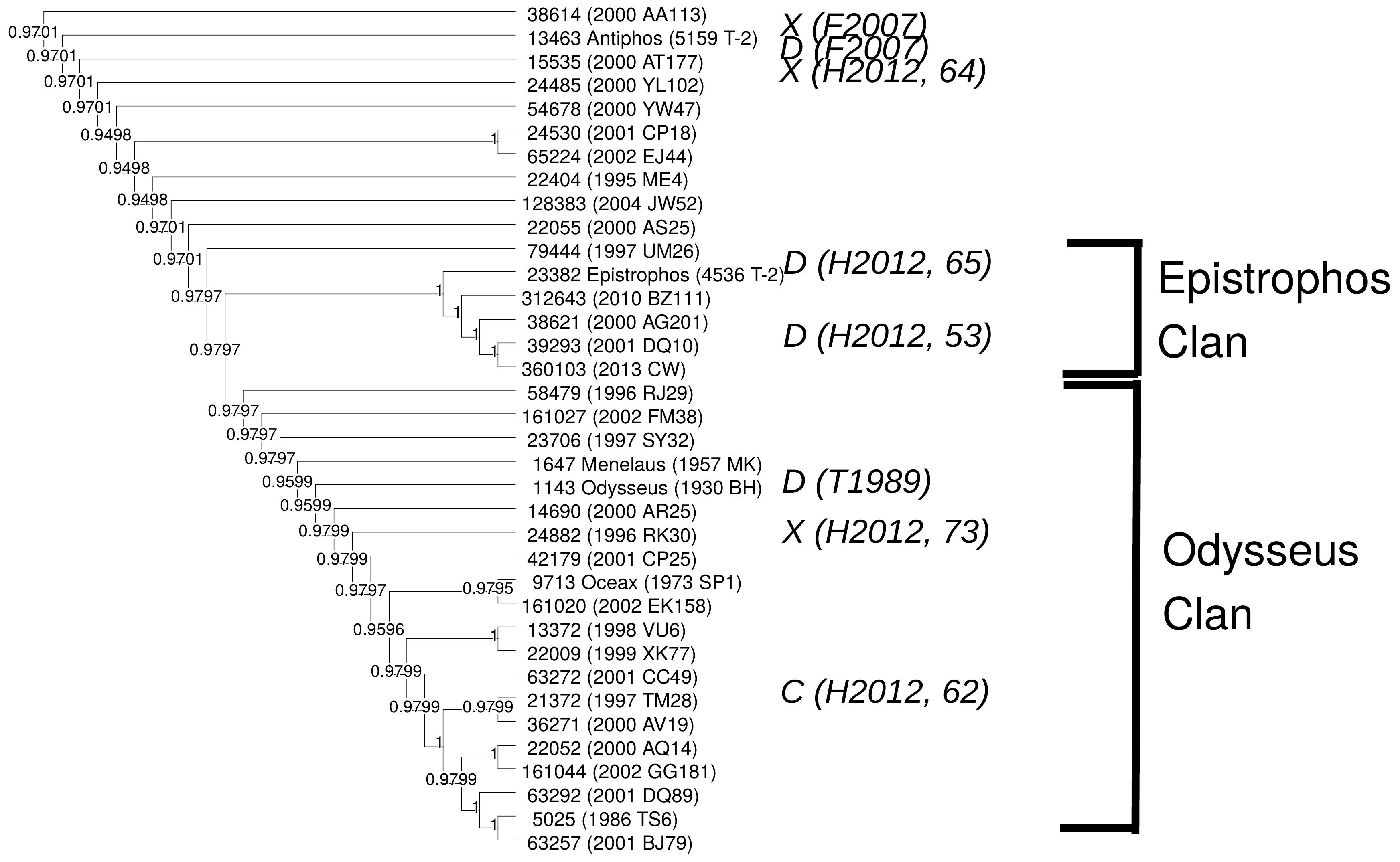}
    \caption{Consensus tree of the L4 Greater Odysseus superclan, including Epistrophos and Odysseus clans. Numbers indicate fraction of 10000 trees where branch is present. \textit{Letters} associate objects with Bus-Demeo taxonomy \citep{Bus2002AsteroidTax, DeMeo2009AsteroidTax}, classified by associated reference T1989: \citet{Tholen1989Taxonomy}; B2004: \citet{Bendjoya2004JTSpectra}; F2007: \citet{Fornasier2007VisSpecTrojans}; H2012, with associated confidence rating: \citet{Hasselmann2012SDSSTaxonomy}..}
    \label{Fig:greaterOdysseus}
\end{figure*}

\begin{figure*}
\centering
\subfloat[1990VU1 clan.\label{fig:1990VU1}]{\includegraphics[width=0.45\textwidth]{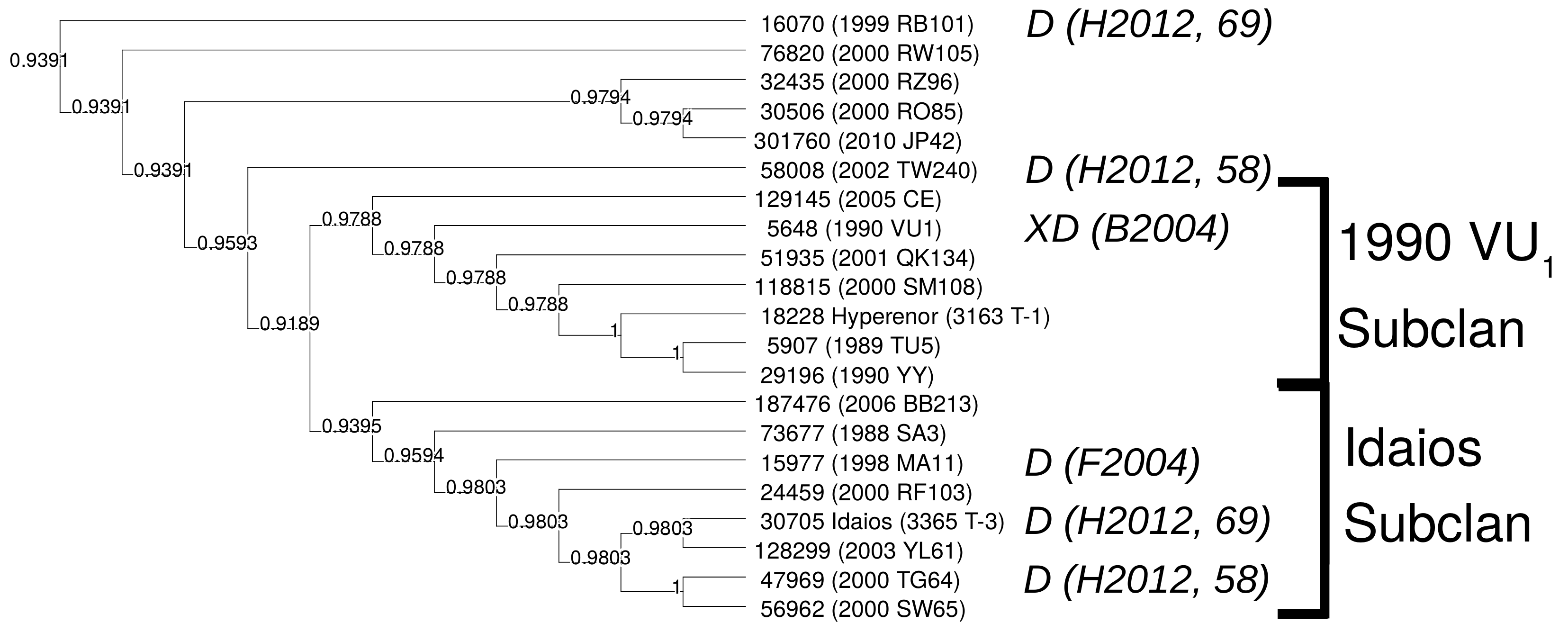}}\hfill
\subfloat[1999RU12 clan.\label{fig:1999RU12}] {\includegraphics[width=0.45\textwidth]{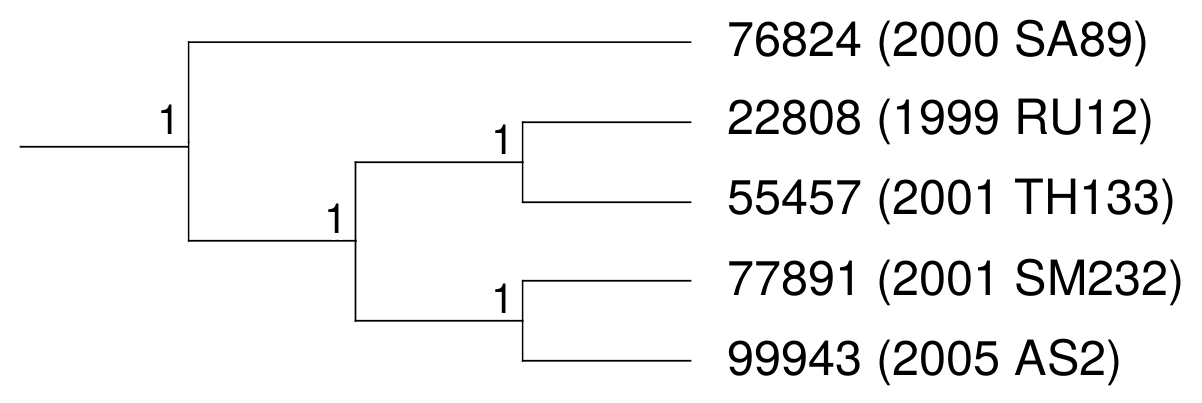}}\hfill
\subfloat[Anchises clan.\label{fig:Anchises}]{\includegraphics[width=0.45\textwidth]{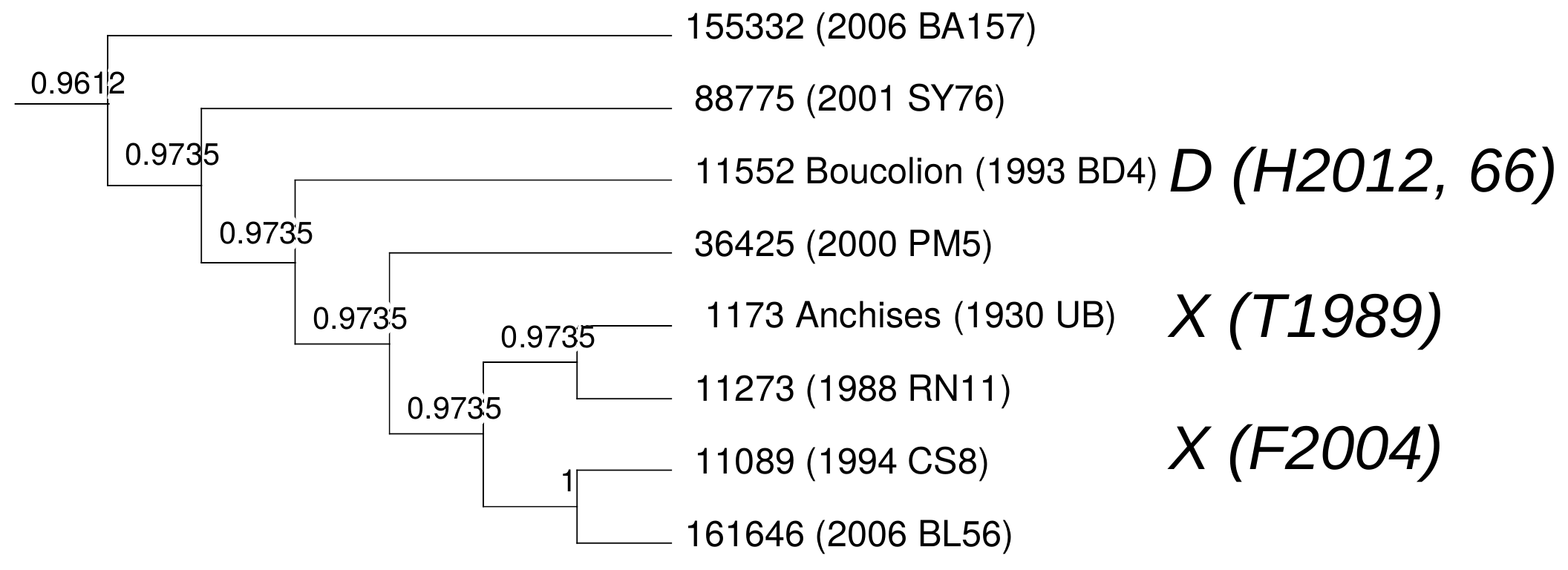}}\hfill
\subfloat[Apisaon clan.\label{fig:Apisaon}]{\includegraphics[width=0.45\textwidth]{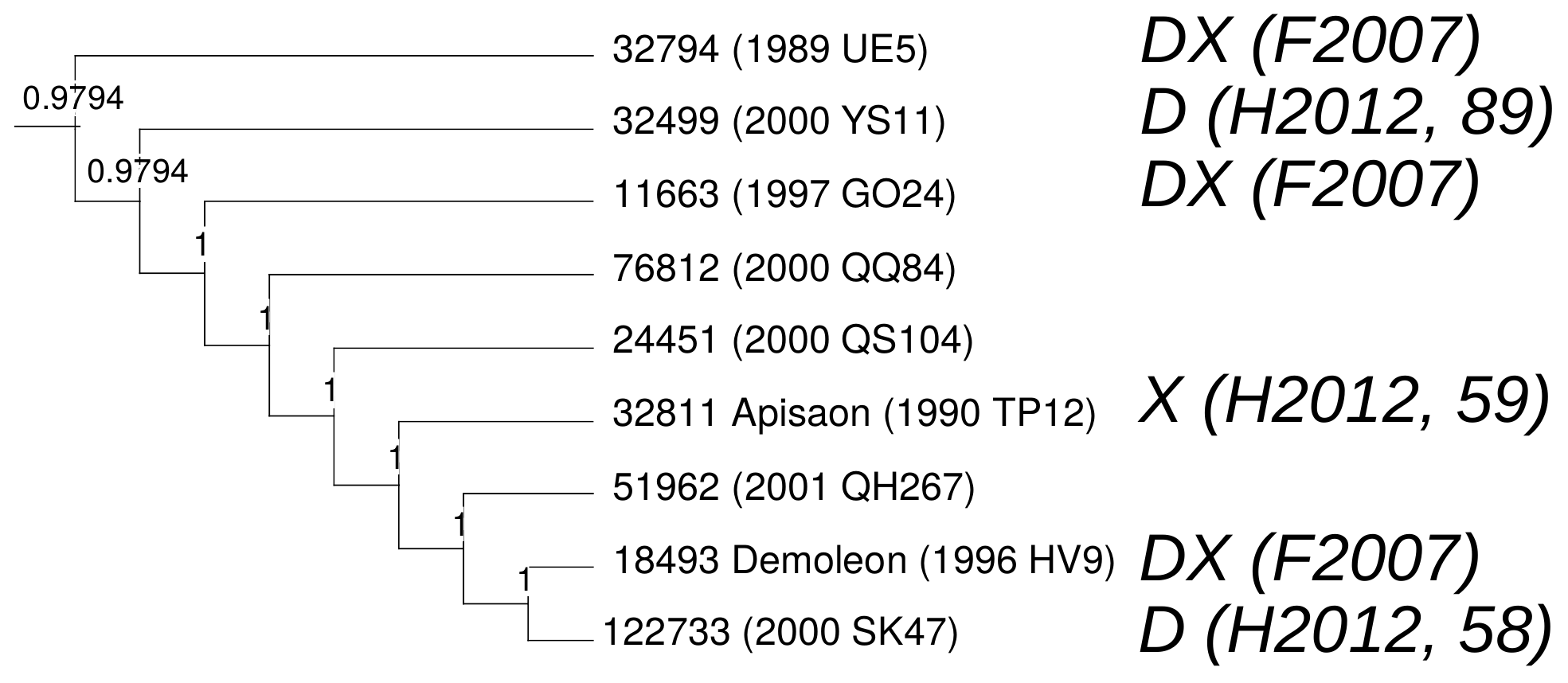}}\hfill
\subfloat[Asteropaios clan.\label{fig:Asteropaios}]{\includegraphics[width=0.45\textwidth]{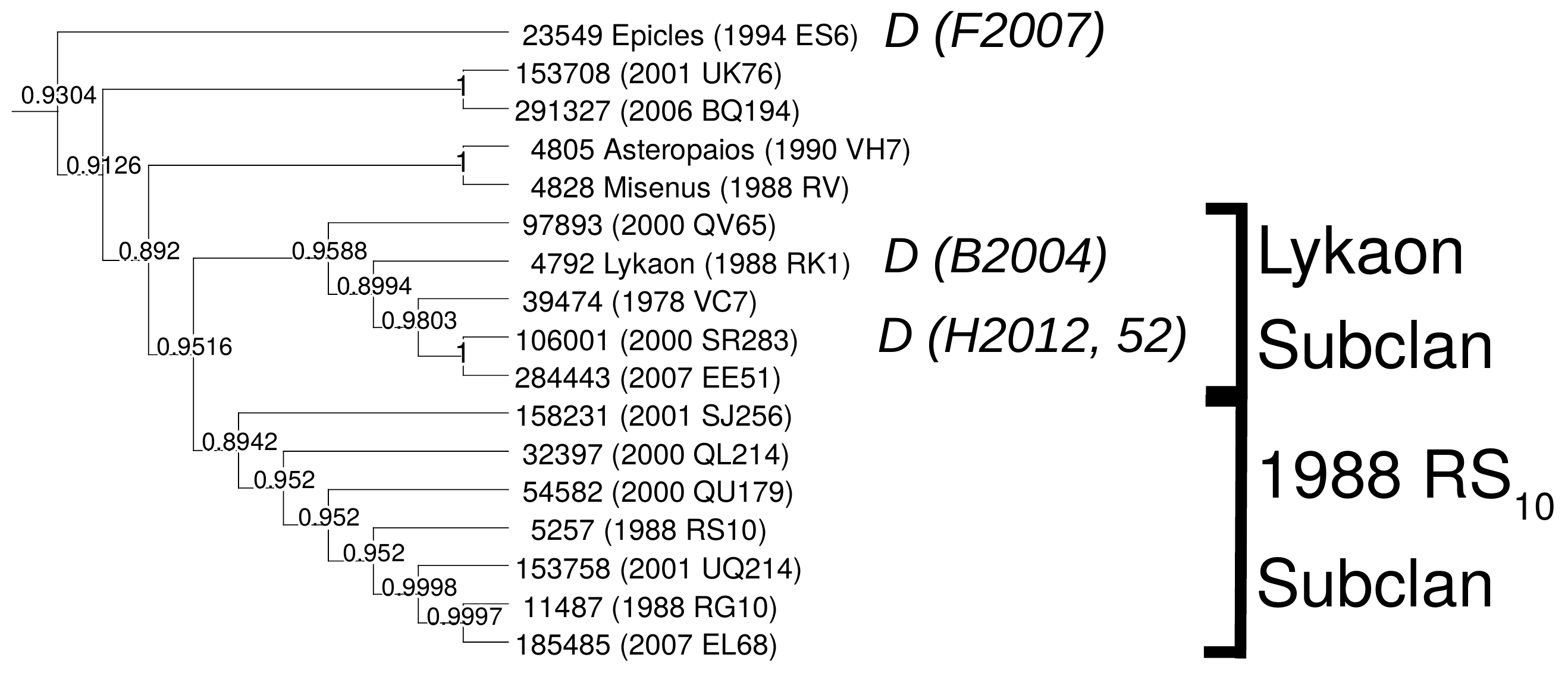}}\hfill
\subfloat[Dolon clan.\label{fig:Dolon}]{\includegraphics[width=0.45\textwidth]{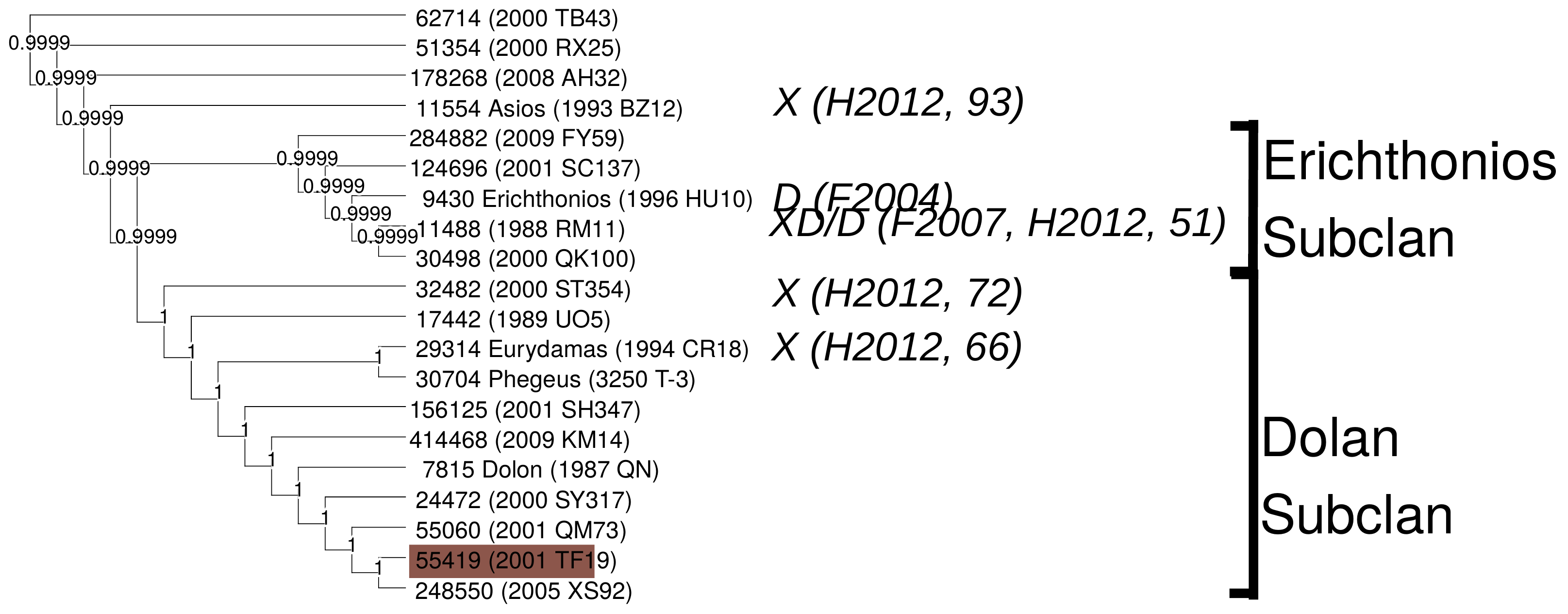}}\hfill
\subfloat[Khryses clan.\label{fig:Khryses}]{\includegraphics[width=0.45\textwidth]{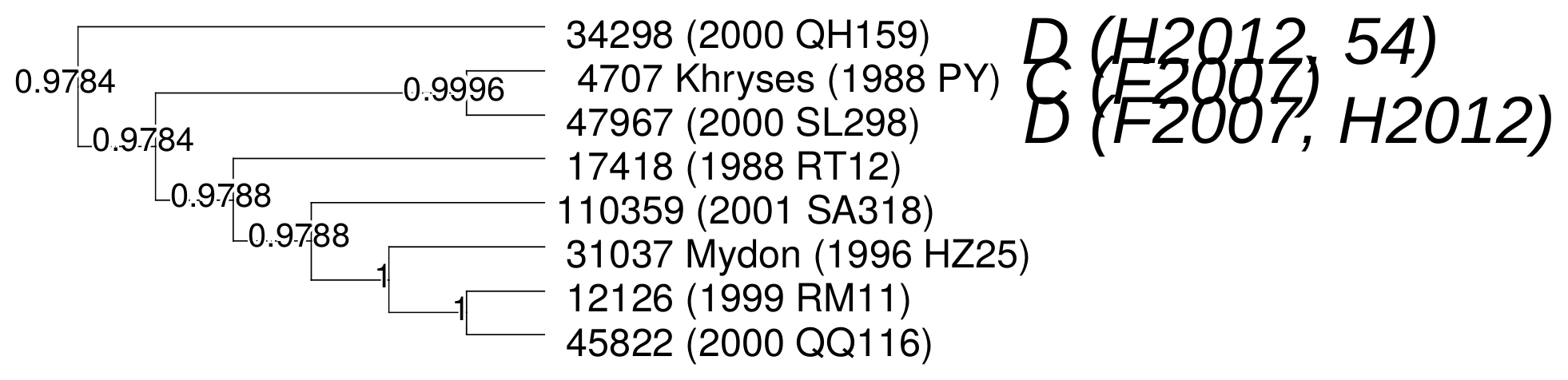}}
\caption{Consensus trees of L5 Trojan clans that are not associated with any superclan. Numbers indicate fraction of 10000 trees where branch is present. \textit{Letters} associate objects with Bus-Demeo taxonomy \citep{Bus2002AsteroidTax, DeMeo2009AsteroidTax}, classified by associated reference T1989: \citet{Tholen1989Taxonomy}; B2004: \citet{Bendjoya2004JTSpectra}; F2007: \citet{Fornasier2007VisSpecTrojans}; H2012, with associated confidence rating: \citet{Hasselmann2012SDSSTaxonomy}.. \textbf{L} indicates objects to be visited by the \textit{Lucy} spacecraft \citep{Levison2017Lucy}. Brown highlights are members of the Ennomos collisional family. } \label{fig:UnassL5clans}
\end{figure*}

\begin{figure*}
	\includegraphics[height=0.89\textheight]{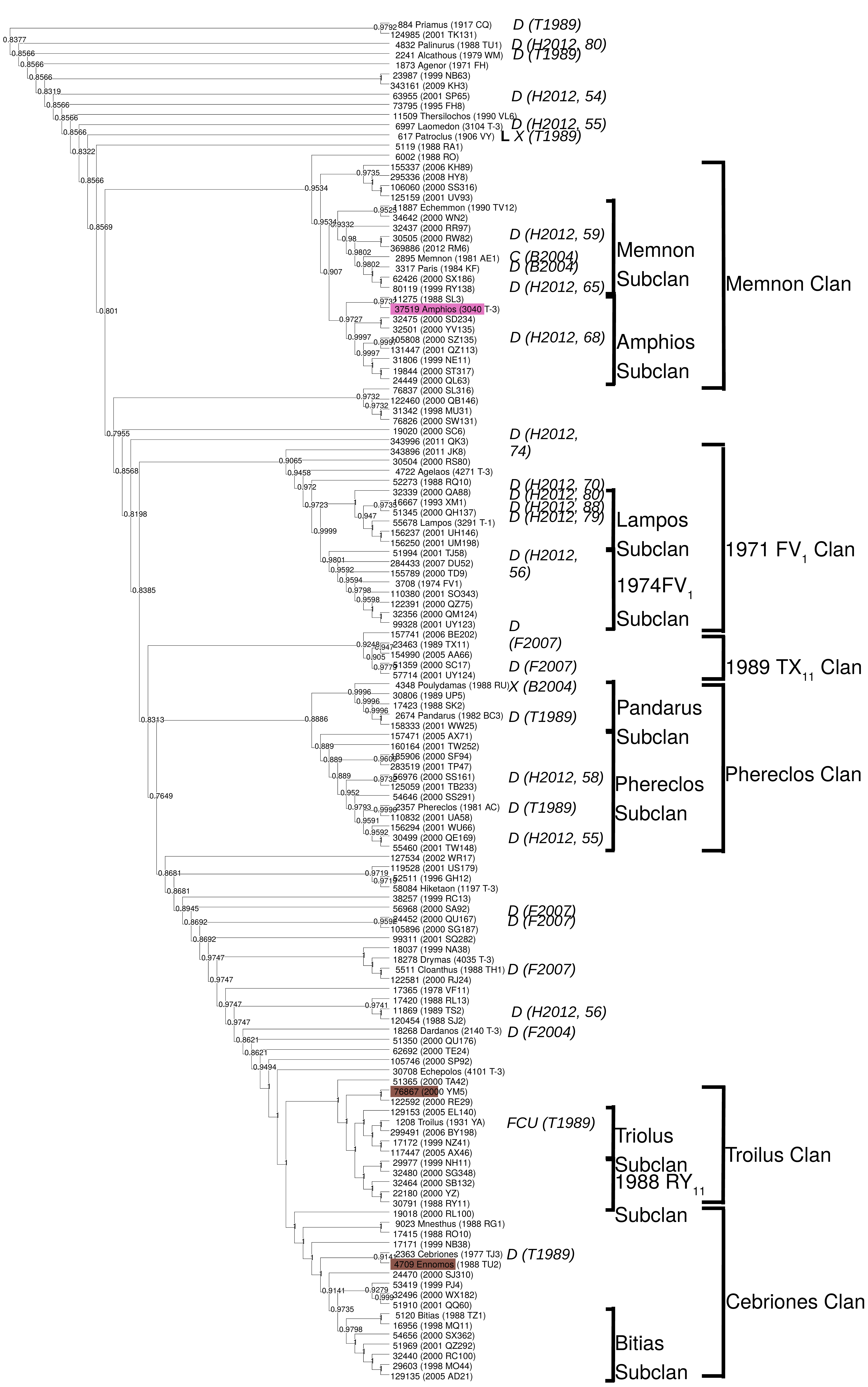}
    \caption{Consensus trees of the L5 Greater Patroclus superclan, including Memnon, 1971 FV$_1$, 1989 TX$_{11}$, Phereclos, Trollus and Cebriones clans. Numbers indicate fraction of 10000 trees where branch is present. \textit{Letters} associate objects with Bus-Demeo taxonomy \citep{Bus2002AsteroidTax, DeMeo2009AsteroidTax}, classified by associated reference T1989: \citet{Tholen1989Taxonomy}; B2004: \citet{Bendjoya2004JTSpectra}; F2007: \citet{Fornasier2007VisSpecTrojans}; H2012, with associated confidence rating: \citet{Hasselmann2012SDSSTaxonomy}.. \textbf{L} indicates objects to be visited by the \textit{Lucy} spacecraft \citep{Levison2017Lucy}. Brown and Purple highlights are members of the Ennomos and 2001 UV$_{209}$ collisional families respectively.}
    \label{Fig:greaterPatroclus}
\end{figure*}

\begin{figure*}
	\includegraphics[width=0.95\textwidth]{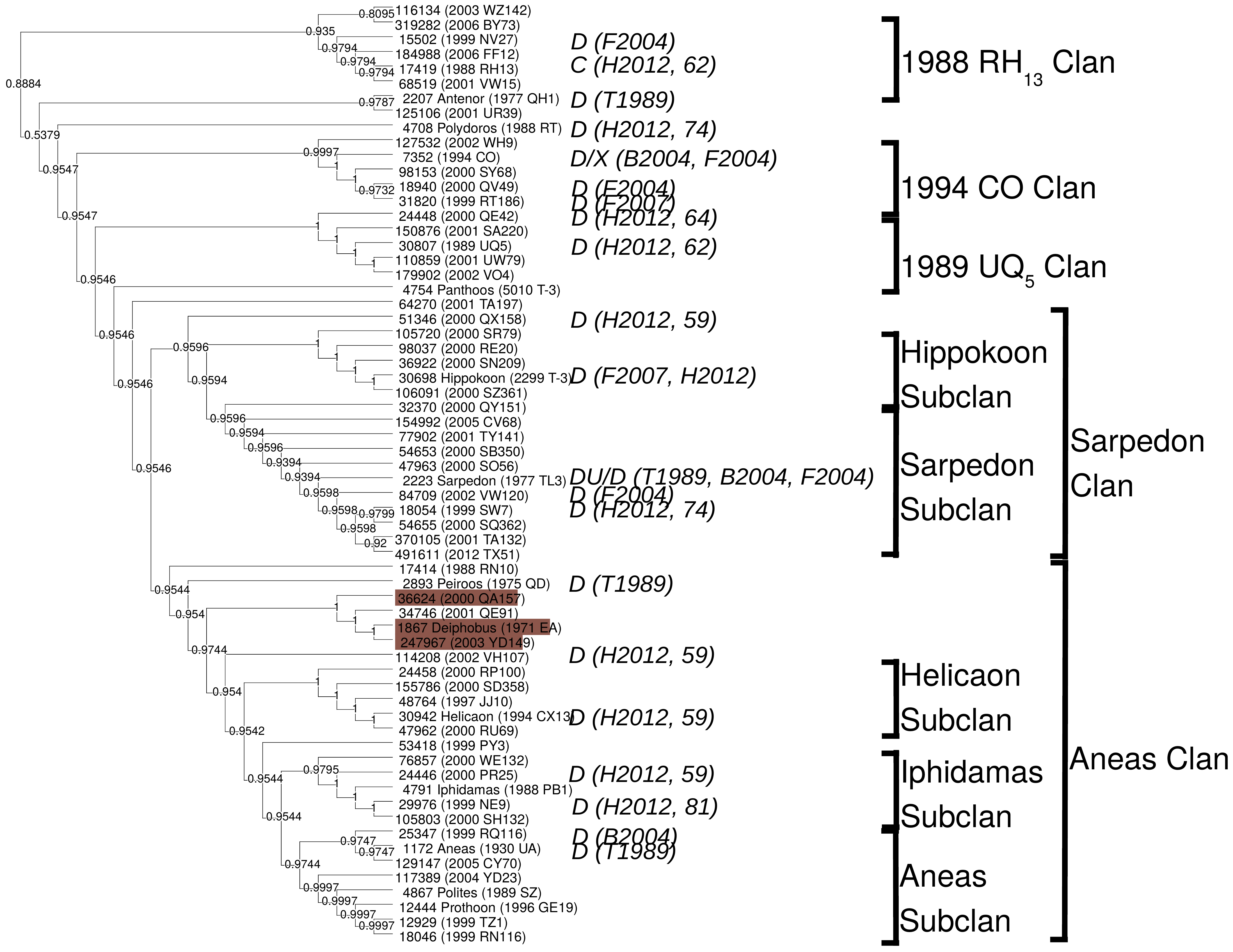}
    \caption{Consensus trees of the L5 Greater Aneas superclan, including 1988 RH$_{13}$, 1994 CO, 1989 UQ$_5$, Sarpedon and Aneas clans. Numbers indicate fraction of 10000 trees where branch is present. \textit{Letters} associate objects with Bus-Demeo taxonomy \citep{Bus2002AsteroidTax, DeMeo2009AsteroidTax}, classified by associated reference T1989: \citet{Tholen1989Taxonomy}; B2004: \citet{Bendjoya2004JTSpectra}; F2007: \citet{Fornasier2007VisSpecTrojans}; H2012, with associated confidence rating: \citet{Hasselmann2012SDSSTaxonomy}.. Brown highlights are members of the Ennomos collisional family.}
    \label{Fig:greaterAneas}
\end{figure*}

\begin{figure*}
	\includegraphics[width=0.95\textwidth]{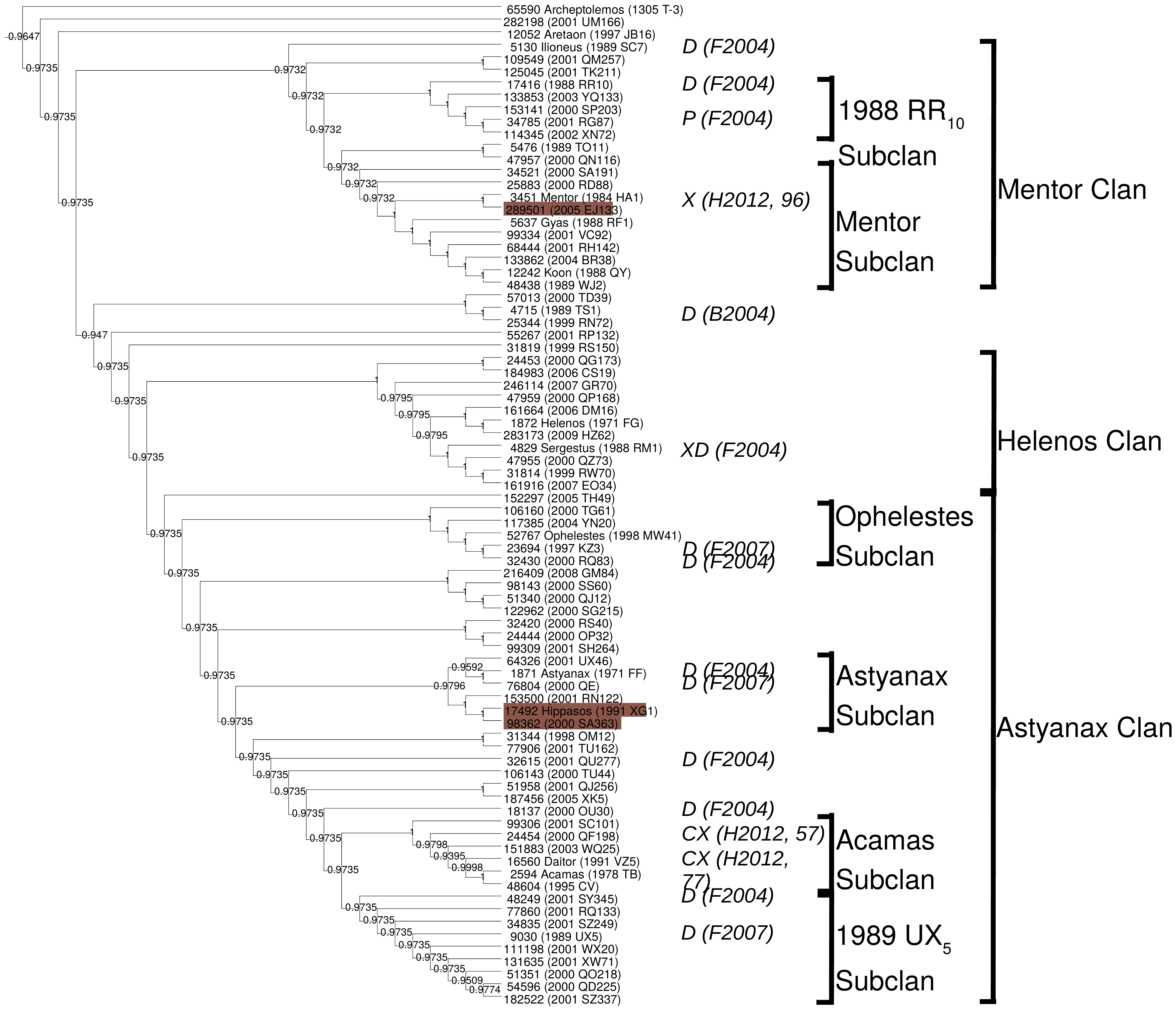}
    \caption{Consensus trees of the L5 Greater Astyanax superclan, including Mentor, Helenos, and Astyanax clans. Numbers indicate fraction of 10000 trees where branch is present. \textit{Letters} associate objects with Bus-Demeo taxonomy \citep{Bus2002AsteroidTax, DeMeo2009AsteroidTax}, classified by associated reference T1989: \citet{Tholen1989Taxonomy}; B2004: \citet{Bendjoya2004JTSpectra}; F2007: \citet{Fornasier2007VisSpecTrojans}; H2012, with associated confidence rating: \citet{Hasselmann2012SDSSTaxonomy}.. Brown highlights are members of the Ennomos collisional family.}
    \label{Fig:greaterAstyanax}
\end{figure*}

The figures here (Figs \ref{fig:UnassL4clans}-\ref{Fig:greaterAstyanax}) show each of the separate superclans, along with the L4 unassociated clans (Fig. \ref{fig:UnassL4clans}) and unassociated L5 clans (Fig. \ref{fig:UnassL5clans}). These are additionally available individually from the PDS. 

We include table \ref{tab:UlyssesClan} as an example of those included in the data archive, available from the PDS. In this dataset, the dispersal velocity calculated from inverse Gauss equations, see Section \ref{SubSec:Method:DisVelo}, to the reference object ($\Delta V_{ref.}$) and to a fictitious cluster center ($\Delta V_{cent.}$) are given for each superclan, clan and subclan independently, for the subset of Jovian Trojans used in this analysis. 

\begin{table}
    \caption{\textbf{Ulysses clan}- $D$: Diameter of the object. From NASA-JPL HORIZONS Solar System Dynamics Database \url{https://ssd.jpl.nasa.gov/} \citep{Giorgini1996JPLSSdatabase}. Where not available, generated from \textit{H} magnitude and mean geometric albedo (0.075).; $\Delta V_{ref}$: dispersal velocity calculated from inverse Gauss equations, see Section \ref{SubSec:Method:DisVelo}, to the reference object;  $\Delta V_{cent.}$: as $\Delta V_{ref}$, with calculations to the fictitious cluster center; $F_{esc}$: Fraction e of clones that escape the Jovian Trojan population in \citep{Holt2020TrojanStability} }
    \centering
\begin{tabular}{lrrrl}
\hline
                 full\_name &    $D$ & $\Delta V_{ref}$ & $\Delta V_{cent.}$ & $F_{esc}$ \\
                            &   km   & $ms^{-1}$        & $ms^{-1}$        &          \\
 \hline
     4834 Thoas (1989 AM2) &  72.33 &             9.83 &              23.99 &  2.20E-01 \\
  5254 Ulysses (1986 VG1) &  76.15 &             0.00 &              17.81 &         - \\
  5264 Telephus (1991 KC) &  68.47 &            34.09 &              16.83 &         - \\
         11396 (1998 XZ77) &  37.11 &            33.67 &              14.19 &         - \\
         13782 (1998 UM18) &  24.97 &            13.89 &              28.86 &  8.90E-01 \\
         16099 (1999 VQ24) &  36.77 &            28.36 &              11.69 &         - \\
         20424 (1998 VF30) &  45.80 &            17.92 &               3.48 &         - \\
         20716 (1999 XG91) &  26.37 &            11.34 &               9.36 &         - \\
          21595 (1998 WJ5) &  35.18 &            12.63 &               6.26 &         - \\
         21599 (1998 WA15) &  28.31 &            48.31 &              28.04 &         - \\
         23958 (1998 VD30) &  46.00 &            18.02 &               5.02 &         - \\
         24501 (2001 AN37) &  24.54 &            17.78 &               1.21 &         - \\
        63195 (2000 YN120) &  24.69 &            35.93 &              18.19 &         - \\
         111819 (2002 DD1) &  19.34 &            17.35 &               9.23 &  3.30E-01 \\
        252173 (2001 DL10) &  15.45 &            40.86 &              20.81 &         - \\
        310027 (2010 AH95) &  11.10 &            36.22 &              16.84 &         - \\
        355768 (2008 RY57) &  11.72 &            10.25 &              10.38 &         - \\
\hline
\end{tabular}
    \label{tab:UlyssesClan}
\end{table}

%width=\columnwidth
%width=0.95\textwidth

% Figure in text

%%%%%%%%%%%%%%%%%%%%%%%%%%%%%%%%%%%%%%%%%%%%%%%%%%

% Don't change these lines
\bsp	% typesetting comment
\label{lastpage}
\end{document}